\documentclass[preprint,12pt]{elsarticle}

\usepackage{lipsum}
\usepackage{physics}
\usepackage{amsfonts,amsmath,amssymb,mathrsfs}
\usepackage{mathtools}
\usepackage{graphicx}
\usepackage{epstopdf}
\usepackage{tikz}
\usetikzlibrary{patterns,snakes,arrows.meta}
\usepackage{algpseudocode}
\usepackage{algorithm}
\usepackage{enumerate}
\usepackage[shortlabels]{enumitem}
\usepackage{subcaption}
\usepackage{hyperref}
\usepackage{cleveref}
\usepackage{mathdots}
\usepackage{changes}

\newtheorem{lemma}{Lemma}
\newtheorem{theorem}{Theorem}
\newenvironment{proof}{\paragraph{Proof:}}{\hfill$\square$}

\newcommand{\dt}{\Delta t}
\newcommand{\e}{\mathrm{e}} 
\newcommand{\id}{\mathrm{id}} 
\newcommand{\ii}{\mathrm{i}} 
\newcommand{\ff}{\mathrm{f}}
\newcommand{\sgn}{\mathrm{sgn}}

\newcommand{\bs}{\boldsymbol{s}}

\newcommand{\tensor}[1]{\underline{\mathbf{#1}}}
\newcommand{\G}[2]{G_{#1,#2}}

\journal{Computer Physics Communications}

\begin{document}

\begin{frontmatter}



\title{Accelerated Inchworm Method with Tensor-Train Bath Influence Functional}


\author{Geshuo Wang} 

\affiliation{organization={Department of Applied Mathematics, University of Washington},
            city={Seattle},
            postcode={98195}, 
            state={WA},
            country={United States}}

\author{Yixiao Sun} 

\affiliation{organization={Department of Mathematics, National University of Singapore},
            addressline={Block S17, 10 Lower Kent Ridge Road}, 
            city={Singapore},
            postcode={119076}, 
            country={Singapore}}

\author{Siyao Yang}
\affiliation{organization={Committee on Computational and Applied Mathematics, Department of Statistics, University of Chicago},
            city={Chicago},
            postcode={60637}, 
            state={IL},
            country={United States}}

\author{Zhenning Cai} 

\affiliation{organization={Department of Mathematics, National University of Singapore},
            addressline={Block S17, 10 Lower Kent Ridge Road}, 
            city={Singapore},
            postcode={119076}, 
            country={Singapore}}

\begin{abstract}
We propose an efficient tensor-train-based algorithm for simulating open quantum systems with the inchworm method, where the reduced dynamics of the open quantum system is expressed as a perturbative series of high-dimensional integrals.
Instead of evaluating the integrals with Monte Carlo methods, we approximate the costly bath influence functional (BIF) in the integrand as a tensor train, allowing accurate deterministic numerical quadrature schemes implemented in an iterative manner.
Thanks to the low-rank structure of the tensor train, our proposed method has a complexity that scales linearly with the number of dimensions.    
Our method couples seamlessly with the tensor transfer method, allowing long-time simulations of the dynamics.
\end{abstract}




\end{frontmatter}



\section{Introduction}
\label{sec_intro}
Real-world quantum systems are rarely isolated, as they inevitably interact with their environment, leading to phenomena such as quantum decoherence and dissipation \cite{grigorescu1998decoherence,schlosshauer2019quantum}. The study of such open quantum systems plays a critical role in various fields, including but not limited to quantum optics \cite{breuer2002theory}, chemical physics \cite{weiss2012quantum}, and quantum computation \cite{nielsen2002quantum}, and necessitates the development of efficient numerical simulation techniques.

In simulating open quantum systems, the primary objective is to investigate the reduced dynamics of the system of interest while properly accounting for the influence of the surrounding environment (or bath). Unlike closed quantum systems, where the evolution is typically Markovian, open quantum systems experience dissipation and memory effects, making their dynamics inherently non-Markovian. These dynamics are formally described by the generalized quantum master equation (GQME), which arises from the Nakajima–Zwanzig projection operator formalism \cite{nakajima1958quantum,zwanzig1960ensemble}. The non-Markovian nature implies that the system’s future evolution depends on its full history, leading to significant computational challenges.

Over the past few decades, numerous methods have been developed to approximate the memory kernel in the Nakajima–Zwanzig equation \cite{amati2022quasiclassical, montoya2016approximate, Kelly2013efficient}. However, accurately evaluating the kernel remains computationally intensive. To alleviate this, Markovian approximations have been introduced to simplify the formalism, resulting in the widely used Lindblad equation \cite{lindblad1976generators,breuer2002theory,cao2025structure}. While effective in the weak system–bath coupling regime, such approximations become inadequate when the coupling strength is large or when memory effects are significant. An alternative perspective to simulate open quantum systems is based on wavefunctions. The multi-configuration time-dependent Hartree (MCTDH) method \cite{wang2003multilayer, beck2000multiconfiguration,meyer1990multi} approximates the wavefunction using time-dependent basis functions, suitable for cases where the environment contains only a small number of modes.
To overcome the exponentially increasing computational cost of wavefunction evolution, tensor network methods have been developed. 

Another major class of techniques for simulating open quantum systems is path integral methods, which describe quantum evolution as an integration of infinite classical paths.
A widely used path integral method is the quasi-adiabatic propagator path integral (QuAPI) \cite{makri1995tensorIINumerical}, which discretizes the full path integral by employing quasi-adiabatic propagator partitioning of the time evolution operator.
The iterative QuAPI (i-QuAPI) algorithm \cite{makri1995tensorITheory,makri1995tensorIINumerical,wang2022differential} introduces a truncation scheme for the extended memory length, significantly reducing computational costs. Compared with MCTDH, this method is more generally applicable, and meanwhile requires more memory, especially for multilevel systems or environments with long memory lengths.
Improvements of this method include the small matrix path integral (SMatPI) and \cite{makri2020smallMatrixPath, makri2021smallMatrixPathIntegralDriven, makri2021smallMatrixPathIntegralExtended,wang2024tree} the blip-summed method \cite{makri2014blip}.
The former replaces large tensors needed by QuAPI with small matrices, enabling the use of extensive memory to account for partial effects of long memory lengths, and the latter utilizes the blip representation of the path integral to allow for ignoring insignificant terms during the summation, showing remarkable reduction of the computational cost in the regime of incoherent dynamics.
The path integral formalism has also motivated some other methods that perform well in different regimes.
For instance, the hierarchical equations of motion (HEOM) \cite{huang2023efficient_yi, shi2018efficient, xu2023performance} is a popular approach for systems with Drude or related spectral density, and the time evolving matrix product operator (TEMPO) method \cite{strathearn2017efficient,strathearn2018efficient,jorgensen2019exploiting,gribben2020exact,bose2022multisite} uses a matrix product state (MPS) to represent the influence functional, applying compression and blip techniques to reduce storage, showing high performance when the MPS representation is memory efficient.

While the path integral methods often suffer from large memory costs, another category of algorithms based on the Dyson series expansion is usually free from such an issue.
The Dyson series \cite{dyson1949radiation} is a perturbative expansion of the evolution operator of the system, in which the summands are integrals with increasing dimensions that are usually represented using Feynman diagrams.
These high-dimensional integrals are often evaluated by Monte Carlo methods, leading to diagrammatic Monte Carlo methods \cite{vanhoucke2010diagrammatic}.
The major drawback of this approach is the numerical sign problem, caused by excessive cancellations of terms to be summed up \cite{loh1990sign}.
The bold-line diagrammatic Monte Carlo method \cite{prokof2007bold, prokof2008bold} can effectively mitigate this issue by grouping certain diagrams into ``bold lines'', thereby effectively reducing the number of diagrams to be summed.
The application of bold-line techniques in open quantum systems is known as the inchworm Monte Carlo method \cite{chen2017inchwormITheory,cai2020inchworm, cai2023numericalAnalysisInchworm}, which has remarkably improved the computational capability of diagrammatic methods and allows for simulations with moderate coupling between the system and the bath \cite{chen2017inchwormIIBenchmarks, yang2021inclusion}, despite being a perturbative theory.
The bottleneck of this approach is still the numerical sign problem, requiring a large number of sample points to evaluate high-dimensional integrals.

The purpose of this work is to tackle the evaluation of these high-dimensional integrals for open quantum systems with bosonic bath. Since the Monte Carlo method may encounter numerical sign problems, we turn to classical numerical integration to discretize the integrals.
In general, this method is not viable in high-dimensional cases, but the structure of the integrand in the inchworm method allows an efficient integration algorithm if the \emph{bath influence functional (BIF)} can be represented by tensor trains (TTs).
Similar ideas have been applied to quantum impurity models \cite{nunez2022learning,erpenbeck2023tensor,guo2024efficient,yu2025inchworm} for deterministic numerical integrations.
For open quantum systems, the BIF is determined by a \emph{two-point correlation (TPC)} function according to Wick's theorem \cite{wick1950evaluation}, and a TT decomposition of the BIF can be derived from the low-rank approximation of the two-point correlation. 
Several studies \cite{liu2024error,huang2024unified} establish theoretical error bounds that guarantee the accuracy of the expected value of observables computed using an approximate two-point correlation function.
Moreover, the TT representing the BIF can be further compressed by rounding techniques \cite{oseledets2011tensor} to reduce memory cost.


Our method offers several advantages over existing approaches \cite{cai2020inchworm,sun2024simulation}. First, by replacing Monte Carlo sampling with numerical quadrature, we achieve deterministic results with controllable precision.
Second, the use of TT approximation allows the computational cost of our method to scale linearly in dimensionality $M$ of the integrals, significantly improving computational efficiency.
Additionally, the linear cost scale enables simulations with larger $M$, extending the reach of real-time simulations in open quantum systems, and our method can also be coupled with the transfer tensor method (TTM), enabling long-time simulations with finite memory lengths. Finally, once the BIF is precomputed, it can be reused when simulating systems with different system-associated Hamiltonians. This feature enhances efficiency, especially when exploring a range of system parameters.

The rest of this paper is organized as follows. In \Cref{sec_inchworm}, we review the inchworm algorithm for open quantum systems \cite{chen2017inchwormITheory,chen2017inchwormIIBenchmarks,cai2020inchworm}. In \Cref{sec_BIF_TT}, we introduce a curse-of-dimensionality-free deterministic numerical integration scheme for high-dimensional perturbation series to be evaluated when simulating open quantum systems. In \Cref{sec_construct_BIF_TT}, we analyze the numerical low-rank structure of TPC matrix and introduce an algorithm to compress the BIF into tensor train. We also analyze the overall computational cost of our approach in this section. In \Cref{sec_numerical_results}, we conduct numerical experiments to study the bond dimensions of the resulting tensor trains and demonstrate the method on the spin-boson model. Finally, \Cref{sec_conclusion} provides concluding remarks and discusses potential directions for future research. The implementation of our method is available at the GitHub repository \cite{BIFTT2025code}.
\section{Open quantum systems and inchworm algorithm}
\label{sec_inchworm}
\subsection{Background}
We study the system-bath open quantum dynamics governed by the von Neumann equation
\begin{equation}
\label{eq_von_Neumann}
    \ii \pdv{\rho}{t} = H\rho - \rho H,
\end{equation}
where $\rho$ is the density matrix, and the total Hamiltonian $H$ takes the form
\begin{equation}
    H = H_0 + W.
\end{equation}
Here $H_0:= H_s \otimes \id_b + \id_s \otimes H_b$ is the unperturbed Hamiltonian, consisting of the system part $H_s$ and the bath part $H_b$, which are operators on the Hilbert spaces associated with the system and the bath, respectively.
The operators $\id_s $ and $\id_b$ are the identity over the spaces of the system and the bath, respectively. 
The term $W = W_s\otimes W_b$ describes the system-bath interaction. 

Our goal is to study the evolution of the expectation of a given observable $O:=O_s\otimes \id_b$, which is assumed to act on the system space only. Specifically, we focus on computing the expectation at any time $t$ given by $\langle O(t) \rangle:= \tr(\rho(t) O )$, where $ \rho(t) = \e^{-\ii H t} \rho(0) \e^{\ii H t}$ is the formal solution of  \cref{eq_von_Neumann}. By further assuming that the system and the bath are initially uncorrelated, \textit{i.e.}, $\rho(0)=\rho_s\otimes \rho_b$, the desired expectation is formulated as  
\begin{equation}
\label{eq_expval_Os}
    \expval{O_s(t)} = \tr_s \left(\rho_s O_s(t)\right), \qquad  O_s(t) = \tr_b(\rho_b \e^{\ii H t} O \e^{-\ii H t})
\end{equation}
where $\tr_s$ and $\tr_b$ are partial trace operator over system and bath space, respectively. In this work, we focus on the case where the bath consists of a number of harmonic oscillators, and the initial density matrix of the bath is in thermal equilibrium $\rho_b = Z^{-1}{\e^{-\beta H_b}}$ with $\beta$ being the inverse temperature and $Z = \tr(\e^{-\beta H_b})$ the normalization factor. 

The difficulty in computing \eqref{eq_expval_Os} lies in the high dimensionality of the bath, which makes the direct computation of the operators $\e^{\pm \ii H t}$ prohibitive. One strategy is to treat the coupling term $W$ as a perturbation and express $O_s(t)$ as a Dyson series \cite{dyson1949radiation,werner2009diagrammatic}, which is an infinite sum of high-dimensional integrals. An approximate evaluation of $O_s(t)$ can then be obtained by evaluating the truncated Dyson series without solving the full density matrix. However, when the system-bath coupling is strong, the Dyson series may suffer from slow convergence as well as large variance when evaluating the integrals via Monte Carlo.
This is known as the numerical sign problem \cite{loh1990sign,cai2023numericalAnalysisInchworm}. 

\subsection{Inchworm method}

To accelerate the convergence of the Dyson series and mitigate the numerical sign problem, the inchworm method \cite{chen2017inchwormITheory,chen2017inchwormIIBenchmarks,cai2020inchworm} has been proposed by performing a resummation such that a large portion of previous calculations can be efficiently reused.
The inchworm method and its variations have been successfully applied in many models such as the quantum spin-boson model \cite{cai2020inchworm,yang2021inclusion,cai2022fast,cai2023bold}, quantum impurity models \cite{dong2017quantum,Antipov2017currents,strand2024inchworm}, Caldeira-Leggett models \cite{wang2025solving} and spin chain models \cite{wang2023real,sun2024simulation}.
Thanks to the resummation, only relatively lower-dimensional integrals need to be evaluated in the renormalized series, mitigating the numerical sign problem when implementing Monte Carlo integration \cite{cai2023numericalAnalysisInchworm}. 
In the inchworm algorithm, we aim to solve the \emph{full propagator} $G(s_\ii, s_\ff)$ for $s_\ii \leqslant s_\ff$, which is an operator on the system space and describes the reduced dynamics of the observable between initial time $s_\ii$ and final time $s_\ff$. In particular, the desired operator $O_s(t)$ is related to the full propagator via
\begin{displaymath}
    O_s(t) = G(-t,t)
\end{displaymath}
for any $t \geqslant 0$. The core idea  of inchworm method is to solve $G(s_\ii, s_\ff)$ in an iterative manner \cite{chen2017inchwormITheory,cohen2015taming}. A practical approach to achieve this is to solve the following integro-differential equation \cite{cai2020inchworm} of $G(s_\ii, s_\ff)$:   
\begin{equation}
\label{eq_inchworm_integro_differential_eq}
\begin{split}
    &\sgn(s_\ff)\pdv{G(s_\ii,s_\ff)}{s_\ff} 
    =
        \ii H_s G(s_\ii,s_\ff) \\ 
      &   + \sum_{\substack{m=1\\ m \text{~is odd}}}^\infty
        \ii^{m+1}
        \int_{s_\ii}^{s_\ff} \int_{s_\ii}^{s_m} \cdots \int_{s_\ii}^{s_2} \left[
        \left( \prod_{j=1}^m \sgn(s_j) \right)
        W_s \mathcal{U}(s_\ii,\bs,s_\ff) \mathcal{L}_b^c (\bs,s_\ff)
        \right]
        \dd s_1 \cdots \dd s_{m-1} \dd s_m
\end{split}
\end{equation}
where $\sgn(\cdot)$ is the sign function. $\mathcal{U}(s_\ii,\bs,s_\ff)$ is a system-associated operator given by 
\begin{equation}
    \mathcal{U}(s_\ii,\bs,s_\ff)
    = G\left(s_m, s_{\ff}\right) W_s G\left(s_{m-1}, s_{m}\right) W_s \cdots W_s G\left(s_1, s_2\right) W_s G\left(s_\ii, s_1\right)
\end{equation}
and the \emph{bath influence functional (BIF)} $\mathcal{L}_b^c(s'_1,s'_2,\cdots,s'_n)$ for some even number $n$ is given by 
\begin{equation}
\label{eq_bif_inchworm}
    \mathcal{L}_b^c (\bs')
    = \sum_{\mathfrak{q} \in \mathcal{Q}_n^c} \prod_{\left(j, k\right) \in \mathfrak{q}} B\left(s'_j, s'_k\right).
\end{equation}
with $\mathcal{Q}_{n}^c$  being the set of all ``connected'' or ``linked'' pairings of integers from 1 to $n$ \cite{stein1978class}. 
More details are available in \ref{app_pairings}.
We point out that asymptotically, the cardinality of $\mathcal{Q}_m^c$ grows as rapidly as $\mathcal{O}((m-1)!!)$. Therefore, computing BIF is the main challenge in the inchworm algorithm.
In BIF, $B(\cdot,\cdot)$ is called the \emph{two-point correlation (TPC)} defined by
\begin{equation}
\label{eq_two_point_correlation}
    B\left(\tau_1, \tau_2\right)
    =\frac{1}{\pi} \int_0^{+\infty} J(\omega)\left[\operatorname{coth}\left(\frac{\beta \omega}{2}\right) \cos (\omega \Delta \tau)-\mathrm{i} \sin (\omega \Delta \tau)\right] \mathrm{d} \omega.
\end{equation}
where $\Delta \tau = \vert \tau_1 \vert - \vert \tau_2 \vert$.
The bath can be simulated by a large number of quantum harmonic oscillators. 
In this case, the spectral density $J(\omega)$ in  \cref{eq_two_point_correlation} takes the form
\begin{equation}
\label{eq_spectral_density}
    J(\omega) = \frac{\pi}{2} \sum_{j=1}^L \frac{c_j^2}{\omega_j} \delta(\omega-\omega_j)
\end{equation}
where we assume that the bath has $L$ harmonic oscillators with frequencies $\{\omega_j\}_{j=1}^L$ and the coupling intensity between the spin and the $j$th harmonic oscillator is characterized by a scalar $c_j$.

To solve the equation \cref{eq_inchworm_integro_differential_eq} numerically, one can use Runge-Kutta methods to solve it as an ordinary differential equation in the $s_{\ff}$ direction. In this work, we use the second-order Heun's method. In each time step, we truncate the series on the right side by a chosen integer $M$ and evaluate the integrals using numerical methods. Upon time discretization with step length $\Delta t$, we aim to evaluate the nodal value $G_{k_1,k_2} \approx G(k_1\Delta t, k_2\Delta t)$ for $-N \leqslant k_1 \leqslant k_2 \leqslant N$ where $N$ is the total number of time steps. Based on \cref{eq_inchworm_integro_differential_eq}, to obtain the value of $G_{k_1,k_2}$, we need the knowledge of all the values of $G_{l_1,l_2}$ with $k_1\leqslant l_1 \leqslant l_2 \leqslant k_2$ to evaluate the integrand. Therefore, one should implement the iteration according to a proper order. In general, one should compute those $G_{k_1,k_2}$ with smaller value of $k_2 - k_1$ first. For example, we can compute all full propagators in the table below from top to bottom and left to right\added{,} as proposed in \cite{wang2025solving}:
\begin{equation}\label{tab:G}
\renewcommand\arraystretch{1.5}
\begin{matrix*}[l]
    \G{-N}{-N} & \G{-N+1}{-N+1} & \cdots & \G{N-2}{N-2} & \G{N-1}{N-1} & \G{N}{N} \\
    \G{-N}{-N+1} & \G{-N+1}{-N+2} & \cdots & \G{N-2}{N-1} & \G{N-1}{N} \\
    \G{-N}{-N+2} & \G{-N+1}{-N+3} & \cdots & \G{N-2}{N} \\
    \vdots &\vdots & \iddots \\
    \G{-N}{N-1} & \G{-N+1}{N} \\
    \G{-N}{N}
\end{matrix*}
\end{equation}
When implementing the numerical scheme of \cref{eq_inchworm_integro_differential_eq}, we need to pay attention to the fact that the full propagator $G(s_\ii,s_\ff)$ has the following discontinuity conditions \cite{cai2020inchworm}
\begin{align}
    \lim_{s_\ff\rightarrow 0^+} G(s_\ii,s_\ff) 
    = O_s \lim_{s_\ff\rightarrow 0^-} G(s_\ii,s_\ff)\text{~for~} s_\ii <0,\\
    \lim_{s_\ii\rightarrow 0^-} G(s_\ii,s_\ff) 
    = \lim_{s_\ff\rightarrow 0^+} G(s_\ii,s_\ff) O_s
    \text{~for~} s_\ff >0.
\end{align}
The discontinuities above imply the initial condition for numerical scheme (first row of \eqref{tab:G}):
\begin{equation*}
    G_{k,k} = \begin{cases}
        \id_s &\text{~if~} k \neq 0 \\
        O_s &\text{~if~} k = 0
    \end{cases}.
\end{equation*}
In addition, the discontinuities also suggest that the numerical approximation of $G(k_1 \Delta t,k_2 \Delta t)$ listed in \eqref{tab:G} has two values for $k_1 = 0$ or $k_2 = 0$, denoted by $G_{k_1,0^{\pm}}$ and $G_{0^{\pm},k_2}$ respectively. One needs to use correct values of $G_{k_1,k_2}$ when interpolating those shorter $G(\cdot,\cdot)$ in $\mathcal{U}$.
Additionally, the full propagator $G(\cdot,\cdot)$ has the following conjugate symmetry property
\begin{equation}
    G(s_\ii,s_\ff) = G(-s_\ff,-s_\ii)^\dagger,
\end{equation}
and the shift invariance property
\begin{equation}
    G(s_\ii,s_\ff) = \begin{cases}
        G(s_\ii+T,s_\ff+T) &\text{~if~}s_\ii > 0, \\
        G(s_\ii-T,s_\ff-T) &\text{~if~}s_\ff < 0
    \end{cases}
\end{equation}
for any $T\geqslant 0$.
These two properties \cite{cai2022fast,cai2023bold} allow the reuse of some computation when constructing $G_{k_1,k_2}$.

\section{Fast numerical integration with tensor-train bath influence functional}
\label{sec_BIF_TT}
The main difficulty of the simulation based on \cref{eq_inchworm_integro_differential_eq} is the high-dimensional integrals over simplices.
These integrals are known as ``path integrals'' \cite{feynman1948space}, a description of quantum mechanics that generalizes the stationary action principle of classical mechanics. Direct computation of a path integral using conventional numerical quadratures suffers from curse of dimensionality. In this section, we propose a fast numerical integration scheme whose computational cost scales linearly with the dimension of the path integrals $m$. The central idea is to assume the bath influence functional $\mathcal{L}_b^c(\bs)$ defined in \eqref{eq_bif_inchworm} admits a certain low-rank factorization. To motivate our algorithm, let us consider a simple case where $\mathcal{L}_b^c(\bs)$ has the following separable form: 
\begin{equation}\label{Lb_rank1}
    \mathcal{L}_b^c(s_1,\cdots,s_m) = L(s_m-s_{m-1})  L(s_{m-1}-s_{m-2}) \cdots  L(s_2-s_1)
\end{equation}
for some kernel function $L(\cdot)$. Under such an assumption, a path integral in \eqref{eq_inchworm_integro_differential_eq} can be reformulated as
\begin{equation*}
    \int_{s_\ii}^{s_\ff} \sgn(s_m) W_s \tilde{G}(s_m,s_\ff)
    \left(
        \int_{s_\ii}^{s_m} \sgn(s_{m-1}) W_s \tilde{G}(s_{m-1},s_m)
        \left(
            \int_{s_\ii}^{s_{m-1}} \cdots \dd s_{m-2}
        \right) \dd s_{m-1}
    \right) \dd s_m,
\end{equation*}
where 
\begin{displaymath}
    \tilde{G}(s_i,s_{i+1}) =  G(s_i,s_{i+1}) L(s_{i+1} - s_i). 
\end{displaymath}
This formulation allows the following fast calculation of the path integral in an iterative manner:
\begin{displaymath}
\mathcal{I}_0 = 1, \qquad
\mathcal{I}_{k+1}(s_{k+1}) = \int_{s_\ii}^{s_{k+1}} \sgn(s_{k+1}) W_s \tilde{G}(s_k, s_{k+1}) \mathcal{I}_k(s_k) \dd s_k.
\end{displaymath}
In other words, if $\mathcal{L}_b^c(\bs)$ satisfies \eqref{Lb_rank1}, the high-dimensional integrals can be regrouped into a sequence of one-dimensional integrals, reducing the computational cost of numerical quadrature from $\mathcal{O}(N^m)$ to only $\mathcal{O}(mN^2)$ where $N$ is the number of grid points. 

However, the bath influence functional $\mathcal{L}_b^c$ in general does not satisfy the separable form \eqref{Lb_rank1}, since it characterizes the non-Markovian effect of the bath on the entire history. In this work, we consider compressing the bath influence functional into a more complicated factorization known as a tensor train (TT) \cite{oseledets2011tensor}, and we call the resulting compressed $\mathcal{L}_b^c$ the bath influence functional tensor train (BIF-TT). With the TT representation, we will show that we are still able to design an efficient way to evaluate the high-dimensional integrals recursively.
In what follows, we will first introduce some basic concepts and operations of tensor trains and then the iterative algorithm to evaluate the high-dimensional path integrals when the BIF is represented by a tensor train.
\subsection{Introduction to TT/MPS}
\label{sec_operations_tt}
Throughout this paper, we use underlined bold capital letters to represent tensors.
For a $d$-tensor $\tensor{A}\in\mathbb{C}^{n_1\times \cdots \times n_d}$,
 we use $\tensor{A}(i_1,\cdots,i_{n_d})$ to represent the element with multi-index $(i_1,\cdots,i_{n_d})$ for $1\leqslant i_j 
 \leqslant n_j$ and $1\leqslant j \leqslant d$.
In particular, a matrix can be regarded as a $2$-tensor. An obvious issue with applying the full tensor representation in numerical simulations is the curse of dimensionality. Naively, storing a tensor of size $n_1 \times \cdots \times n_d$ requires memory allocated for $n_1 \cdots n_d$ numbers, which grows exponentially as $d$ increases.

A popular approach to reduce the memory cost of a high-dimensional tensor is to utilize the underlying low-rank structure of the tensor and represent it as a tensor train (TT). The tensor-train decomposition was first mathematically introduced with theoretical guarantee by Oseledets in 2009 \cite{oseledets2009new,oseledets2009compact}, but the idea can be dated back to the 1990s in the study of quantum many body systems \cite{white1992density,white1993density} where TT was better known as matrix product states (MPS). In the TT representation, a tensor $\tensor{A} \in \mathbb{C}^{n_1\times \cdots \times n_d}$ takes the form 
\begin{equation}\label{TT representation}
    \tensor{A} 
    = \tensor{A}^{(1)} \times^1 \tensor{A}^{(2)} \times^1 \cdots \times^1 \tensor{A}^{(d)}
\end{equation}
where $\tensor{A}^{(j)} \in \mathbb{C}^{r_{j-1}\times n_j \times r_j}$ is called a tensor core and $(r_0,r_1,\cdots,r_d)$ are known as the TT ranks or bond dimensions with $r_0 = r_d = 1$. The tensor contraction ``$\times^1$'' between $\tensor{A}^{(j-1)}$ and $\tensor{A}^{(j)}$ sums over the last index of $\tensor{A}^{(j-1)}$ and the first index of $\tensor{A}^{(j)}$. Consequently, \eqref{TT representation} is formed elementwisely as
\begin{equation*}
    \tensor{A}(i_1,i_2,\cdots,i_d) 
    = \sum_{\alpha_1=1}^{r_1}\sum_{\alpha_2=1}^{r_2}\cdots \sum_{\alpha_{d-1}=1}^{r_{d-1}}\tensor{A}^{(1)}(1,i_1,\alpha_1)\tensor{A}^{(2)}(\alpha_1,i_2,\alpha_2) \cdots \tensor{A}^{(d)}(\alpha_{d-1},i_d,1).
\end{equation*}
A graphical interpretation of the TT structure in~\eqref{TT representation} is provided by the tensor diagrams in \Cref{fig:tt-diagram}, which are a useful tool for visualizing tensor networks and will be employed later to illustrate the algorithm.
\begin{figure}[ht!]
    \centering
    \subfloat[Tensor contraction]{
    \begin{tikzpicture}
        \draw (-1,0) -- (2.2,0);
        \draw (0,0) -- (0,1);
        \draw [line width=1.5] (0,0) -- (1.5,0);
        \draw[fill=white,line width = 1.2] (0,0) circle (0.3);
        \node at (0,0) {$\tensor{A}$};
        \draw[fill=white,line width = 1.2] (1.2,0) circle (0.3);
        \node at (1.2,0) {$\tensor{B}$};
        \node at (-0.5,0.22) {$i_1$};
        \node at (0.6,0.22) {$k$};
        \node at (1.8,0.22) {$j_2$};
        \node at (0.2,0.6) {$i_2$};
        \node at (2.5,0) {$=$};
        \draw (3,0) -- (5,0);
        \draw (4,0) -- (4,1);
        \draw[fill=white,line width = 1.2] (4,0) circle (0.3);
        \node at (4,0) {$\tensor{C}$};
        \node at (3.5,0.22) {$i_1$};
        \node at (4.2,0.6) {$i_2$};
        \node at (4.6,0.22) {$i_2$};
    \end{tikzpicture}
    }
    \qquad  
    \subfloat[Tensor train]{
    \begin{tikzpicture}
        \draw (0,0) -- (3.5,0);
        \draw (4.3,0) -- (5.2,0);
        \draw (0,0) -- (0,1);
        \draw (1.3,0) -- (1.3,1);
        \draw (2.6,0) -- (2.6,1);
        \draw (5.2,0) -- (5.2,1);
        \draw[fill=white,line width = 1.2] (0,0) circle (0.4);
        \node at (0,0) {$\tensor{A}^{\!(1)}$};
        \draw[fill=white,line width = 1.2] (1.3,0) circle (0.4);
        \node at (1.3,0) {$\tensor{A}^{\!(2)}$};
        \draw[fill=white,line width = 1.2] (2.6,0) circle (0.4);
        \node at (2.6,0) {$\tensor{A}^{\!(3)}$};
        \draw[fill=white,line width = 1.2] (5.2,0) circle (0.4);
        \node at (5.2,0) {$\tensor{A}^{\!(d)}$};
        \node at (0.2,0.6) {$i_1$};
        \node at (1.5,0.6) {$i_2$};
        \node at (2.8,0.6) {$i_3$};
        \node at (5.4,0.6) {$i_d$};
        \node at (0.6,0.22) {$\alpha_1$};
        \node at (1.9,0.22) {$\alpha_2$};
        \node at (3.2,0.22) {$\alpha_3$};
        \node at (4.4,0.22) {$\alpha_{d-1}$};
        \node at (3.9,0) {$\cdots$};
    \end{tikzpicture}
    }
    \caption{Tensor diagrams illustrating \eqref{TT representation}. (a) Tensor contraction $\tensor{A}\times^1 \tensor{B}$ between a 3-dimensional tensor $\tensor{A}(i_1,i_2,i_3)$ and a 2-dimensional tensor $\tensor{B}(j_1,j_2)$. Each tensor is denoted as a core with legs representing the tensor indices. By connecting the legs of same size, the tensor contraction produces a new 3-dimensional tensor $\tensor{C}$; (b) TT representation for a $d$-dimensional tensor defined in Eq.~\eqref{TT representation}. Each vertical solid leg represents a index $i_k$, and each horizontal solid leg represents an auxiliary index $\alpha_k$. A horizontal connection between two cores indicates contraction over the associated auxiliary indices.}
    \label{fig:tt-diagram}   
\end{figure}

There are several existing methods for compressing a high-dimensional tensor into the TT format. A classical approach is the well-known TT-SVD \cite{oseledets2011tensor}, which directly applies SVD to matricizations of the full tensor. However, since the tensor size grows exponentially with the dimension, TT-SVD suffers from the curse of dimensionality in both storage and computation. Another widely used method is TT-cross \cite{oseledets2010ttcross}, which constructs a TT representation by selecting pivot interpolation points. While effective in many settings, TT-cross often requires multiple sweeps across dimensions to identify near-optimal pivots, making the overall computational complexity difficult to control. Similar challenges arise in optimization-based approaches~\cite{bradley2020modeling,han2018unsupervised,novikov2021tensor}, owing to the highly non-convex nature of the associated optimization problems. Finally, there exist optimization-free methods based on randomized linear algebra \cite{peng2024tensor,hur2023generative} for computing tensor-train representations in density estimation problems; however, these approaches typically rely on structural assumptions that the target tensors, such as the BIF considered here, may not satisfy. In this work, we propose an optimization-free compression algorithm leveraging the definition of BIF to be specified in \Cref{sec_construct_bif_tt}.

The direct advantage of the TT representation is that its memory cost scales linearly with respect to the dimensionality as $\mathcal{O}(dnr^2)$ where $n = \max_j n_j$ and $r = \max_j r_j$. Moreover, many tensor operations such as elementwise summation and multiplication based on TT representation also have linear computational complexity with respect to the dimensionality (see more details on the complexity in \ref{app:tt operations}). These properties enable the design of an efficient algorithm for compressing BIF into TT format.

\subsection{Sequential evaluation of high-dimensional integrals}
\label{sec_TT_integral}
\begin{figure}
    \centering
\includegraphics[width=0.7\linewidth]{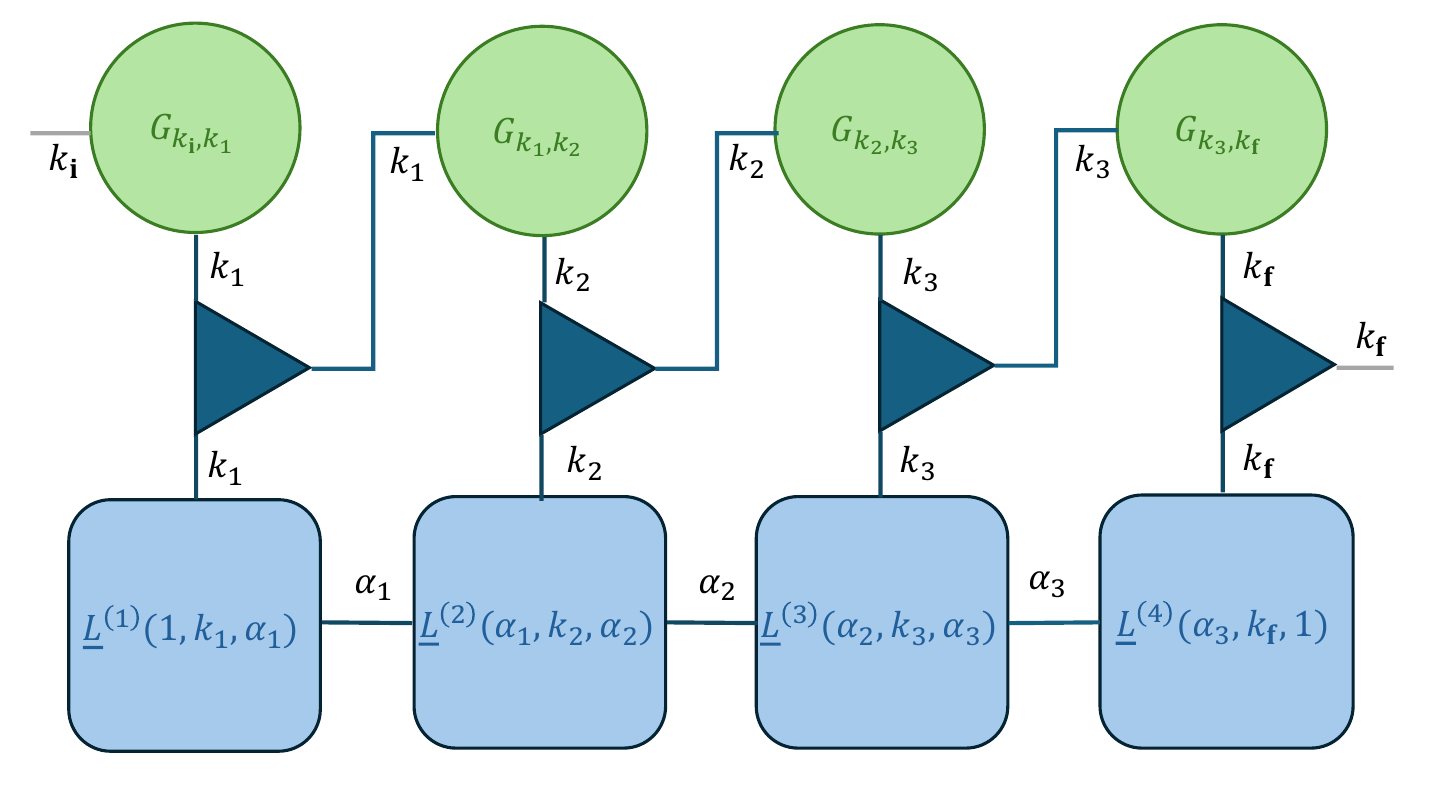}
    \caption{Tensor diagram illustrating the desired integral $I_m$ \cref{eq_integral_to_compute} with BIF-TT for $m=3$. Constant matrix $W_s$ are omitted in the diagram for simplicity.}
    \label{fig:example_biftt_m3}
\end{figure}

We will postpone the construction of BIF-TT to \Cref{sec_construct_BIF_TT}. For now, let us assume the bath influence functional is already in TT format, and we will use this subsection to elaborate a deterministic numerical integration scheme to evaluate the $m$-dimensional integrals appearing on the right-hand side of \cref{eq_inchworm_integro_differential_eq}:
\begin{equation}
\label{eq_integral_to_compute}
\begin{split}
    I_m 
    =  \int_{s_\ii}^{s_\ff} \int_{s_\ii}^{s_m} \cdots \int_{s_\ii}^{s_2} \left[
        \left( \prod_{j=1}^m \sgn(s_j) \right)
        W_s \mathcal{U}(s_\ii,\bs,s_\ff) \mathcal{L}_b^c (\bs,s_\ff)
        \right]
        \dd s_1 \cdots \dd s_{m-1} \dd s_m.
\end{split}
\end{equation}
Under a discrete mesh, assume the $(m+1)$-dimensional tensor formed by the values of BIF on grids $\mathcal{L}_b^c(k_1\dt,\cdots,k_m\dt,k_\ff\dt)$ for indices $k_\ii\leqslant k_1\leqslant \cdots \leqslant k_m \leqslant k_\ff$ with $k_\ii \Delta t = s_\ii$ and $k_\ff\Delta t=s_\ff$ is approximated by a TT \eqref{TT representation}:
\begin{equation}\label{bif-tt}
     \tensor{L}_{m+1} = \tensor{L}^{(1)} \times^1 \tensor{L}^{(2)} \times^1 \cdots \times^1 \tensor{L}^{(m+1)} 
\end{equation}
with rank $(r_0,r_1,\cdots,r_{m+1})$ for an odd integer $m$, whose tensor diagram representation is given as \Cref{fig:tt-diagram}(b). Inserting the TT format of $\tensor{L}_{m+1}$, the desired integral $I_m$ can be represented as a tensor network shown in \Cref{fig:example_biftt_m3}. In the diagram, green circles represent the full propagators $G(k_{j-1},k_{j})$, blue cubes denote cores of BIF-TT $\tensor{L}_{4}$, and dark blue triangles represent Kronecker delta 3-tensors, which have value 1 if all three legs take the same index and have value 0 otherwise. The tensor contractions are visually indicated by lines connecting the respective objects. Accordingly, the integration process is visually interpreted as a right-to-left construction of the diagram. For simplicity, the interaction operators $W_s$ are omitted in the tensor diagram.

\begin{figure}
    \centering
    \includegraphics[width=0.9\linewidth]{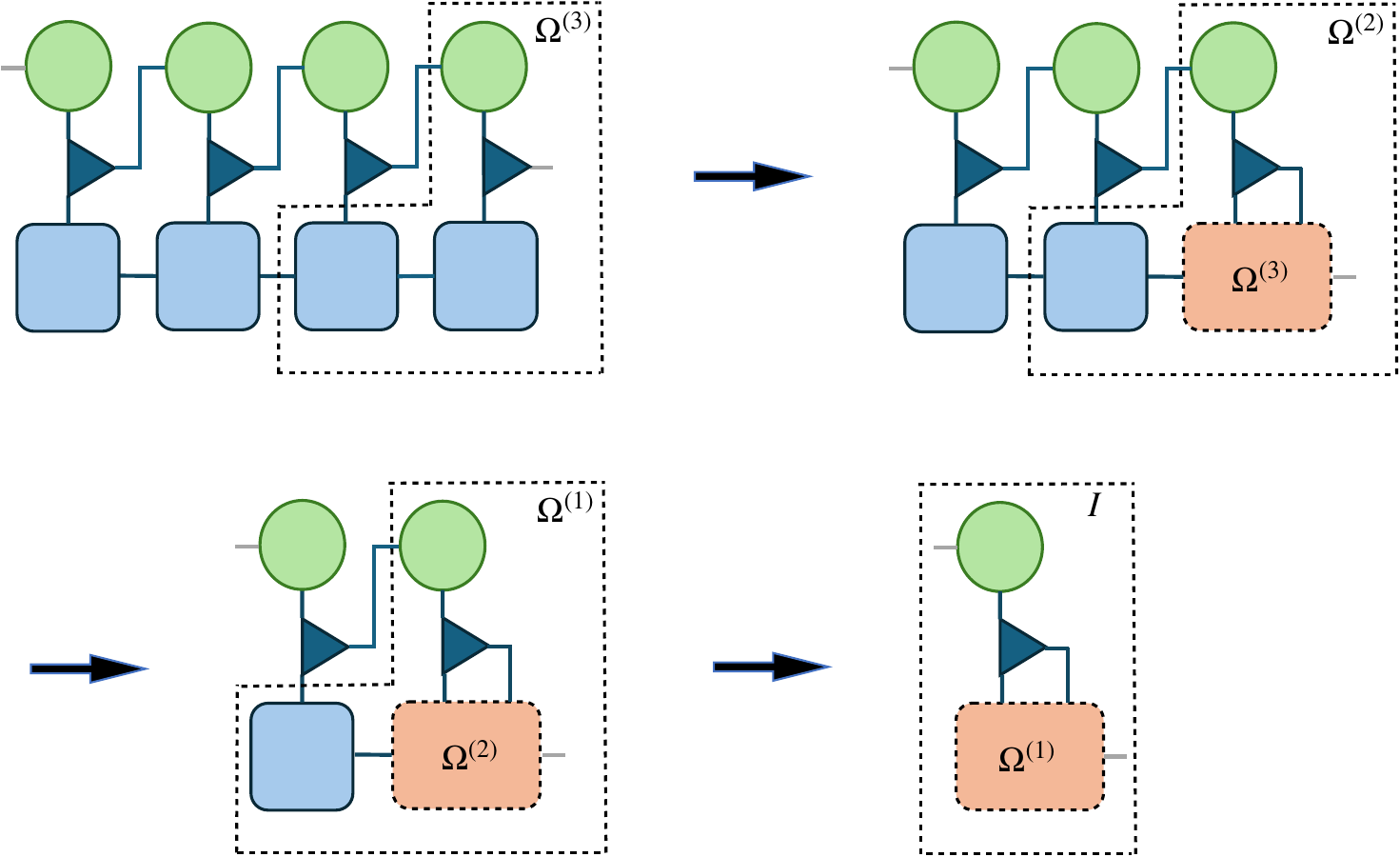}
    \caption{Tensor diagrams illustrating the iterative computation \cref{eq_biftt_iteration} of $I_m$ for $m=3$.}
    \label{fig:integral_iteration}
\end{figure}

The tensor-train structure of the BIF allows us to compute the $m$-dimensional integral illustrated in \Cref{fig:example_biftt_m3} via sequential evaluations of $m$ one-dimensional numerical integration based on the composite trapezoidal rule in an iterative manner shown by the tensor diagrams in \Cref{fig:integral_iteration}. In each iteration, we leverage the low-rank structure of BIF-TT to form a four-dimensional tensor $\Omega^{(j)}$ defined by
\begin{subequations}
\label{eq_biftt_iteration}
\begin{equation}
\label{eq_omega_middle}
     \Omega^{(j)}(k_j, \alpha_{j-1}) = \sum_{\alpha_j} \tensor{L}^{(j)}(\alpha_{j-1},k_j,\alpha_j) I^{(j)}(k_j,\alpha_j) W_s,
\end{equation}
for $j=m-1,m-2,\cdots,1$ where 
\begin{equation}
      \label{eq_I_middle}
    I^{(j-1)}(k_{j-1},\alpha_{j-1}) = \sum_{k_j=k_{j-1}}^{k_\ff} w_{k_j} \sgn(k_j) \Omega^{(j)}(k_j,\alpha_{j-1})G_{k_{j-1},k_j}. 
\end{equation}
Here we need to set $k_j=k_\ii, k_\ii+1, \cdots, -1, 0^-, 0^+, 1, \cdots, k_\ff-1, k_\ff$
and $\alpha_j=1,2,\cdots,r_j$ and compute the corresponding values. In particular, we compute $\Omega^{(m)}$ as the initial iteration by 
\begin{equation}
         \label{eq_omega_start}
    \Omega^{(m)}(k_m,\alpha_{m-1}) = W_s G_{k_m,k_\ff} W_s
    \sum_{\alpha_m} \tensor{L}^{(m)}(\alpha_{m-1},k_m,\alpha_m)
    \tensor{L}^{(m+1)}(\alpha_m,k_\ff,1).
\end{equation}
The desired integral $I_m$ in \cref{eq_integral_to_compute} is then identical to  $I^{(0)}(k_0,\alpha_0)$ in the final iteration if we set $k_0=k_\ii$. 
\end{subequations}

We would like to comment here about the choice of weights $w_{k_j}$.
It comes from the numerical quadrature for an integral from $[k_{j-1}\Delta t, k_\ff \Delta t]$.
In the implementation, we choose to apply the composite trapezoidal rule for the quadratures.
When $k_\ii\geqslant 0^+$ or $k_\ff\leqslant 0^-$,
the values of $w_{k_j}$ are chosen as follows:
\begin{equation}
    w_{k_j} 
    = \begin{cases}
        0 & \text{~if~} k_{j-1}=k_\ff ,\\
        \frac{1}{2}\Delta t & \text{~if~} k_{j-1}<k_\ff \text{~and~} k_j = k_{j-1}, k_\ff ,\\
        \Delta t & \text{~if~} k_{j-1}<k_\ff \text{~and~} k_{j-1}<k_j<k_\ff.
    \end{cases}
\end{equation}
When $k_\ii<0$ and $k_\ff>0$ 
, we need some careful treatment for the values of $G$'s.
Since we have decomposed the high-dimensional integral into one-dimensional integrals, we only need to split the integral domain if the integral domain crosses the origin. 
As a result of applying trapezoidal rules on each interval, we should regard the value of $G_{k_{j-1},0}$ by
\begin{equation}
    G_{k_{j-1},0} = \begin{cases}
        G_{0^+,0^+} &\text{if~}k_{j-1}=0^-, \\
        \frac{1}{2}\left(G_{k_{j-1},0^+}
        + G_{k_{j-1},0^-}\right) &\text{if~}k_{j-1}<0^-.
    \end{cases}
\end{equation}

With BIF-TT, we can achieve exact deterministic integration based on composite trapezoidal rule with computational complexity scaling linearly with the dimensionality $m$ of the high-dimensional integral $I_m$. A detailed analysis of the overall computational complexity will be given in the next section.


\section{Tensor-train compression of bath influence functional}
\label{sec_construct_BIF_TT}
We now elaborate how to compress the bath influence functional into a tensor train.

\subsection{Two-point correlation matrix}
\label{sec_tpc_matrix}
%
Before considering the $m$-variable function $\mathcal{L}_b^c(\bs)$ for a general $m$, we first study the simplest case $m=2$.
In this case, $\mathcal{L}_b^c$ is exactly the two point correlation function $B$ as defined in \eqref{eq_two_point_correlation}.
For the purpose of numerical computation, we introduce a grid for the two variables of $B$.
We focus on its values on grid points $(\tau_1,\tau_2)\in \{(k_1\Delta t, k_2 \Delta t) \ | \  k_1,k_2 = -N,-N+1,\cdots,N-1,N\}$.
The values of $B(\tau_1,\tau_2)$ on grid points form a two-point correlation matrix $\mathbf{B}$ with
\begin{equation}
\label{eq_tpc_matrix_fftpc_matrix}
    \mathbf{B} = \frac{1}{\pi} \int_{0}^\infty J(\omega)
    \mathcal{B}(\omega) \dd \omega,
\end{equation}
where $\mathcal{B}(\omega)$ is a matrix with elements defined by
\begin{equation}
\label{eq_fixed_frequency_tpc_matrix}
    \mathcal{B}_{k_1,k_2}(\omega) = 
        \coth\left(\frac{\beta\omega}{2}\right)
        \cos(\omega \Delta k \Delta t)
        - \ii \sin(\omega \Delta k \Delta t)
\end{equation}
for fixed $\omega$ and $\Delta k = \vert k_1 \vert - \vert k_2 \vert$.
Based on the definition, the two-point correlation matrix $\mathbf{B}$ is a $(2N+1)\times (2N+1)$ complex matrix.
It is clear that the matrices $\mathbf{B}$ and $\mathcal{B}(\omega)$ are Hermitian. The following lemma states the low-rankness of $\mathcal{B}(\omega)$ with a fixed frequency $\omega$:
\begin{lemma}
\label{lemma_rank2}
    For a fixed nonzero frequency $\omega$ and positive time step $\Delta t$,
     the matrix $\mathcal{B}(\omega)$ defined by \cref{eq_fixed_frequency_tpc_matrix} has rank 2.
\end{lemma}
The proof of \Cref{lemma_rank2} can be found in \ref{app_proof}.
\Cref{lemma_rank2} shows that $\mathcal{B}$ has a low-rank structure.
Since the two-point correlation matrix $\mathbf{B}$ is a ``combination'' of $\mathcal{B}$'s as shown in \cref{eq_tpc_matrix_fftpc_matrix}, we may expect that the matrix $\mathbf{B}$ also has a low-rank structure.
For the matrix $\mathbf{B}$, we have the following theorem for its rank:
\begin{theorem}
\label{thm_rank_bound}
    Let $r$ be the rank of the two-point correlation matrix $\mathbf{B}$, we have $r\leqslant N+1$.
    In particular, if harmonic bath is modeled by $L$ quantum harmonic oscillators, \emph{i.e.}, the spectral density is given by \cref{eq_spectral_density}, we have 
    \begin{equation}
    \label{eq_rank_upper_bound}
        r\leqslant \min\{N+1,2L\}.
    \end{equation}
\end{theorem}
The proof of \Cref{thm_rank_bound} can be found in \ref{app_proof}.
In the numerical simulation, the numerical rank of matrix $\mathbf{B}$ is in general lower than the bound given by \Cref{thm_rank_bound}.
We will see this later in our numerical tests in \Cref{sec_numerical_results}.
In other words, the columns (rows) of matrix $\mathbf{B}$ have some extra dependency from the numeric viewpoint.
Although it is difficult to clearly estimate the numerical rank of $\mathbf{B}$, we may try to understand the source of this extra dependency.

We first give an intuitive understanding in a simplified model where $J(\omega) = \frac{\pi}{2}\left(\delta(\omega_0)+\delta(\omega_0+\Delta\omega)\right)$.
Then $\mathbf{B} = \frac{1}{2}\left(\mathcal{B}(\omega_0) + \mathcal{B}(\omega_0+\Delta \omega)\right)$.
From the definition of $\mathcal{B}(\omega)$, we may observe $\mathcal{B}(\omega_0+\Delta \omega) = \mathcal{B}(\omega_0) + \Delta \mathcal{B}$ with $\Vert \Delta \mathcal{B}\Vert \sim \Delta \omega$.
Therefore, $\Vert \mathbf{B} - \mathcal{B}(\omega_0) \Vert \sim \Delta \omega$. 
Based on \Cref{lemma_rank2}, $\mathcal{B}(\omega_0)$ is a rank-2 matrix.
Therefore, $\mathbf{B}$ is close to a rank-2 matrix if $\Delta \omega$ is small.
The small difference $\mathbf{B} - \mathcal{B}(\omega_0)$ will only produce some small perturbation in the singular values so $\mathbf{B}$ may be regarded as a rank-2 matrix numerically.

The toy model inspires us to consider the contribution of $\mathcal{B}(\omega)$ with similar $\omega$ as a whole in \cref{eq_tpc_matrix_fftpc_matrix}.
If we split the domain $[0,\infty)$ into subintervals and focus on a simple subinterval $[\omega_j,\omega_{j+1}]$,
 we have the following observation.
\begin{theorem}
\label{thm_low_rank}
For a smooth spectral density $J(\omega)$ and a small interval $[\omega_j,\omega_{j+1}]$,
we define two matrices
\begin{equation}
    \mathfrak{B}_1 = \frac{1}{\pi} \int_{\omega_j}^{\omega_{j+1}} J(\omega) \mathcal{B}(\omega) \dd \omega,
    \quad
    \mathfrak{B}_2 = \frac{1}{\pi} J(\bar{\omega}) \mathcal{B}(\bar{\omega}) (\omega_{j+1}-\omega_j)
\end{equation}
with $\bar{\omega} = \frac{\omega_j + \omega_{j+1}}{2}$,
 it holds that 
\begin{equation}
    \Vert \mathfrak{B}_1 - \mathfrak{B}_2 \Vert_{F} \leqslant C (\omega_{j+1}-\omega_j)^3,
\end{equation}
where the constant $C$ depends on the spectral density $J$ and the function $\mathcal{B}$.
\end{theorem}
The proof of \Cref{thm_low_rank} can be found in \ref{app_proof}.
Based on \Cref{thm_low_rank}, when $J$ is smooth, the matrix $\mathbf{B}$ can be written as the sum of $\mathfrak{B}_1$ while an estimation of $\mathbf{B}$, say $\mathbf{B}_{\Delta \omega}$, is given by the sum of $\mathfrak{B}_2$. 
In this way, the matrix $\mathbf{B}$ is close to the matrix $\mathbf{B}_{\Delta \omega}$, whose rank is no more than $2\times\text{(number of intervals)}$.
When we have a non-smooth $J$ such as \cref{eq_spectral_density}, we can also expect that the two-point correlation matrix is close to a relatively low rank matrix.

Another source of the low-rank property comes from the fact that $\lambda(\omega)-1=\coth(\frac{\beta\omega}{2})-1$ decays to 0 exponentially when $\omega$ is large.
Therefore, for the matrices $\mathcal{B}$'s for high frequencies themselves may be very close to a rank-1 matrix as we can observe in \cref{eq_rank_2_FF_TPCM}.

In the numerical experiments, instead of estimating the numerical rank based on some theory, we directly apply the singular value decomposition and truncate the singular values based on a fixed threshold.
We will present some examples in \Cref{sec_numerical_results}.


\subsection{Construction of BIF-TT}
\label{sec_construct_bif_tt}
In this subsection, we construct the TT representation of the bath influence functional $\mathcal{L}_b^c(\bs)$ for even values of $m$.

In \Cref{sec_tpc_matrix}, 
we discussed the low-rank property of the two-point correlation matrix $\mathbf{B}$. Assuming $\operatorname{rank} (\mathbf{B}) = r$, the matrix $\mathbf{B}$ can then be written as a two-core tensor train
\begin{equation}
\label{eq_tpc_tt}
    \mathbf{B} = \tensor{B}^{(1)}\times^1 \tensor{B}^{(2)}
\end{equation}
where $\tensor{B}^{(1)}\in\mathbb{C}^{1\times (2N+1) \times r}$ and  $\tensor{B}^{(2)}\in\mathbb{C}^{r\times (2N+1) \times 1}$. 
The tensors $\tensor{B}^{(1)},\tensor{B}^{(2)}$ are essentially matrices as each of them only has two effective indices.
This decomposition can be obtained by the singular value decomposition (SVD).
One can further reduce $r$ with truncated SVD, at the cost of $\mathcal{O}(N^3)$. 
Similar to the matrix $\mathbf{B}$ defined in \Cref{sec_tpc_matrix}, we only consider the values of the bath influence functional on the grid points $(s_1,\cdots,s_m) \in \{(k_1\dt,\cdots,k_m\dt); k_j = -N,-N+1,\cdots,N-1,N\text{~for~} j=1,\cdots,m\}$.
The values of the bath influence functional on these points form an $m$-core tensor $\tensor{L}_m$ with
\begin{equation}
    \tensor{L}_m(k_1,\cdots,k_m) = \mathcal{L}_b^c(k_1\dt,\cdots,k_m\dt)
    \text{~for~}k_j = -N,-N+1,\cdots,N-1,N,
\end{equation}
which is the sum of Hadamard (elementwise) products of matrices $\mathbf{B}$ (see \cref{eq_bif_inchworm}).
To start with, we give an example for the construction of BIF-TT when $m=4$ and $m=6$.
\begin{equation}
\label{eq_construction_L4_L6}
\begin{split}
    \tensor{L}_4 &= \mathbb{TT}(\mathbf{B},4,[1,3])
    \odot
    \mathbb{TT}(\mathbf{B},4,[2,4]);\\
    \tensor{L}_6 &= 
    \mathbb{TT}(\mathbf{B},6,[1,4])
    \odot
    \mathbb{TT}(\mathbf{B},6,[2,5])
    \odot
    \mathbb{TT}(\mathbf{B},6,[3,6]) \\
    &+\mathbb{TT}(\mathbf{B},6,[1,4])
    \odot
    \mathbb{TT}(\mathbf{B},6,[2,6])
    \odot
    \mathbb{TT}(\mathbf{B},6,[3,5]) \\
    &+\mathbb{TT}(\mathbf{B},6,[1,3])
    \odot
    \mathbb{TT}(\mathbf{B},6,[2,5])
    \odot
    \mathbb{TT}(\mathbf{B},6,[4,6]) 
    \\
    &+\mathbb{TT}(\mathbf{B},6,[1,5])
    \odot
    \mathbb{TT}(\mathbf{B},6,[2,4])
    \odot
    \mathbb{TT}(\mathbf{B},6,[3,6]). 
\end{split}
\end{equation}
Here 
\begin{equation}
    \label{eq_TT_index_augmentation}
    \mathbb{TT}(\tensor{X},D,[k_1,\cdots,k_d])
\end{equation}
extends a given $d$-dimensional TT $\tensor{X}$ to a $D$-dimensional TT $\tensor{Z}$ where $d<D$ such that the value of $\tensor{Z}$ only depends on indices $j_{k_1},\cdots, j_{k_d}$ and is identical to $\tensor{X}$ if these indices are same as indices of $\tensor{X}$. ``$\odot$'' denotes the Hadamard product, and ``$+$'' is the elementwise summation between TTs.  We again note that the Hadamard product and elementwise summation between TTs has complexity that is linear in the dimensionality of the TTs (see more details in \ref{app:tt operations}).

In the construction of $\tensor{L}_4$, we construct two tensor trains $\tensor{K}_1, \tensor{K}_2$
\begin{equation}
    \tensor{K}_1 = \mathbb{TT}(\mathbf{B},4,[1,3]),
    \quad
    \tensor{K}_2 = \mathbb{TT}(\mathbf{B},4,[2,4]).
\end{equation}
The elements of $\tensor{K}_1,\tensor{K}_2$ are actually given by
\begin{equation}
    \tensor{K}_1(k_1,k_2,k_3,k_4) = \mathbf{B}_{k_1,k_3},
    \quad
    \tensor{K}_2(k_1,k_2,k_3,k_4) = \mathbf{B}_{k_2,k_4}.
\end{equation}
The BIF-TT $\tensor{L}_4$ is then given by the Hadamard product of $\tensor{K}_1$ and $\tensor{K}_2$.
Similarly, in the construction of $\tensor{L}_6$, we need to compute the Hadamard products of three tensor trains. An extra step to sum up the four terms is also required.

The Hadamard product and summations in \cref{eq_construction_L4_L6} will significantly increase the bond dimensions, leading to high memory cost of the produced BIF. In our implementation, we will specify the maximum bond dimension or a rounding accuracy, and apply TT rounding algorithm \cite{oseledets2011tensor,holtz2012manifolds,lee2018fundamental}, which performs truncated SVD sequentially for the cores to reduce the bond dimensions to the target rank or rounding accuracy.

In general, based on the formula \cref{eq_bif_inchworm} and the tensor operations introduced in \Cref{sec_operations_tt}, we can design a method to construct the BIF-TT in the following four steps: 
\begin{enumerate}[start=1,label={(\bfseries BIF\arabic*)}]
    \item \label{steps_tt1} Extend the tensor $\mathbf{B}$ represented as \cref{eq_tpc_tt} to a tensor train with $m$ cores according to \cref{eq_tt_extension_values}. The original cores should be placed correspondingly for each pair of $(j,k)$ appearing in \cref{eq_bif_inchworm}.
    \item \label{steps_tt2} For each $\mathfrak{q} \in \mathcal{Q}_m^c$, compute the Hadamard product of corresponding $m/2$ tensor trains in \ref{steps_tt1}.
    \item \label{steps_tt3} Sum up all tensor trains generated for all $\mathfrak{q} \in \mathcal{Q}_m^c$ in \ref{steps_tt2}.
    \item Compress the tensor train during \ref{steps_tt2} and \ref{steps_tt3} if needed.
\end{enumerate}

This simple process allows the construction of tensor trains for bath influence functionals for different even $m$.
The construction starts from the two-point correlation tensor train \cref{eq_tpc_tt} and applies Hadamard products and summations of tensor trains to obtain the resulting tensor trains.
The whole process can be summarized by \Cref{algo:tt-bif}.
\begin{algorithm}[ht]
  \caption{Compression of BIF-TT}\label{algo:tt-bif}
  \begin{algorithmic}[1]
\State \textbf{Input:} Two-point correlation matrix $\mathbf{B}$ defined in \cref{eq_two_point_correlation}. Dimension $m$ (even number). Upper bound for bond dimension $R$.
\State \textbf{Output:} BIF-TT $\tensor{L}_m$ defined in \cref{bif-tt}. 
\State Initialize $\tensor{L}_m \gets 0$.
 \For{$i$ from $1$ to $|\mathcal{Q}^c_m|$}  \Comment{Set $\mathcal{Q}^c_m$ is defined in \ref{app_pairings}}
 \State Initialize $\tensor{K}_i \gets 1$.
  \For{$j$ from $1$ to $m/2$}
\State Compute TT Hadamard product  $\tensor{K}_i \gets \tensor{K}_i \odot \mathbb{TT}(\mathbf{B},m,\mathfrak{q}_{i,j})$, where $\mathbb{TT}(\cdot,\cdot,\cdot)$ is TT extension operation \cref{eq_TT_index_augmentation}.
   \EndFor
   \State Perform TT rounding to $\tensor{K}_i$ to trim bonds with size larger than $R$ to $R$.
\State Compute TT sum $\tensor{L}_m \gets \tensor{L}_m + \tensor{K}_i$.
 \EndFor
 \State Perform TT rounding to $\tensor{L}_m$ to trim bonds with size larger than $R$ to $R$.
   \end{algorithmic}
\end{algorithm}

Inspired by the inclusion-exclusion principle for inchworm method \cite{boag2018inclusion,yang2021inclusion}, we can further reduce the computational cost for TT construction by an iterative procedure with respect to $m$.
For example, we can construct $\tensor{L}_6$ by $\tensor{L}_4$ and $\mathbf{B}$:
\begin{equation}
\label{eq_construnction_L6}
    \begin{split}
        \tensor{L}_6
        =\ & \mathbb{TT}(\textbf{B},6,[1,3])
        \odot \mathbb{TT}(\tensor{L}_4,6,[2,4,5,6]) \\
        +\ & \mathbb{TT}(\textbf{B},6,[1,4])
        \odot \mathbb{TT}(\tensor{L}_4,6,[2,3,5,6]) \\
        +\ & \mathbb{TT}(\textbf{B},6,[1,5])
        \odot \mathbb{TT}(\tensor{L}_4,6,[2,3,4,6]) \\
        +\ & \mathbb{TT}(\textbf{B},6,[1,4])
        \odot \mathbb{TT}(\textbf{B},6,[2,6]) \odot \mathbb{TT}(\textbf{B},6,[3,5])
    \end{split}
\end{equation}
In this construction, we delete the arc connecting with the first point. 
If the remaining diagram is still inchworm-proper, the remaining diagram (with two arcs) has been computed in $\tensor{L}_4$.
We need to construct the whole diagram from the beginning only if the remaining diagram is not inchworm proper.
The first three terms in \cref{eq_construnction_L6} construct a partial sum in $\mathcal{L}_b^c(\bs)$ for $m=6$ from the one with $m=4$ by inserting a pair connecting the first point and all middle points, respectively.
For the last term, when we remove the pair $(1,4)$, the remaining set $\{(2,6),(3,5)\}$ is not inchworm proper.
Therefore, in the last term, we construct it based on the idea of \ref{steps_tt2}.
Although it seems to save little in the construction of $\tensor{L}_6$ compared to \cref{eq_construction_L4_L6}, this idea to construct BIF-TTs does help reduce the computational cost especially when $m$ is large since $\tensor{L}_{m-2}$ is actually the sum of many terms.

In the construction of $\tensor{L}_m$, for $\mathfrak{q}\in\mathcal{Q}_m^c$, there exists a unique index $j$ such that $(1,j)\in\mathfrak{q}$. 
Consider the set $\mathfrak{q}\setminus \{(1,j)\}$, and relabel the indices to $\{1,2,\cdots,m-2\}$ to obtain a pairing $\mathfrak{q}_0 \in \mathcal{Q}_{m-2}$, there are two possible cases: 
\begin{itemize}
    \item if $\mathfrak{q}_0 \in \mathcal{Q}_{m-2}^c$, then the term $\prod_{(j,k)\in \mathfrak{q}} B(s_j,s_k)$ is included in the term $B(s_1,s_j) \mathcal{L}_b^c(s_2,\cdots,s_{j-1},s_{j+2},\cdots,s_m)$;
    \item if $\mathfrak{q}_0 \notin \mathcal{Q}_{m-2}^c$, then we need to construct the value corresponding to the pairing $\mathfrak{q}$ based on \ref{steps_tt2}.
\end{itemize}


To better demonstrate the advantage, we also present the procedure for the case $m=8$ in the appendix.
For $m=8$, the original inchworm BIF has 27 terms, while our method only requires computation and summation of 12 terms.
Furthermore, when $m=10$, the number of terms in our method is 66 versus 248 terms in the original summation.

\subsection{Computational cost}
\label{sec_cost}
In this subsection, we discuss the overall computational cost of the method. In the method we have proposed, the first step is the construction of BIF-TT, summarized in \Cref{algo:tt-bif}. 
In the procedures, performing $m/2$ Hadamard products (Line 7) in each iteration for outer loop produces a TT $\tensor{K}_i$ with a maximum bond dimension of $r^{m/2}$ and the complexity is $\mathcal{O}(mr^m N)$. The TT rounding for $\tensor{K}_i$ has the major cost on the right-to-left orthogonalization, which has complexity at $\mathcal{O}(mr^{3m/2})$ for sufficiently large $m$. After $\tensor{K}_i$ is rounded, the sum of $|\mathcal{Q}^c_m|$ TTs eventually gives a TT with the largest bond dimension $m!!R$. The final step of rounding then has complexity $\mathcal{O}(m(m!!R)^3)$. Overall, the computational complexity is
\begin{equation}
    \mathcal{O}\left(  m(m!!)r^{3m/2} +  m(m!!R)^3   \right).
\end{equation}
The iterative method demonstrated in \eqref{eq_construnction_L6} can effectively reduce the computational cost.
At present, we are unable to provide a precise complexity analysis of this iterative construction because we do not yet have an estimate for the number of terms appearing in the construction, as provided at the end of \Cref{sec_construct_bif_tt}.

For an odd $m$, we assume that the $(m+1)$-core BIF-TT representing the tensor $\tensor{L}_{m+1}$ has bond dimensions $(r_1,\cdots,r_m)$.
We now estimate the computational cost for the evaluation of the $m$-dimensional integral \eqref{eq_integral_to_compute} using the algorithm introduced in \Cref{sec_TT_integral}.
We begin with equation \cref{eq_omega_start}, where each $\Omega^{(m)}(k_m,\alpha_{m-1})$ requires summation over $r_m$ terms, and this has to be done for all $k_m$ and $\alpha_{m-1}$.
Hence, the computational cost of \cref{eq_omega_start} is estimated as $\mathcal{O}(r_{m-1}r_m(k_\ff-k_\ii))$.
Similarly, the computational cost for $I^{(m-1)}$ is 
\begin{equation}
\label{eq_compcost_I_start}
    \mathcal{O}\left(\sum_{k_{m-1}=k_\ii}^{k_\ff}r_{m-1}\left(k_\ff-k_{m-1}\right)\right)
    = \mathcal{O}(r_{m-1}(k_\ff-k_\ii)^2).
\end{equation}
By the same approach, one can find that the computational costs of \cref{eq_omega_middle} and \cref{eq_I_middle} are $\mathcal{O}(r_{j-1}r_j(k_\ff-k_\ii))$ and $\mathcal{O}(r_{j-1}(k_\ff-k_\ii)^2)$, respectively.
Thus, the overall complexity for the evaluation of this $m$-dimensional integral is
\begin{equation}
\label{eq_compcost_single_integral}
    \mathcal{O}\left(
        \sum_{j=1}^{m}
        \left(r_{j-1}r_j (k_\ff-k_\ii)
        + r_{j-1}(k_\ff-k_\ii)^2\right)
    \right)
    \lesssim \mathcal{O}\left(
        m(k_\ff-k_\ii)(k_\ff-k_\ii+R)R
    \right),
\end{equation}
where $R=\max\{r_1,\cdots,r_m\}$.
With the BIF-TT, the computational cost for each integral scales linearly with respect to the integral dimension.
If we do not employ the BIF-TT form, we would have to evaluate values of the integrand on $\mathcal{O}\left({(k_\ff-k_\ii)^m/{m!}}\right)$ points to compute the same integral.

Note that for each pair of $k_\ii$ and $k_\ff$, the integral \eqref{eq_integral_to_compute} must be calculated for all $m = 1,3,\cdots,\bar{M}$, and to find $G(-N\Delta t, N\Delta t)$, we need the full propagator for all $k_\ii$ and $k_\ff$ satisfying $-N \leqslant k_\ii \leqslant k_\ff \leqslant N$.
Therefore, the total computational cost of the algorithm is
\begin{equation}
    \mathcal{O}\left(
        \sum_{k_\ii=-N}^N \sum_{k_\ff=k_\ii}^N
        \sum_{\substack{m=1\\m\text{~is odd}}}^{\bar{M}}m(k_\ff-k_\ii)(k_\ff-k_\ii+R)R
    \right)
    = \mathcal{O}
    \left(
        \bar{M}^2
        N^3 R(N+R)
    \right).
\end{equation}



\section{Numerical results}
\label{sec_numerical_results}
In this section, we first introduce the spin-boson model, a simple but fundamental open quantum system model. Numerical experiments will be carried out for the simulation of the spin-boson model.

\subsection{Spin-boson model}
\label{subsec_spin_boson_model}
The spin-boson model depicts a single spin immersed in a quantum harmonic bath, where the evolution of the density matrix of the coupled system is governed by the dimensional von Neumann equation \eqref{eq_von_Neumann}:
\begin{equation} \label{eq:vonNeumann}
\mathrm{i} \hbar \frac{\mathrm{d}}{\mathrm{d}t} \rho(t) = [H_s + H_b + W, \rho(t)],
\end{equation}
where
\begin{displaymath}
    H_s = \epsilon \hat{\sigma}_z + \Delta \hat{\sigma}_x, \quad
    H_b = \sum_{j=1}^L \frac{1}{2}\left(\frac{\hat{p}_j^2}{m_j} + m_j \omega_j^2 \hat{q}_j^2\right), \quad
    W = \hat{\sigma}_z \otimes \left( \sum_j c_j \hat{q}_j \right).
\end{displaymath}
In this model, $\hat{\sigma}_x$, $\hat{\sigma}_z$ are Pauli matrices, and $m_j$, $\hat{p}_j$ and $\hat{q}_j$ are the mass, momentum operator and position operator of the $j$th harmonic oscillator, respectively.
The coefficient $\epsilon$ represents the energy difference between two spin states, and $\Delta$ denotes the frequency of the spin flipping. The parameter $\omega_j$ is the frequency of the $j$th harmonic oscillator while $c_j$ stands for the coupling intensity between the spin and the $j$th harmonic oscillator.
For simplicity, we consider the dimensionless form of the spin-boson model by the following changes of variables:
\begin{gather*}
t' = t\Delta /\hbar, \quad \epsilon' = \epsilon/\Delta, \quad \hat{q}_j' = \hat{q}_j \sqrt{m_j \Delta} / \hbar, \quad \omega_j' = \hbar \omega_j/\Delta, \\
\hat{p}_j' = \hat{p}_j / \sqrt{m_j \Delta}, \quad c_j' = \hbar c_j / \sqrt{m_j \Delta^3}, \quad \rho'(t') = \left( \frac{\hbar}{\sqrt{m_j \Delta}} \right)^L \rho(t).
\end{gather*}
After the above change of variables and dropping all primes, the Hamiltonians take the dimensionless form
\begin{equation}\label{eq_spin_boson_model}
    H_s = \epsilon \hat{\sigma}_z + \Delta \hat{\sigma}_x, \quad
    H_b = \sum_j \frac{1}{2}\left(\hat{p}_j^2 + \omega_j^2 \hat{q}_j^2\right), \quad
    W = \hat{\sigma}_z \otimes \left( \sum_j c_j \hat{q}_j \right).
\end{equation}
Here the dimensionless $\omega_j$ and $c_j$ are identical to those appearing in \cref{eq_spectral_density}, and the flipping frequency $\Delta$ is set to 1. The inverse temperature is also expressed in dimensionless form, namely $\beta = \Delta/(k_B T)$ where $k_B$ is the Boltzmann constant, throughout the numerical experiments in \Cref{subsec_bond_dim_biftt} - \Cref{subsec_ttm}.
 
In our numerical test, we set the dimensionless frequencies and coupling intensities based on the Ohmic spectral density \cite{makri1999linear,chen2017inchwormIIBenchmarks} by
\begin{equation}\label{eq_spectral_density_paras}
    \begin{aligned}
        \omega_l &=-\omega_c \ln \left(1-\frac{l}{L}\left[1-\exp \left(-\omega_{\max } / \omega_c\right)\right]\right),\\
         c_l &=\omega_l \sqrt{\frac{\xi \omega_c}{L}\left[1-\exp \left(-\omega_{\max } / \omega_c\right)\right]},
    \end{aligned}
\end{equation}
where the number of harmonic oscillators $L$, the primary frequency of the harmonic oscillators $\omega_c$, and the maximum frequency $\omega_{\text{max}}$ are set as
$$L = 400, \quad \omega_c=2.5, \quad \omega_{\text{max}}=4\omega_c.$$

Based on the above choice of parameters, the two-point correlation function is proportional to the Kondo parameter $\xi$. Hence, the bath influence functional $\mathcal{L}_b^c(\bs)$ is proportional to $\xi^{m/2}$. 
This fact allows us to compute the BIF-TT only for $\xi=1$ and utilize the result across all models with different $\xi$, which, as a result, significantly reduces the computational cost associated with constructing the BIF-TT for each individual case.

\subsection{Bond dimensions of BIF-TT}
\label{subsec_bond_dim_biftt}
Before carrying out computations of the spin-boson model, we will first study the BIF-TT constructed using the algorithm in \Cref{sec_construct_bif_tt}, focusing on its rank under various physical and numerical settings.

In \Cref{thm_rank_bound} we established an upper bound for the rank of the two-point correlation matrix $\mathbf{B}$, given by $r\leqslant \min\{N+1, 2L\}$. However, {our \texttt{MATLAB} experiments show that the actual rank $r$ is often significantly lower than this theoretical upper bound. For instance, the growth of $\text{rank}(\mathbf{B})$ with increasing number of time steps $N$ is shown in \Cref{fig:rank_dt01}, where we vary the inverse temperature parameter $\beta$ over 5, 2, and 1, while fixing time step size $\dt = 0.1$. 
As shown in the figure, the rank of $\mathbf{B}$ is much lower than the theoretical bound $\min\{N+1,2L\}$ given by \Cref{thm_rank_bound}.
In fact, $\text{rank}(\mathbf{B})$ is more closely related to the maximum simulation time $N\Delta t$. It can be seen from \Cref{fig:rank_dt010204} that for the same $N\dt$ with different time step $\Delta t=0.1,0.2,0.4$, the ranks of $\mathbf{B}$ are similar.
We also show an example of the ``numerical low-rank'' property of matrix $\mathbf{B}$, as discussed at the end of \Cref{sec_tpc_matrix}.
Specifically, we fix $N = 500$ and $\dt = 0.1$, and plot the decay of the singular values of Hermitian $\mathbf{B} \in \mathbb{R}^{1001 \times 1001}$ for $\beta = 5, 2, 1$ in \Cref{fig:spectrum_B}. 
As shown, the singular values $\lambda_i$ decay rapidly, 
confirming that $\mathbf{B}$ exhibits strong numerical compressibility.



\begin{figure}
    \centering
    \begin{subfigure}[b]{0.48\textwidth} 
        \centering
\includegraphics[width=1.1\textwidth]{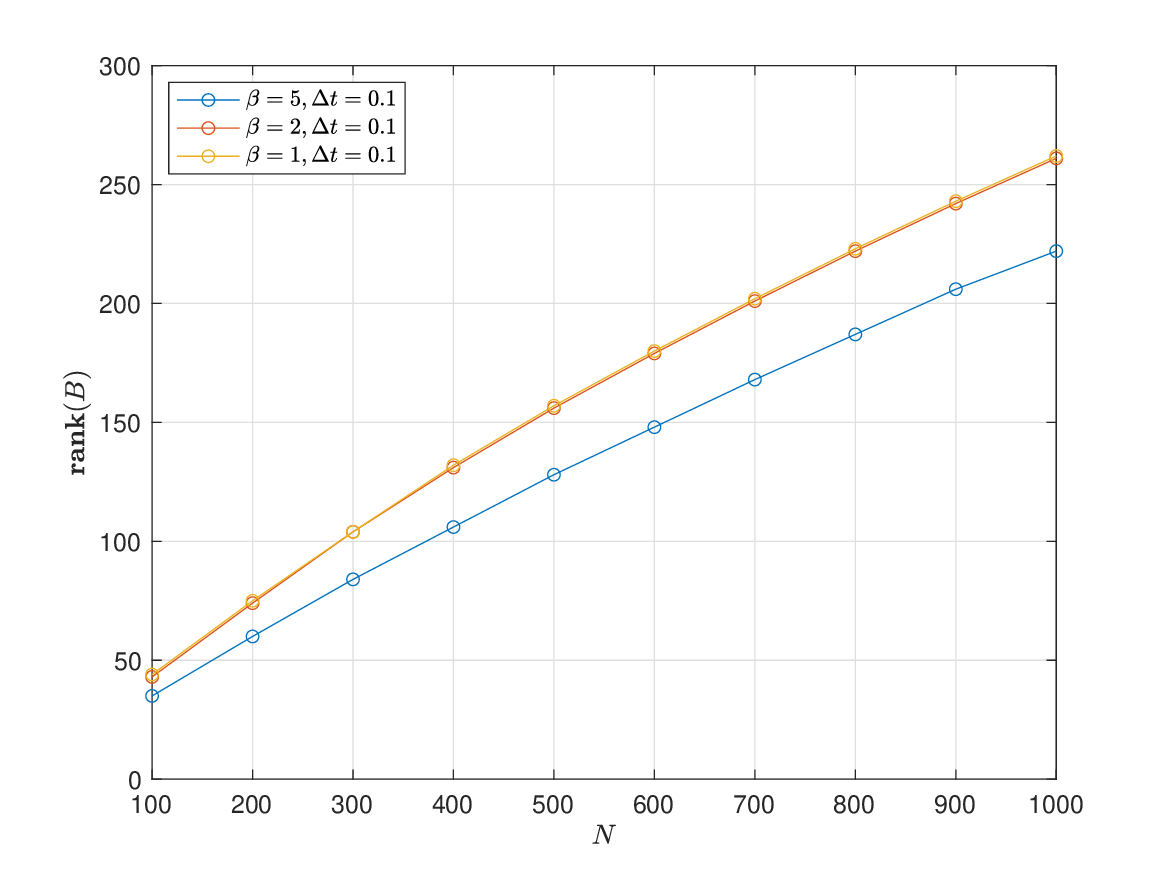}
        \caption{Growth of $\text{rank}(\mathbf{B})$ with respect to $N$.}
        \label{fig:rank_dt01}
    \end{subfigure}
    \hfill 
    \begin{subfigure}[b]{0.48\textwidth} 
        \centering
        \includegraphics[width=1.1\textwidth]{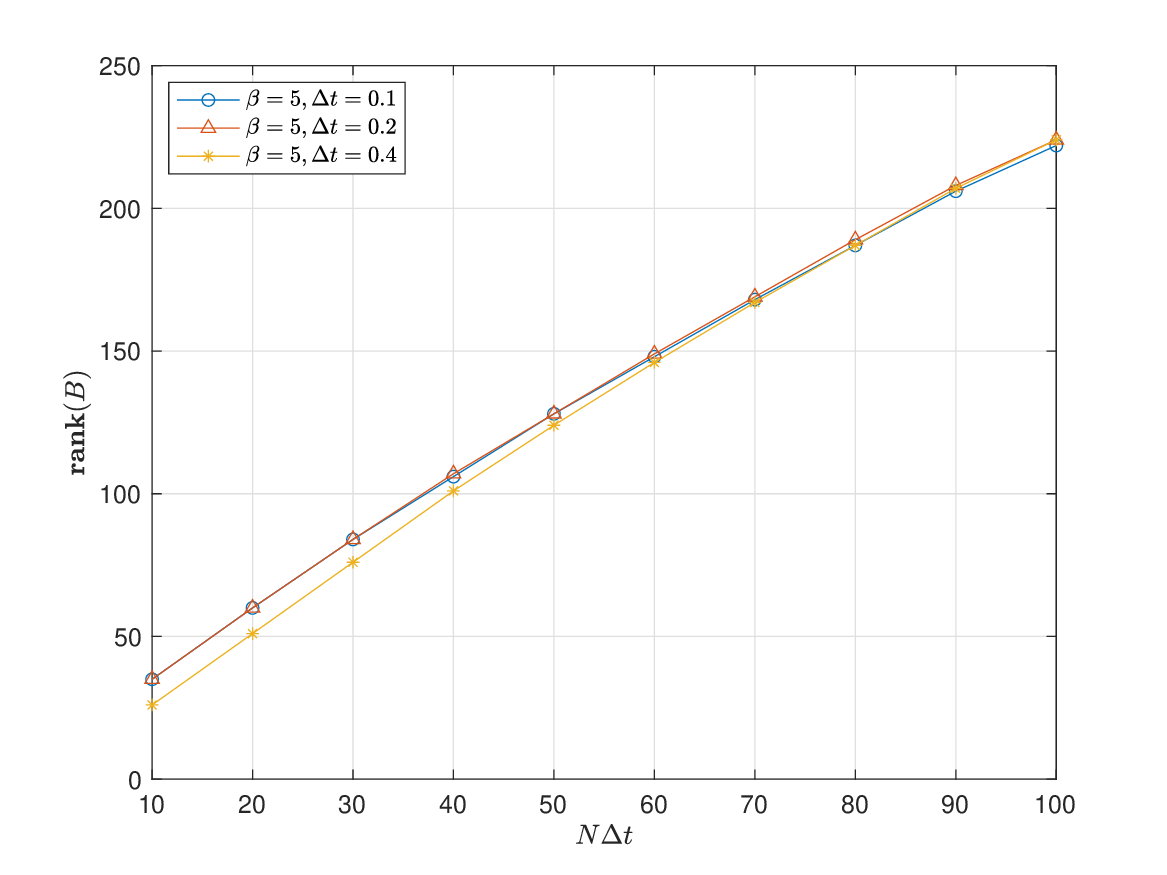}
        \caption{Growth of $\text{rank}(\mathbf{B})$ with respect to $N\dt$.}
        \label{fig:rank_dt010204}
    \end{subfigure}
    \caption{Numerical rank of the two-point correlation matrix $\mathbf{B}$.}
    \label{fig:rankB}
\end{figure}

\begin{figure}
    \centering
    \includegraphics[width=.6\textwidth]{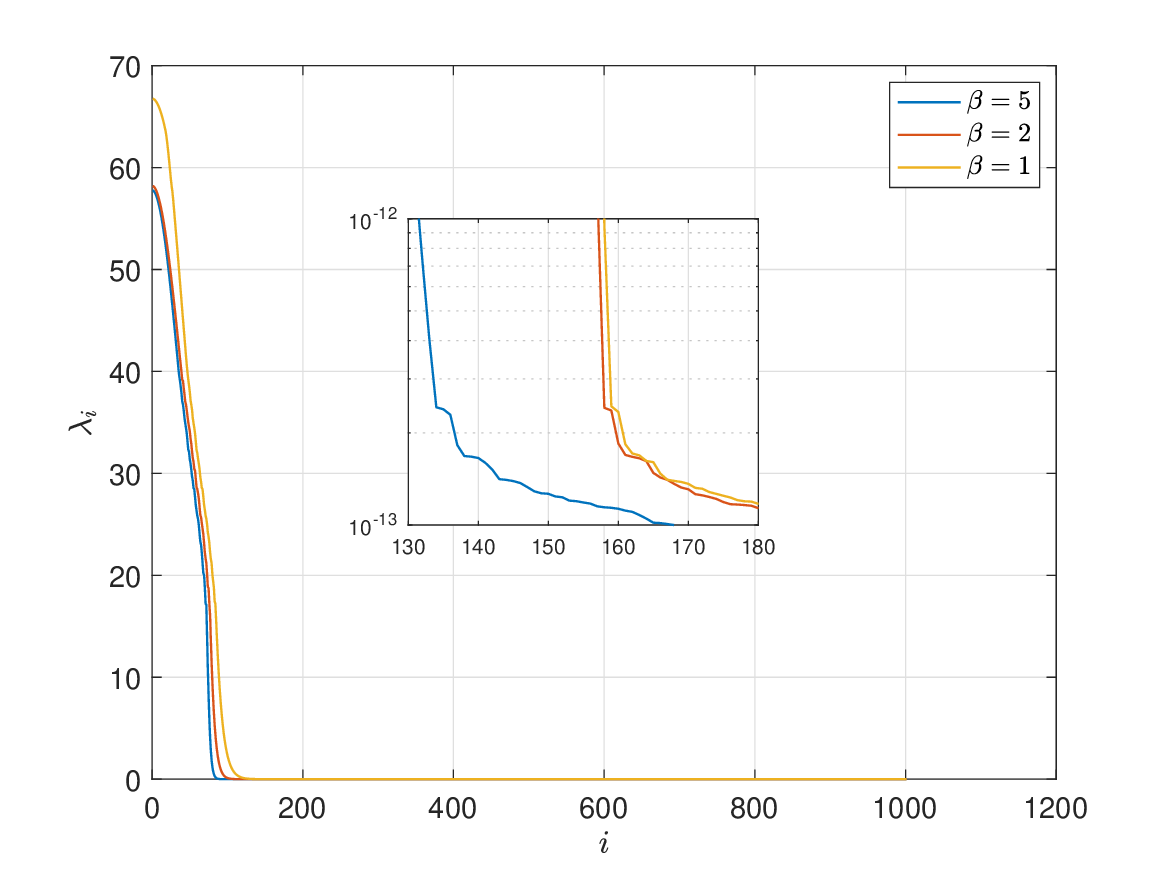}
    \caption{Decay of the singular values $\lambda_i$ of the two-point correlation matrix $\mathbf{B}$. }
    \label{fig:spectrum_B}
\end{figure}

We now consider the TT-rounded bath influence functional $\hat{\tensor{L}}_{m+1}$, whose accuracy is governed by the TT-rounding tolerance $\eta$, satisfying
$\Vert \tensor{L}_{m+1} - \hat{\tensor{L}}_{m+1} \Vert_F \leqslant \eta \Vert \tensor{L}_{m+1} \Vert_F$.
Fixing the time step size at $\dt = 0.2$, we examine the effect of different inverse temperatures $\beta = 5, 2, 1$ and TT-rounding tolerances $\eta = 1.0 \times 10^{-4},\ 1.0 \times 10^{-6}$ on the bond dimension of $\hat{\tensor{L}}_{m+1}$.
Due to memory constraints, more time steps $N$ are considered for $m=3$ compared to the case $m=5$.
In \Cref{fig:bonddim_m3_dt02,fig:bonddim_m5_dt02},
 the solid lines correspond to the results without TT-rounding and the dashed lines represent the results with different TT-rounding tolerances.
As shown in \Cref{fig:bonddim_m3_dt02,fig:bonddim_m5_dt02}, the maximum bond dimension increases approximately 
quadratically in $N$.
Additionally, we observe that higher temperatures (corresponding to smaller $\beta$) result in larger bond dimensions. This suggests that thermal fluctuations lead to increased complexity in the TT representation of the BIF.
 For the relative error, defined as $\Vert \tensor{L}_{m+1} - \hat{\tensor{L}}_{m+1}\Vert_F / \Vert \tensor{L}_{m+1}\Vert_F$, we notice that it exactly matches the rounding tolerance $\eta$ as shown in \Cref{fig:rel_error_m3,fig:rel_error_m5}, confirming the numerical stability of the approximation. For the absolute error, defined as $\Vert \tensor{L}_{m+1} - \hat{\tensor{L}}_{m+1}\Vert_F$, we observe a clear decreasing trend with increasing $\beta$, as shown in \Cref{fig:abs_error_m3,fig:abs_error_m5}. 
\begin{figure}
\centering
\begin{subfigure}[b]{0.48\textwidth}
    \includegraphics[width = 1.1\textwidth]{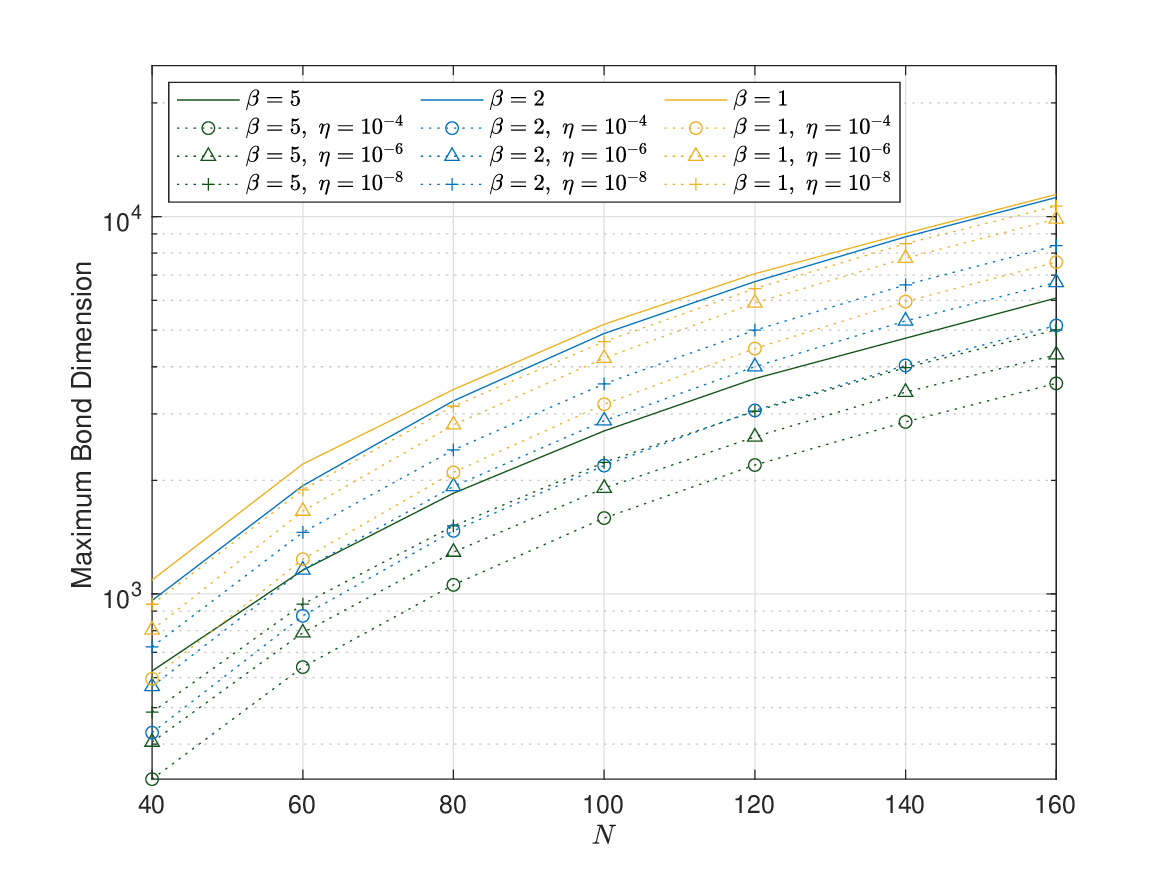}
    \caption{Maximum bond dimension: $m=3$.}
    \label{fig:bonddim_m3_dt02}
    \end{subfigure}
\hfill
\begin{subfigure}[b]{0.48\textwidth}
        \includegraphics[width = 1.1\textwidth]{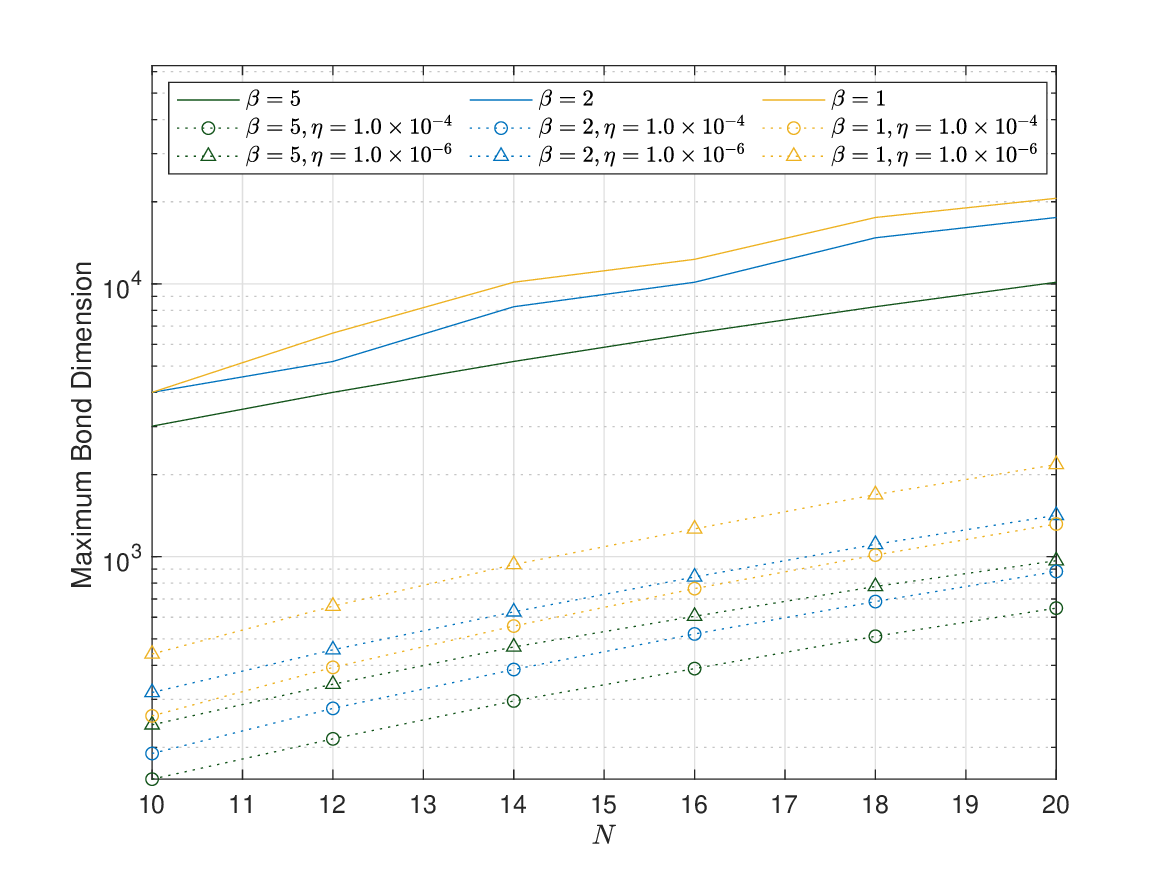}
    \caption{Maximum bond dimension: $m=5$.}
    \label{fig:bonddim_m5_dt02}
    \end{subfigure}
    
\vspace{0.5em}

 \begin{subfigure}[b]{0.48\textwidth}
    \includegraphics[width = 1.1\textwidth]{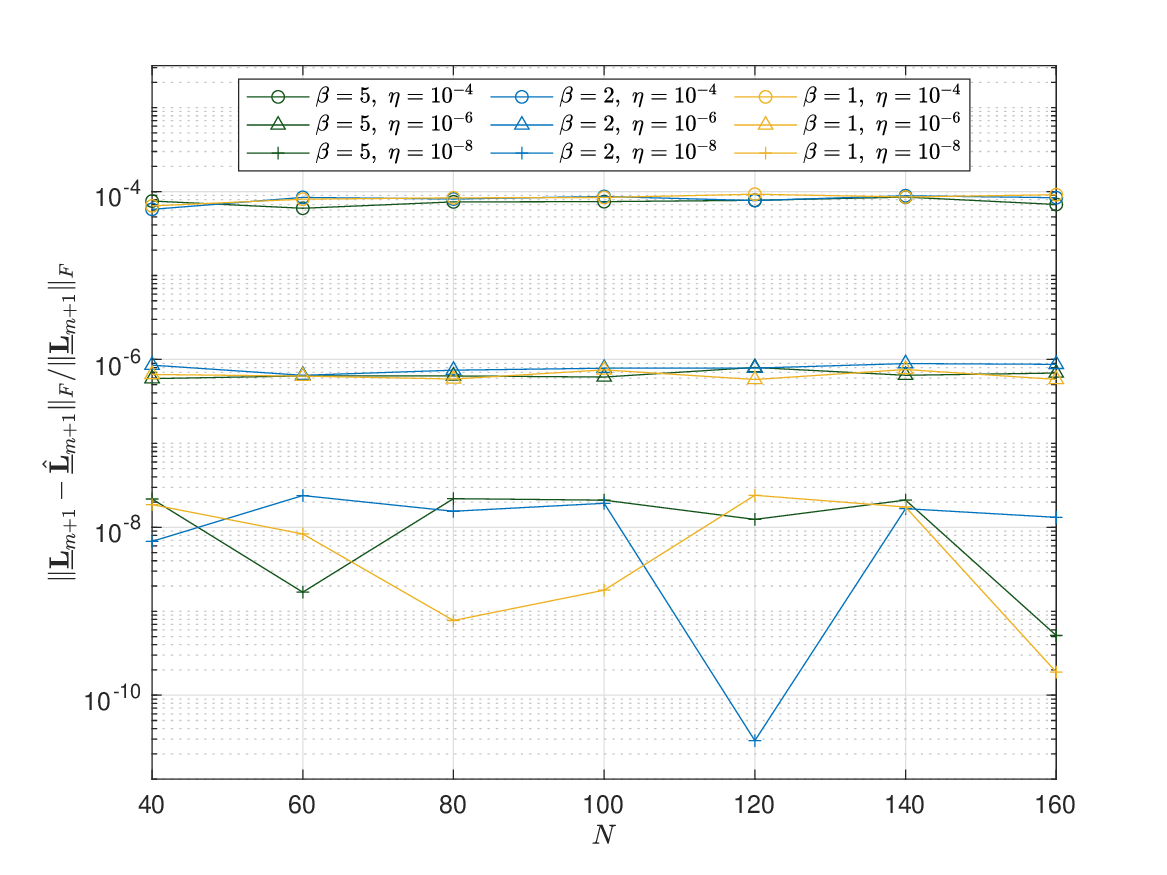}
    \caption{Relative error: $m=3$.}
    \label{fig:rel_error_m3}
    \end{subfigure}
    \hfill
    \begin{subfigure}[b]{0.48\textwidth}
        \includegraphics[width = 1.1\textwidth]{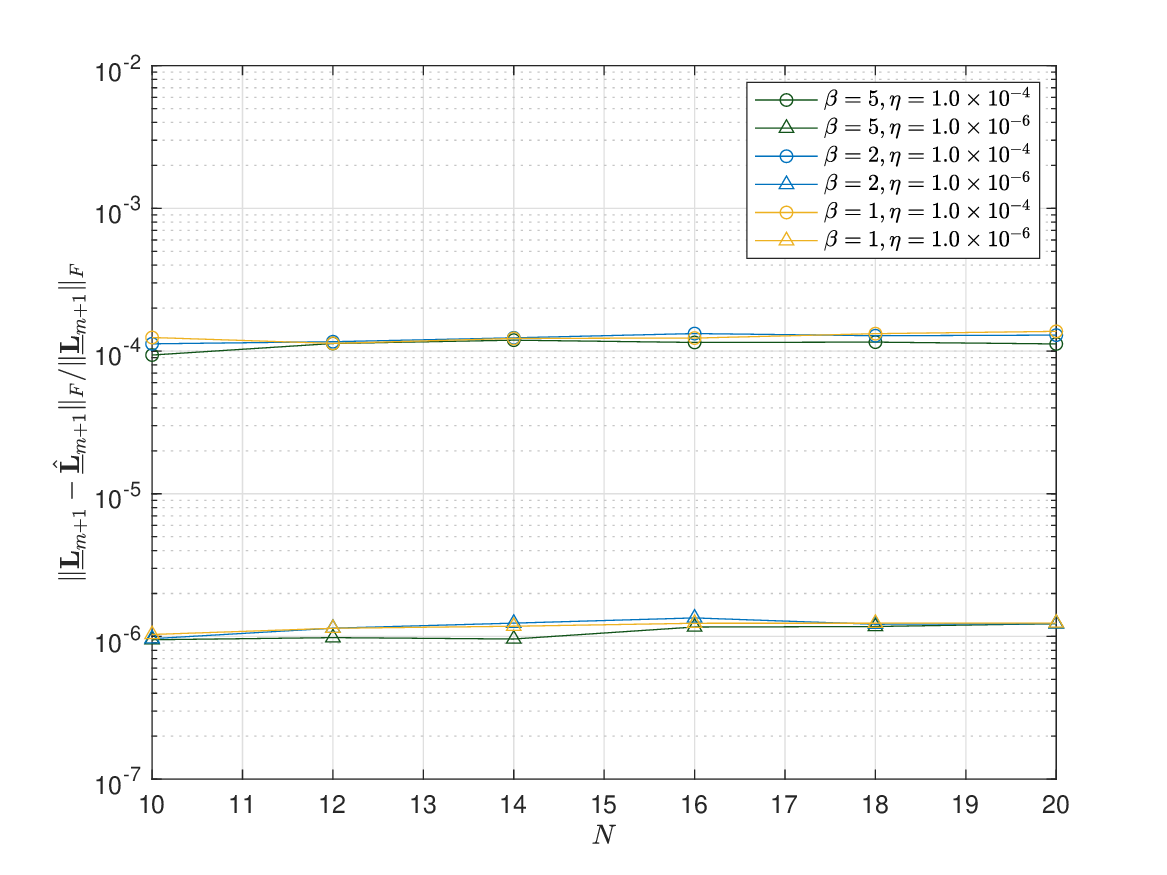}
        \caption{Relative error: $m=5$.}
        \label{fig:rel_error_m5}
    \end{subfigure}
    \vspace{0.5em}
     \begin{subfigure}[b]{0.48\textwidth}
        \includegraphics[width = 1.1\textwidth]{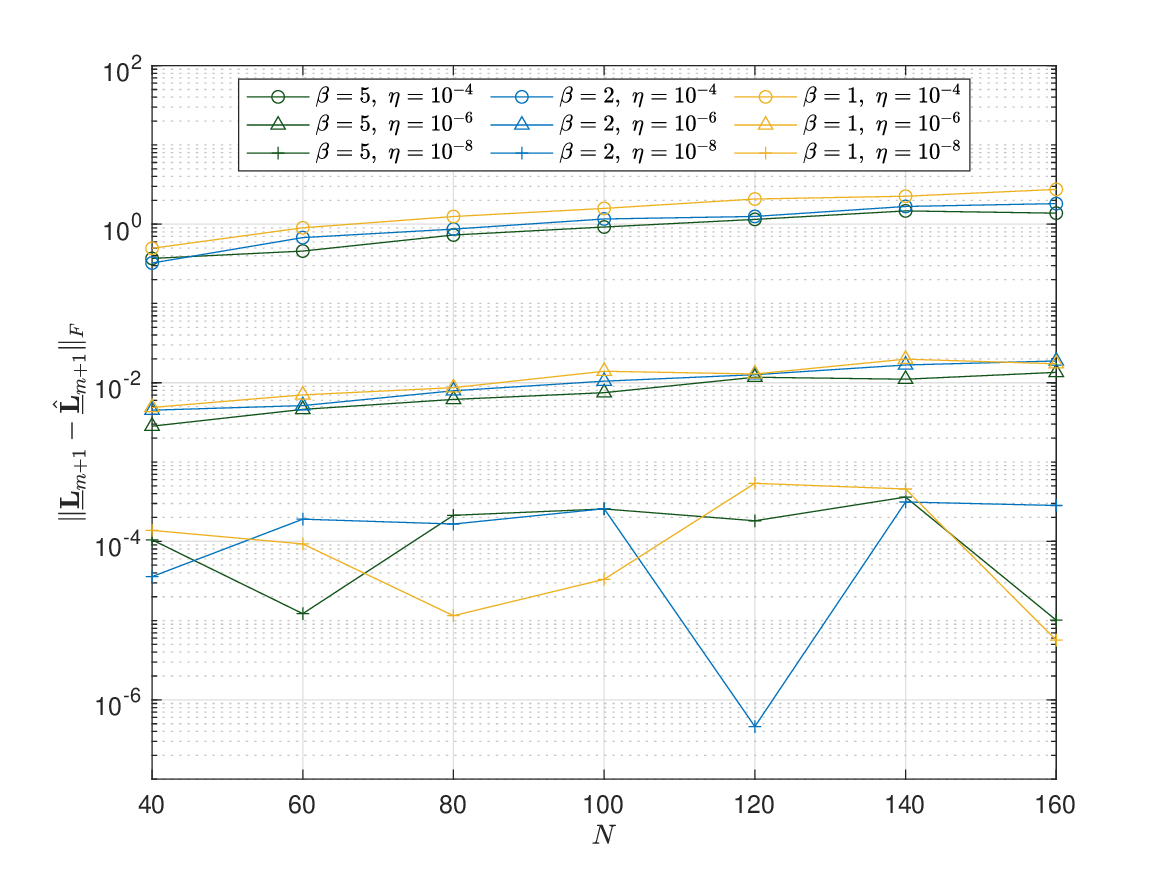}
        \caption{Absolute error: $m=3$.}
        \label{fig:abs_error_m3}
    \end{subfigure}
    \hfill    
    \begin{subfigure}[b]{0.48\textwidth}
        \includegraphics[width = 1.1\textwidth]{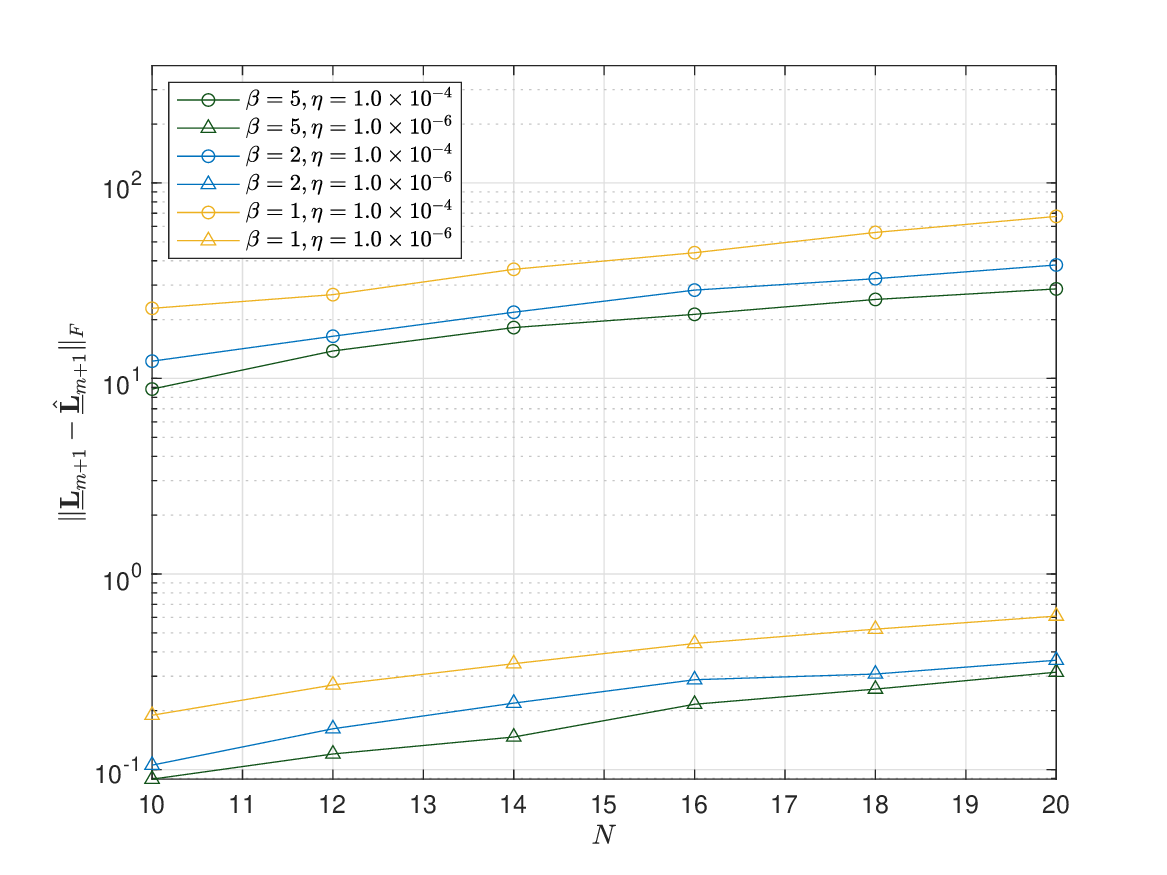}
        \caption{Absolute error: $m=5$.}
        \label{fig:abs_error_m5}
    \end{subfigure}
    \caption{Properties of rounded TT $\hat{\tensor{L}}_{m+1}$ for different $\beta=1,2,5$ and rounding tolerance $\eta$ when $m=3,5$ and $\dt = 0.2$.}
    \label{fig:Lb_dt02}
\end{figure}

\subsection{Convergence tests of the inchworm method with BIF-TTs}
\label{subsec_conv_inchworm}
In practice, the infinite series in \cref{eq_inchworm_integro_differential_eq} is truncated for computational feasibility.
Specifically, the summation over odd $m$ is restricted to the range $m = 1, 3, \ldots, M$, where $M$ is chosen to balance accuracy and computational cost.
The second-order composite trapezoidal rule is adopted to evaluate the $M$-dimensional integral.

To show the performance enhancement of the numerical integration with BIF-TT, we compare the computational times for the methods with and without using the TT structure.
The results are plotted in \Cref{fig:single_integral_time_n10,fig:single_integral_time_n20}. Here, $N$ denotes the number of time steps used to discretize the integration interval $[0, N\Delta t]$.
Generally speaking, for larger values of $N$ and $M$, the classical method without TT, which requires $\mathcal{O}(N^M)$ operations, is less affordable due to the faster growth of the computational cost, while our method \eqref{eq_biftt_iteration} shows a much flatter curve, demonstrating its clear superiority for high-dimensional integrations.
To further evaluate the scalability of the BIF-TT integration, we also test its performance for larger $N$. \Cref{fig:single_integral_biftt_time} presents the evaluation time per single $M$-dimensional integral with BIF-TT for $N = 100, 150, 200$. In \Cref{fig:single_integral_biftt_ram}, we plot the corresponding storage requirements of the rounded BIF-TT tensors $\hat{\tensor{L}}_M$ with a maximum bond dimension of 500, which also scale moderately with respect to $M$ and $N$.

\begin{figure}
    \centering
     \begin{subfigure}[b]{0.45\textwidth} 
        \centering
        \includegraphics[width=1.1\textwidth]{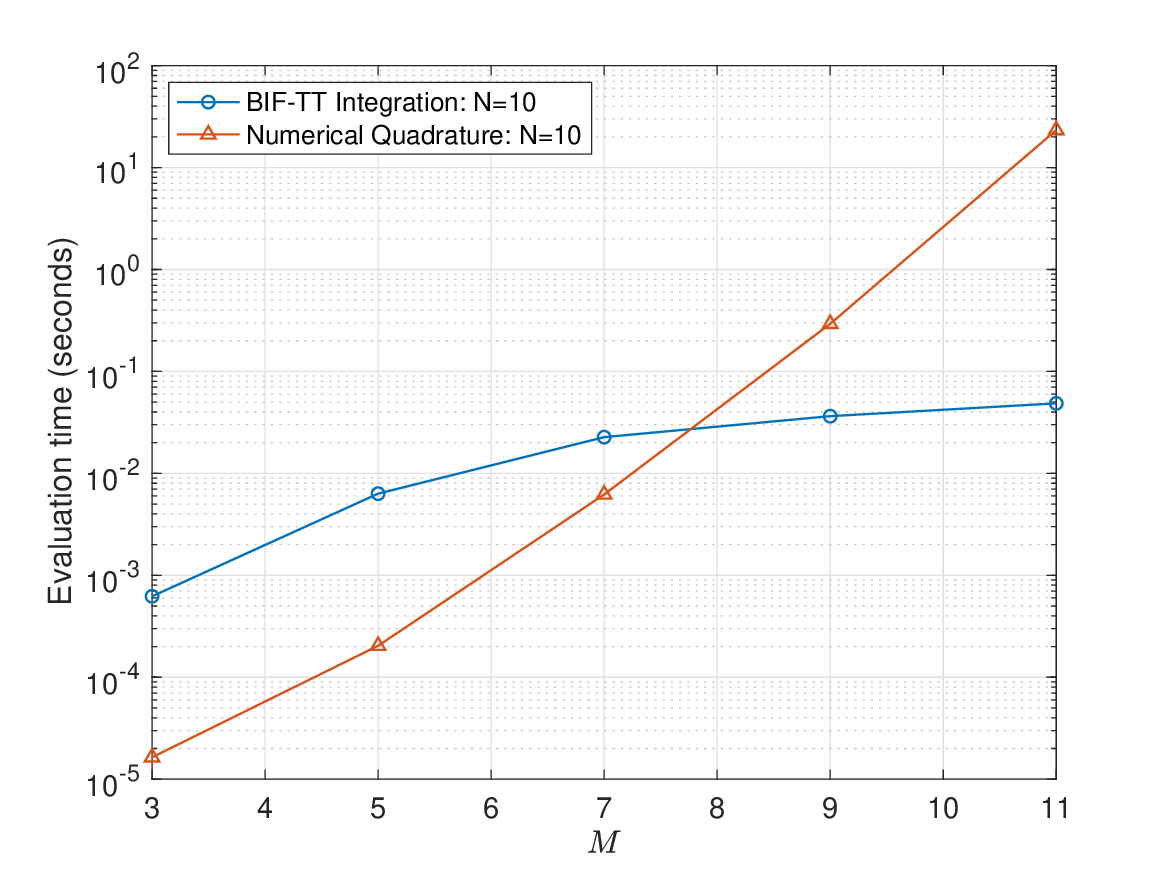}
    \caption{Comparison of evaluation time for a single $M$-dimensional integral using BIF-TT integration and direct numerical quadrature for $N = 10$.}
    \label{fig:single_integral_time_n10}
    \end{subfigure}
\hfill
\begin{subfigure}[b]{0.45\textwidth} 
        \centering
    \includegraphics[width=1.1\textwidth]{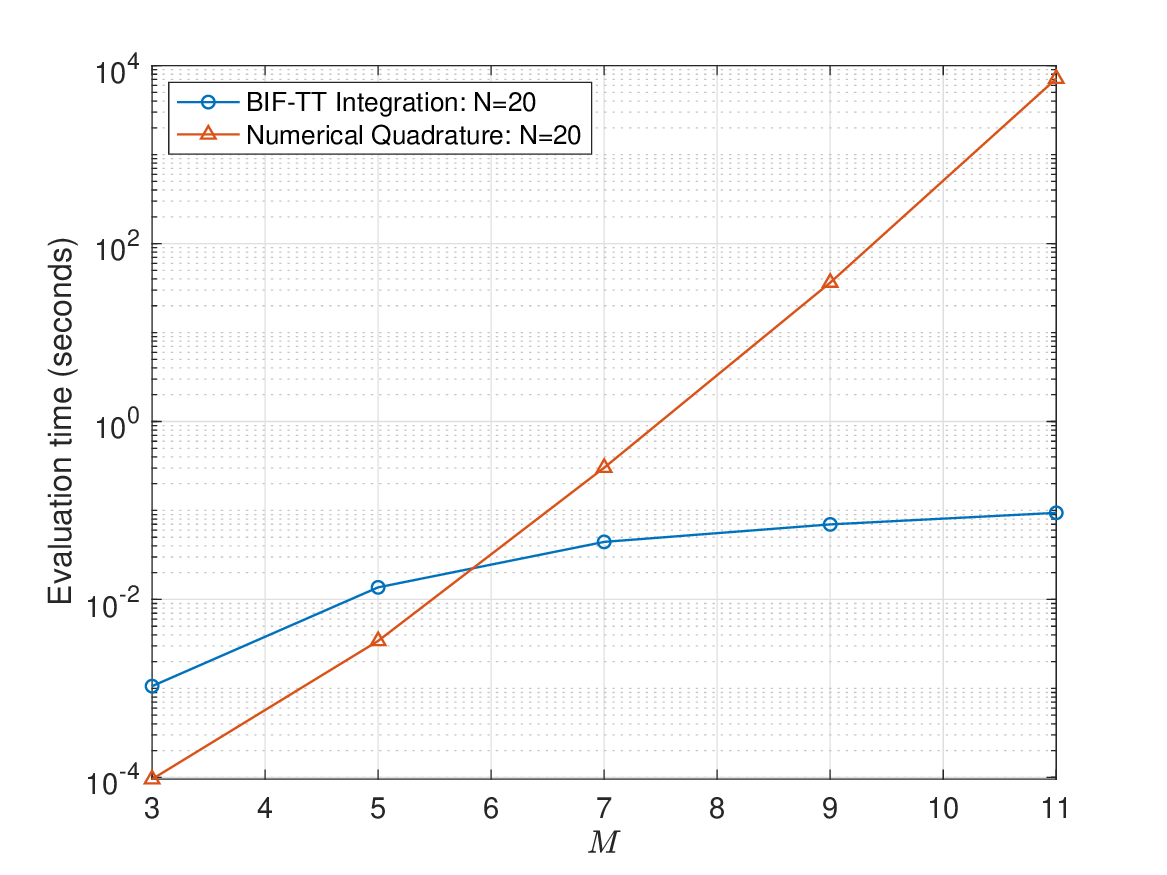} 
    \caption{Comparison of evaluation time for a single $M$-dimensional integral using BIF-TT integration and direct numerical quadrature for $N = 20$.}
\label{fig:single_integral_time_n20}
\end{subfigure}\\
\begin{subfigure}[b]{0.45\textwidth} 
        \centering
        \includegraphics[width=1.1\textwidth]{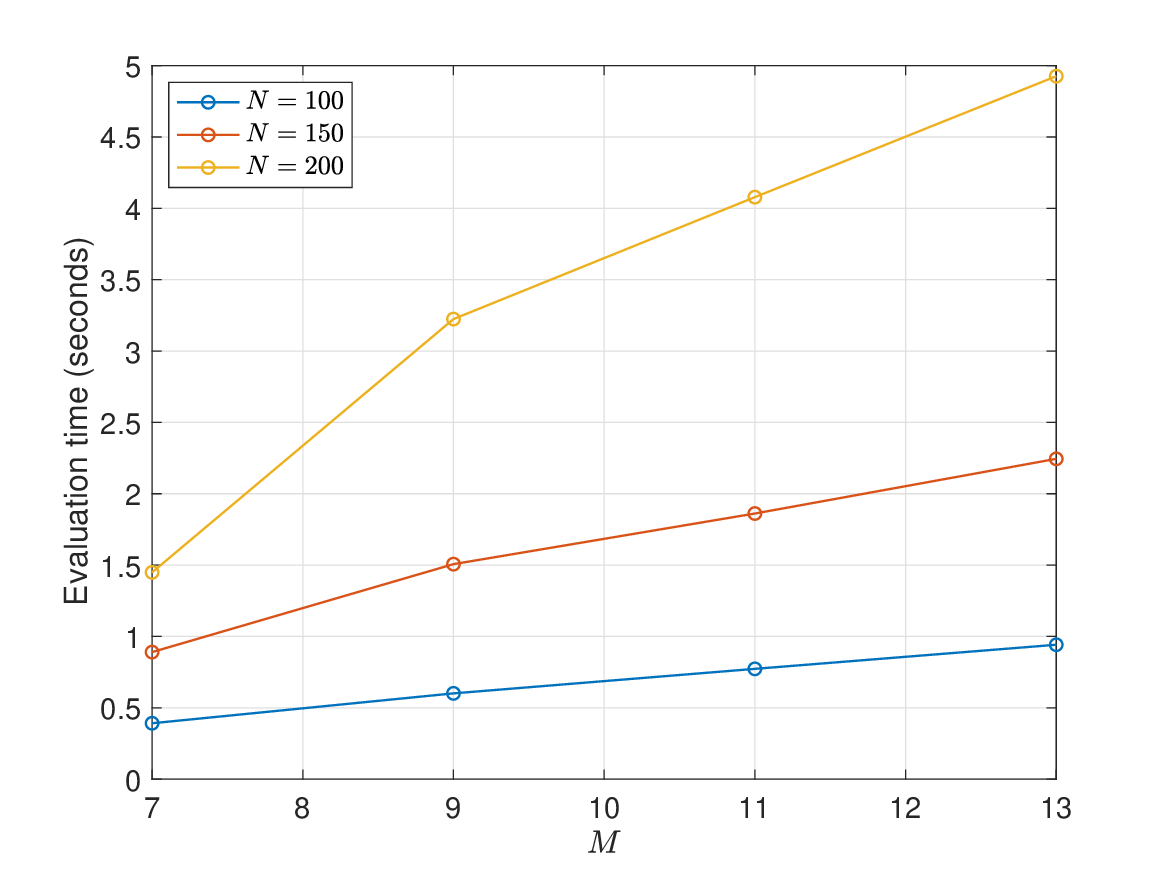} 
    \caption{Evaluation time per single $M$-dimensional integral using BIF-TT integration for $N = 100, 150, 200$.}
\label{fig:single_integral_biftt_time}
\end{subfigure}
\hfill
\begin{subfigure}[b]{0.45\textwidth} 
        \centering
    \includegraphics[width=1.1\textwidth]{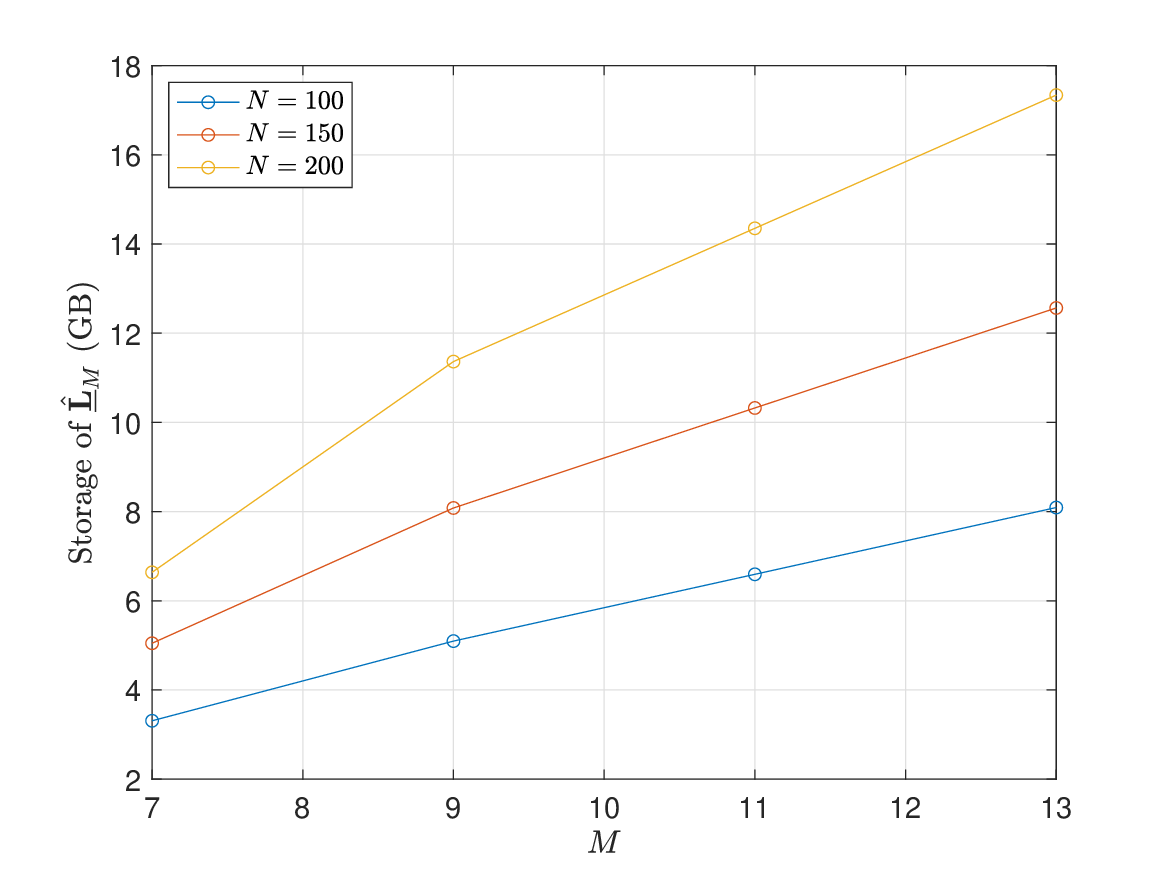} 
    \caption{Storage requirements (in GB) of rounded BIF-TT tensors $\hat{\tensor{L}}_M$ with a maximum bond dimension of 500 for $N = 100, 150, 200$.}
\label{fig:single_integral_biftt_ram}
\end{subfigure}
\caption{Evaluation time and storage analysis of BIF-TT integration for single $M$-dimensional integrals.}
\label{fig:single_integral_time_m13}
\end{figure}

We now present the convergence test of the inchworm method using the BIF-TTs constructed in the previous section.
Consider the model parameters
$$\Delta = 1.0, \quad \epsilon = 1.0, \quad \beta = 5.0$$
and investigate the convergence behavior. For the cases $M=1,3,5$,
the maximum bond dimension of the BIF-TT $\tensor{L}_{M+1}$ is $3r^3 + r$ where $r$ is the rank of matrix $\mathbf{B}$ as defined in \cref{eq_tpc_matrix_fftpc_matrix}. Even without rounding, this bond dimension is manageable within the scope of our experiments. Therefore, for these values of $M$, we will carry out simulations with exact BIF-TTs, allowing us to check the convergence with respect to other parameters without introducing TT-rounding errors.

We represent the observables $\langle \sigma_z(t)\rangle$ computed using time step size $\Delta t= 0.1, 0.2, 0.4$ as $\langle \sigma_z(t)\rangle_{0.1}$, $\langle \sigma_z(t)\rangle_{0.2}$, and $\langle \sigma_z(t)\rangle_{0.4}$, respectively. 
Due to the lack of an exact solution,  the numerical convergence order $p$ is estimated by
\begin{equation}
\label{eq:convergence_order}
    p = \log_2 \frac{\norm{\langle \sigma_z(t)\rangle_{0.4}-\langle \sigma_z(t)\rangle_{0.2}}_2}{\norm{\langle \sigma_z(t)\rangle_{0.2}-\langle \sigma_z(t)\rangle_{0.1}}_2}
\end{equation}
where \(\norm{\cdot}_2\) denotes the Euclidean norm. The results are displayed in \Cref{table_convergence_dt}. Taking $M=1$, $3$, and $5$ with no TT-rounding applied, a second-order convergence in time step $\Delta t$ can be observed. For the case $M=5$, we also display the figures in \Cref{fig:conv_dt} illustrating the evolution of $\langle\sigma_z(t)\rangle$, showing that stronger coupling between the system and the bath leads to less quantum fluctuations.
Here we would like to note that these examples cannot be accurately simulated by taking a Markovian approximation, even for $\xi = 0.2$. In the weak coupling regime, the dynamics of the reduced density matrix is usually approximated by the Lindblad equation \cite{lindblad1976generators}, whose numerical solutions are also presented in \Cref{fig:conv_dt}. While the differences between the Lindblad solutions and ours become smaller for weaker couplings, the non-Markovian effect is evident in all the four cases. This illustrates the necessity of using a tensor train to represent the BIF.

\begin{table}[!ht]
    \centering
    \caption{Convergence order $p$ \eqref{eq:convergence_order} of the inchworm method for various $\xi$ and $M$.}
    \label{table_convergence_dt}
    \begin{tabular}{c|c|c|c}
        &  $\xi = 0.2$ & $\xi = 0.4$ & $\xi = 0.8$\\ \hline
        $M=1$ & $p=2.1853$ & $p = 2.4965$ & $p = 2.5835$ \\
        $M=3$ & $p=2.0963$ & $p=2.3707$ & $p=2.5754$\\
        $M=5$ & $p=2.0874$ & $p= 2.3366$ & $p = 2.3621$
    \end{tabular}
\end{table}

\begin{figure}[!ht]
    \centering
    \begin{subfigure}[b]{0.45\textwidth} 
        \centering
        \includegraphics[width=1.1\textwidth]{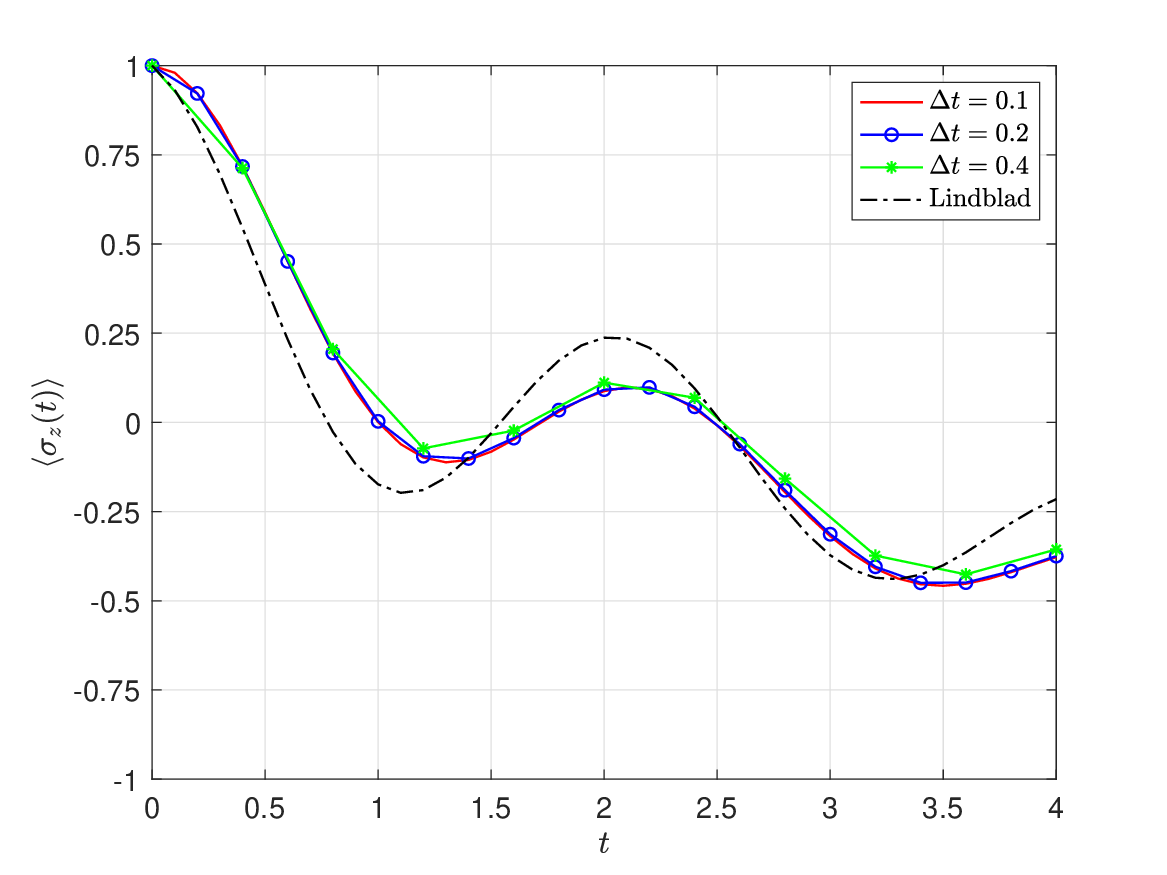}
        \caption{$\xi = 0.2$}
        \label{fig:conv_dt_xi02}
    \end{subfigure}
    \hfill 
    \begin{subfigure}[b]{0.45\textwidth} 
        \centering
        \includegraphics[width=1.1\textwidth]{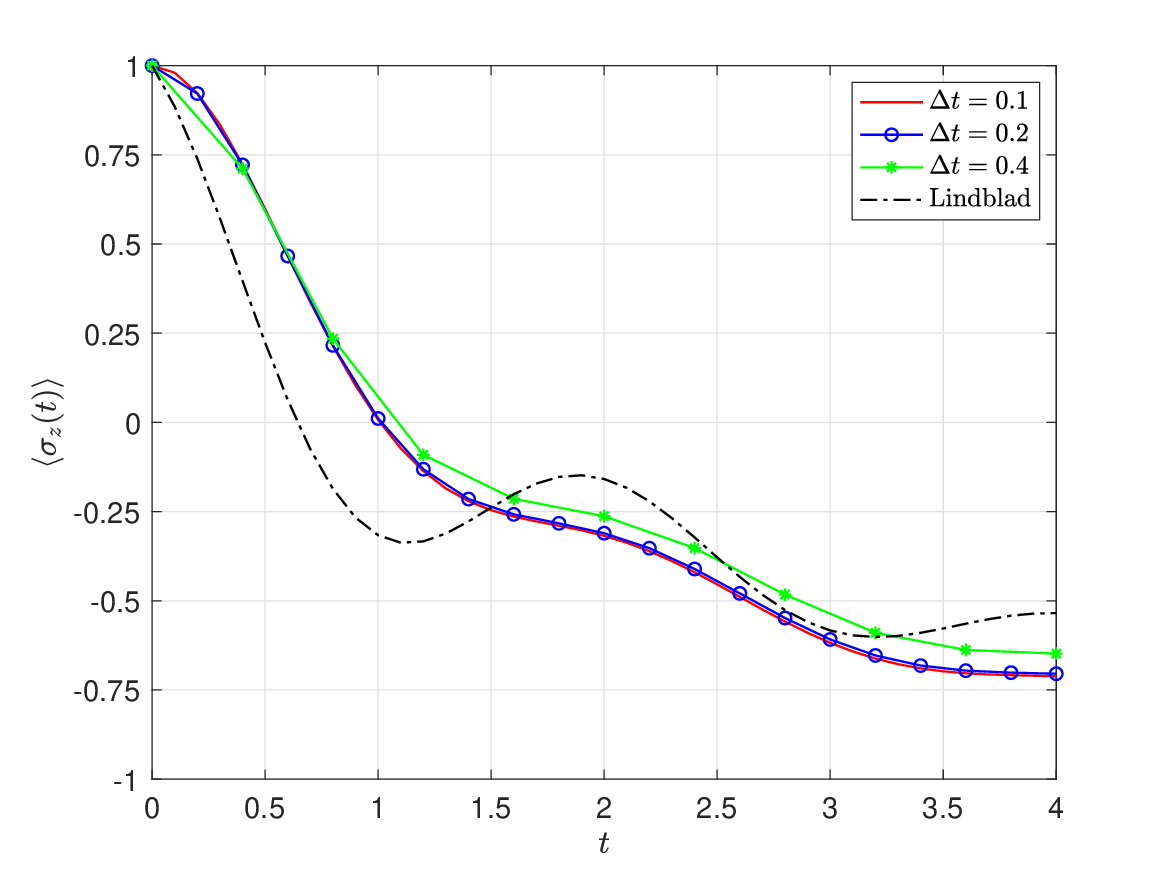}
        \caption{$\xi = 0.4$}
        \label{fig:conv_dt_xi04}
    \end{subfigure}
    \vspace{0.5em}
    \begin{subfigure}[b]{0.45\textwidth} 
        \centering
        \includegraphics[width=1.1\textwidth]{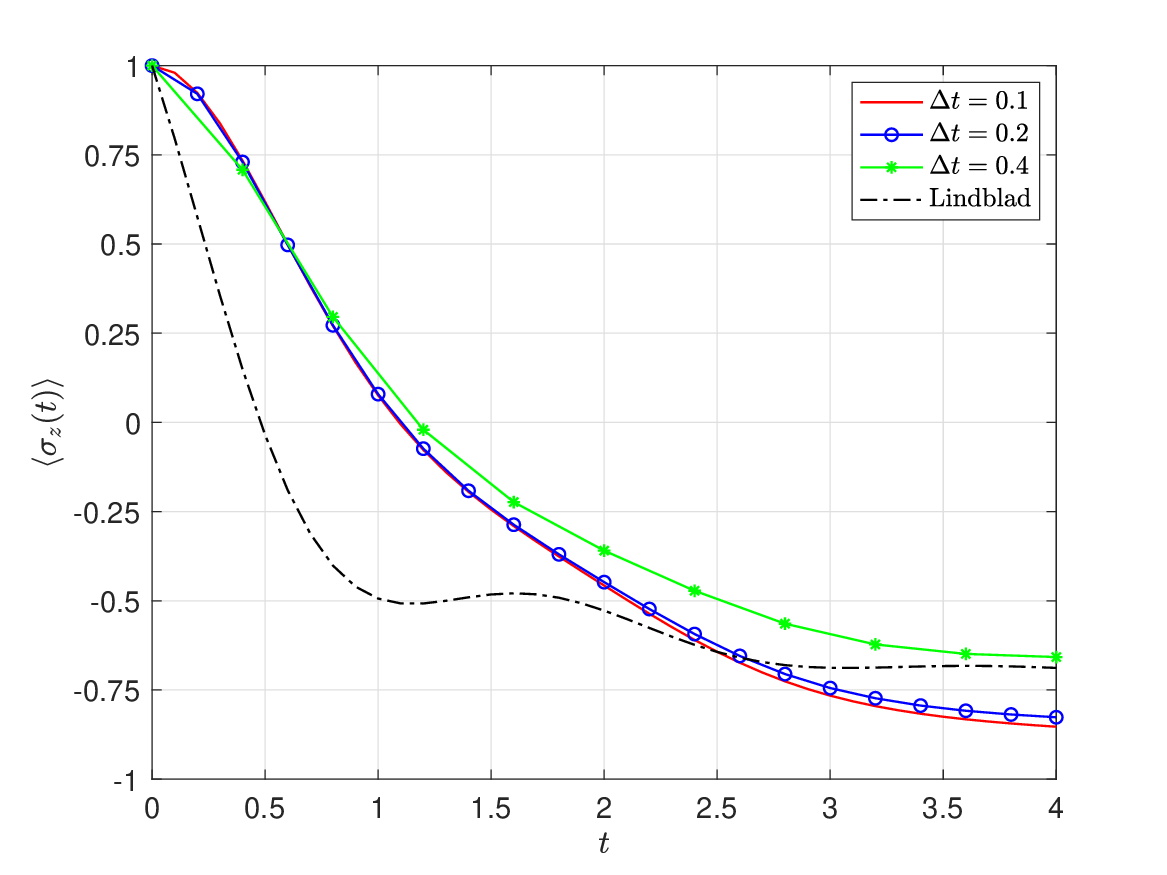}
        \caption{$\xi = 0.8$}
        \label{fig:conv_dt_xi08}
    \end{subfigure}
    \hfill
    \begin{subfigure}[b]{0.45\textwidth} 
        \centering
        \includegraphics[width=1.1\textwidth]{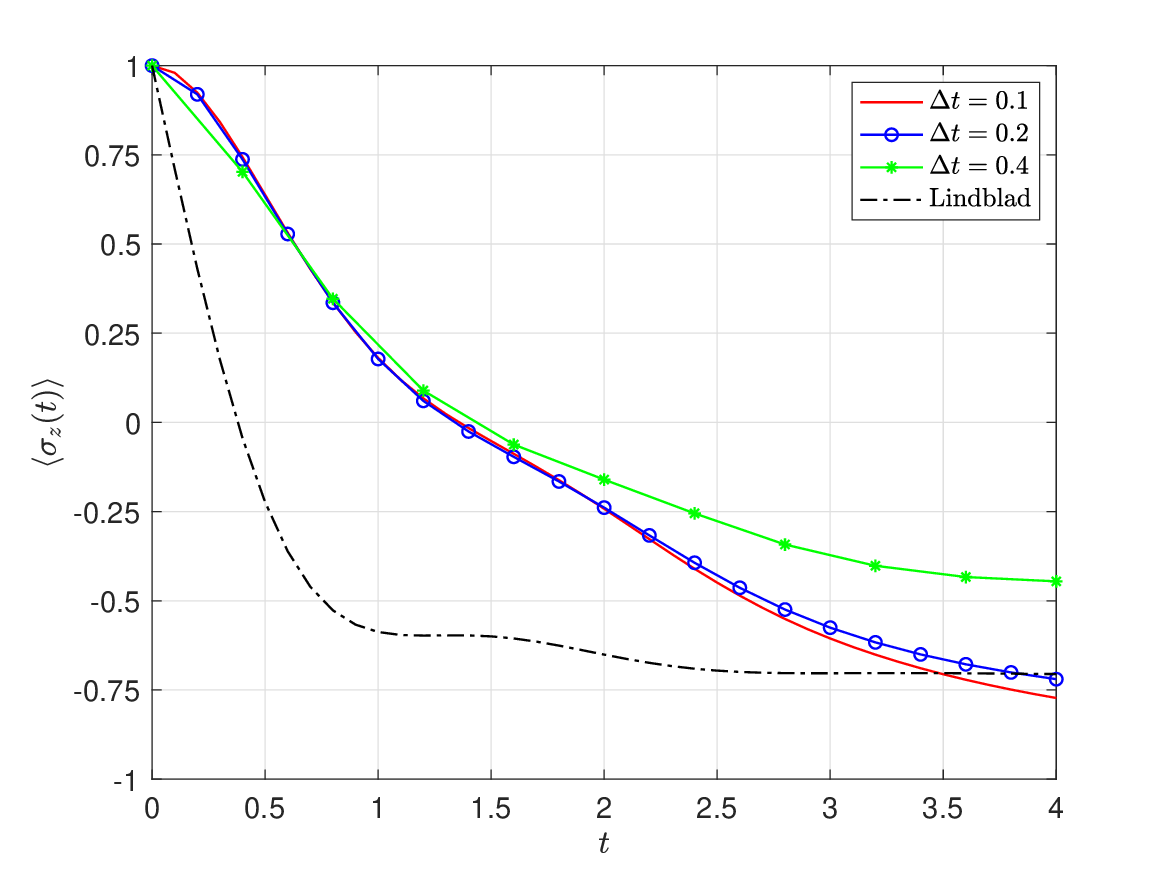}
        \caption{$\xi = 1.2$}
        \label{fig:conv_dt_xi12}
    \end{subfigure}
    \caption{Convergence of inchworm method with respect to $\Delta t$ for different $\xi$.}
    \label{fig:conv_dt}
\end{figure}

We now study the effect of TT-rounding in the simulation of spin-boson models.
Here, by choosing $M = 3$, $\xi = 0.2$ and $\Delta t = 0.2$, we can obtain results with different rounding parameters $\eta$ for $\beta = 5$ and $\beta = 1$ for $N = 40$ time steps, as shown in  \Cref{fig:trace_round_m3_N40_dt02}.
The four curves in both figures (\Cref{fig:trace_round_m3_xi02_beta5_N40_dt02} and \Cref{fig:trace_round_m3_xi02_beta1_N40_dt02}) are nearly indistinguishable, indicating that TT-rounding introduces negligible numerical error in the observable, and the BIF-TT representation remains accurate and reliable throughout the real-time propagation.
This insensitivity to rounding tolerance also gives us greater freedom to control the storage cost of the BIF-TT without decreasing observable accuracy.

\begin{figure}
    \centering
    \begin{subfigure}[b]{0.48\textwidth}
        \includegraphics[width = 1.1\textwidth]{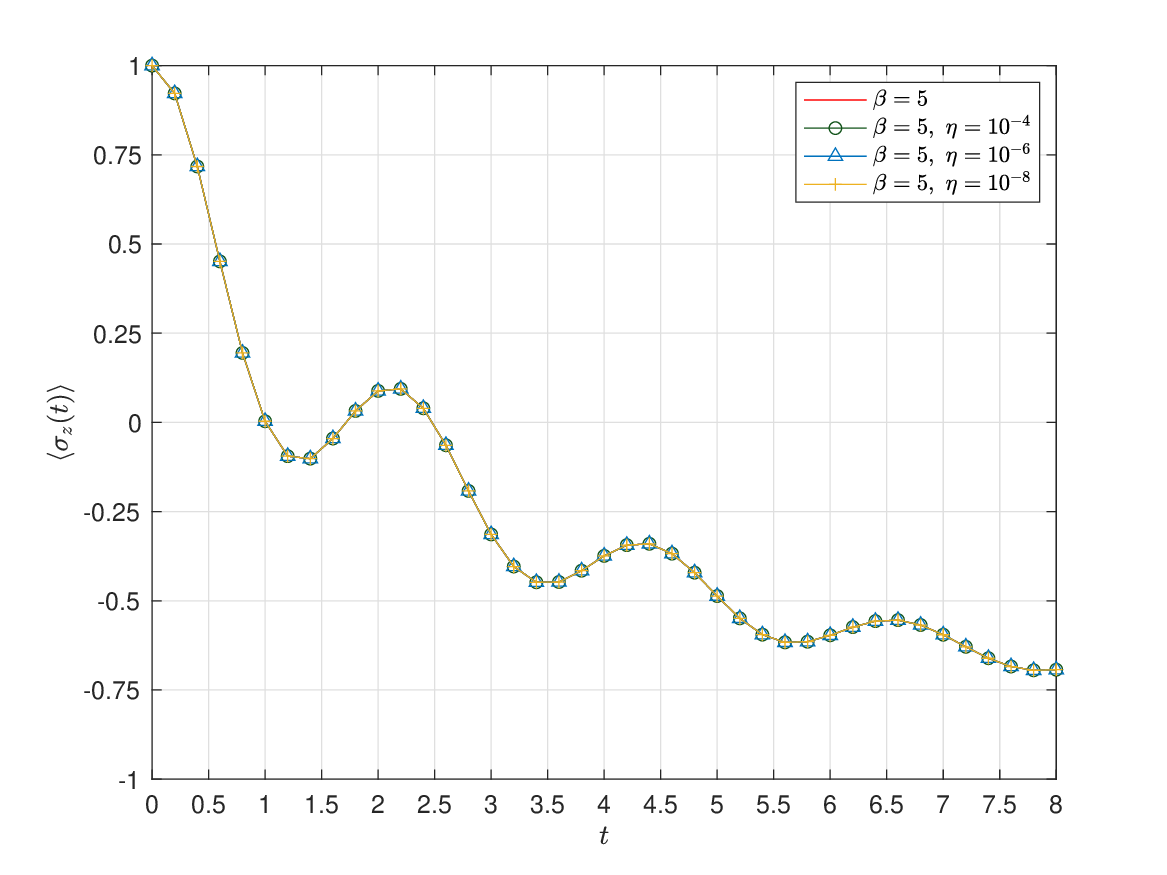}
        \caption{ $\beta = 5$}
        \label{fig:trace_round_m3_xi02_beta5_N40_dt02}
    \end{subfigure}
    \hfill
    \begin{subfigure}[b]{0.48\textwidth}
        \includegraphics[width = 1.1\textwidth]{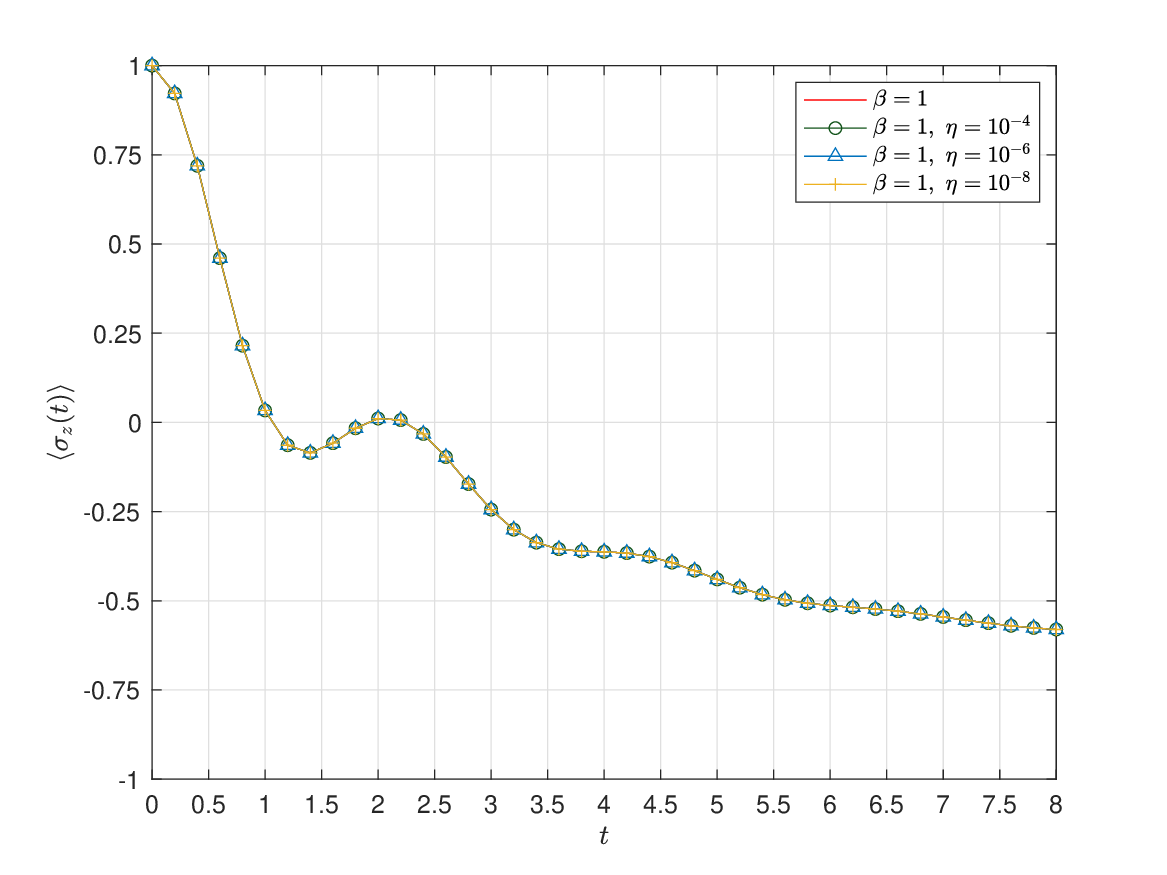}
        \caption{ $\beta = 1$}
        \label{fig:trace_round_m3_xi02_beta1_N40_dt02}
    \end{subfigure}
    \caption{Evolution of $\langle \sigma_z (t) \rangle$ with $M=3$, $\dt = 0.2$, and rounding tolerances $\eta = 10^{-4},\ 10^{-6},\ 10^{-8}$.}
    \label{fig:trace_round_m3_N40_dt02}
\end{figure}

One alternative way of TT-rounding is to limit the memory usage of the TT instead of controlling the accuracy.
This approach is more useful when the value of $M$ is large, where a small rounding tolerance may lead to large bond dimensions.
In the following example, we pick $M = 5$ and $\Delta t = 0.1$ and simulate the spin-boson model with BIF-TTs rounded such that the memory usage is no more than 4 gigabytes.
For $m = 5$, the bond dimensions of $\tensor{L}_6$ without TT-rounding are $[4r,4r^2,(3r^3+r),4r^2,4r]$ for $r = \operatorname{rank}(\mathbf{B})$.
In our case, the numerical rank of $\mathbf{B}$ is 16, requiring approximately 31GB of memory to store $\tensor{L}_6$ with tensor elements being double-precision complex numbers.
However, the rounded BIF-TT takes only less than 2GB of memory, and this storage cost grows only linearly with respect to $M$.
For various values of $\xi$, the comparison of numerical results with and without TT-rounding is exhibited in \Cref{fig:compare_inchworm_m5_beta5}, where the results of the inchworm method with classical numerical quadrature are also provided as references.
Again, all lines are almost on top of each other, demonstrating sufficient accuracy even with a significantly lower memory cost.

\begin{figure}
    \centering
    \begin{subfigure}[b]{0.32\textwidth} 
        \centering
        \includegraphics[width=1.1\textwidth]{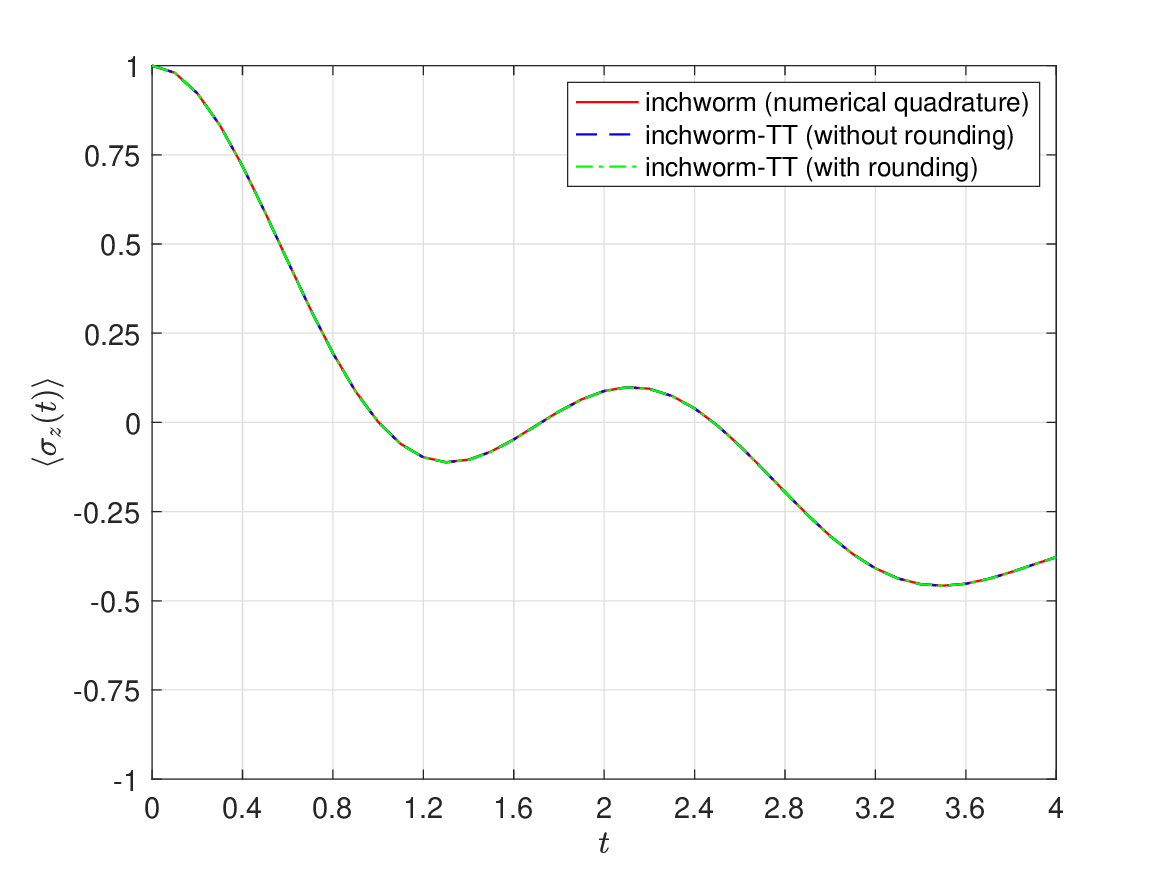}
        \caption{$\xi = 0.2$}
        \label{fig:compare_inchworm_m5_beta5_xi02}
    \end{subfigure}
    \hfill 
    \begin{subfigure}[b]{0.32\textwidth} 
        \centering
        \includegraphics[width=1.1\textwidth]{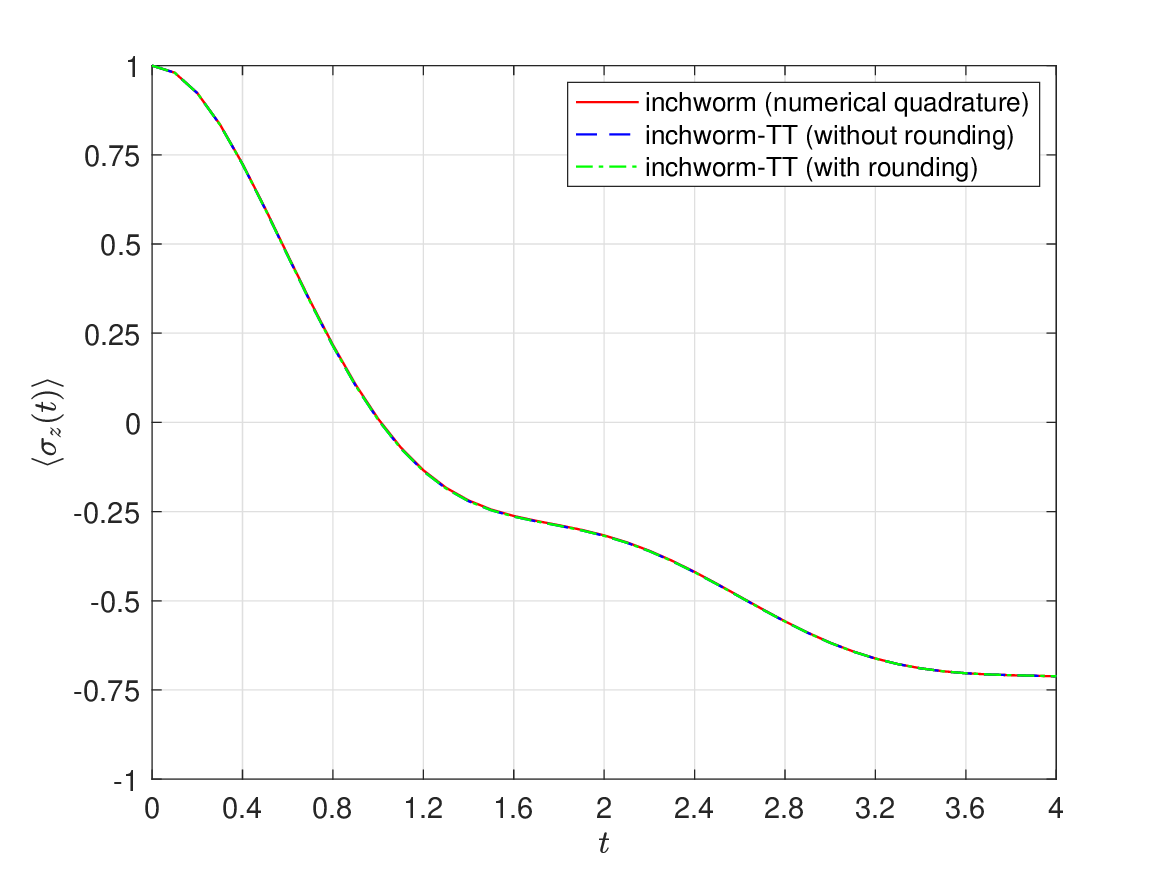}
        \caption{$\xi = 0.4$}
        \label{fig:compare_inchworm_m5_beta5_xi04}
    \end{subfigure}
    \hfill
    \begin{subfigure}[b]{0.32\textwidth} 
        \centering
        \includegraphics[width=1.1\textwidth]{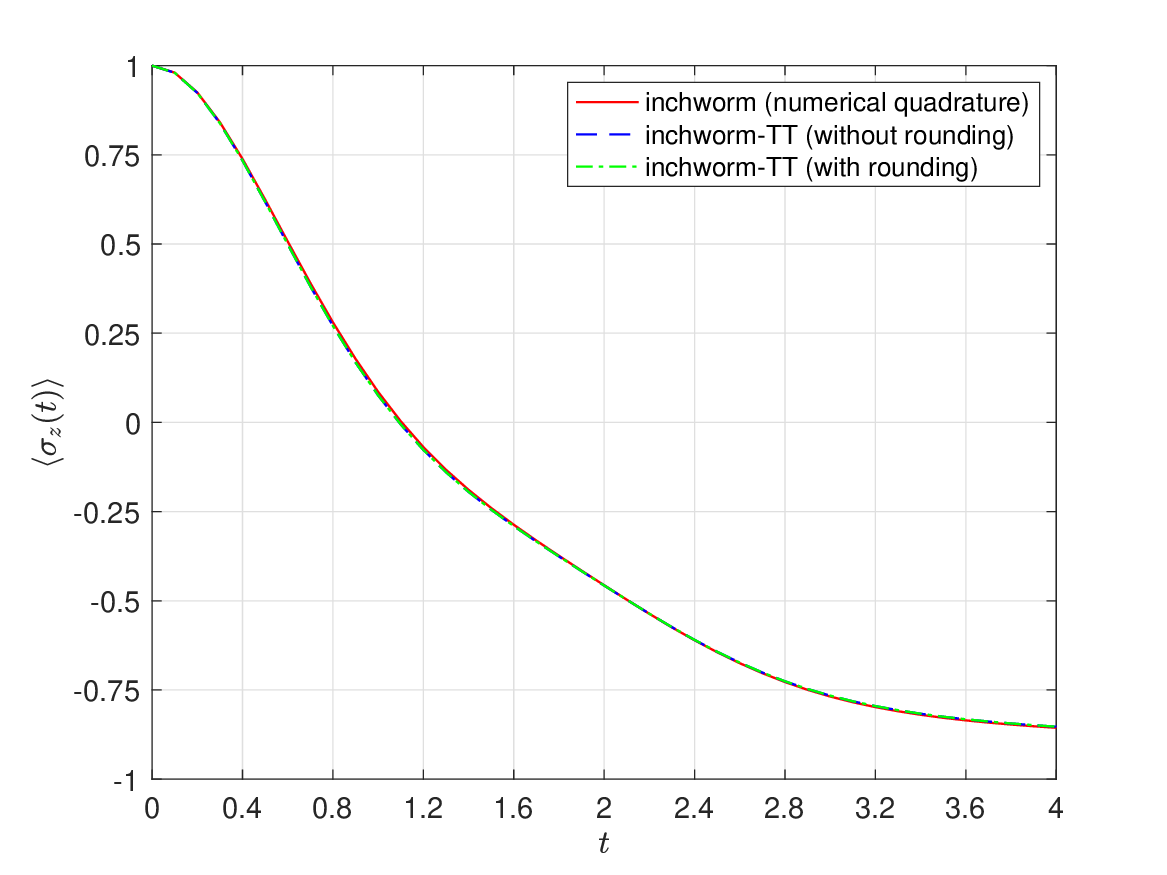}
        \caption{$\xi = 0.8$}
        \label{fig:compare_inchworm_m5_beta5_xi08}
    \end{subfigure}
    \caption{Comparison of results with and without TT-rounding. The solution without using TTs is also plotted as references.}
    \label{fig:compare_inchworm_m5_beta5}
\end{figure}

The last test in this subsection investigates the convergence with respect to $M$.
For cases where $M \geqslant 7$, to control the storage cost, we apply TT-rounding with maximum bond dimension $500$ during the construction of BIF-TT.
Since the inchworm method is a numerical strategy based on the series expansion and should be categorized as a perturbative method, we expect that larger $M$ is needed for stronger coupling between the system and the bath.
Such a phenomenon can be clearly observed in \Cref{fig:conv_m}, where the Kondo parameter $\xi$ represents the coupling intensity.
Nevertheless, for all the four cases in \Cref{fig:conv_m}, results converge with respect to $M$, and for a larger $M$, the accuracy can be maintained for a longer time.
For even longer simulations, instead of further increasing $M$, we will turn to another approach, to be presented in the next subsection, to control the computational cost.

\begin{figure}
    \centering
    \begin{subfigure}[b]{0.45\textwidth} 
        \centering
        \includegraphics[width=1.1\textwidth]{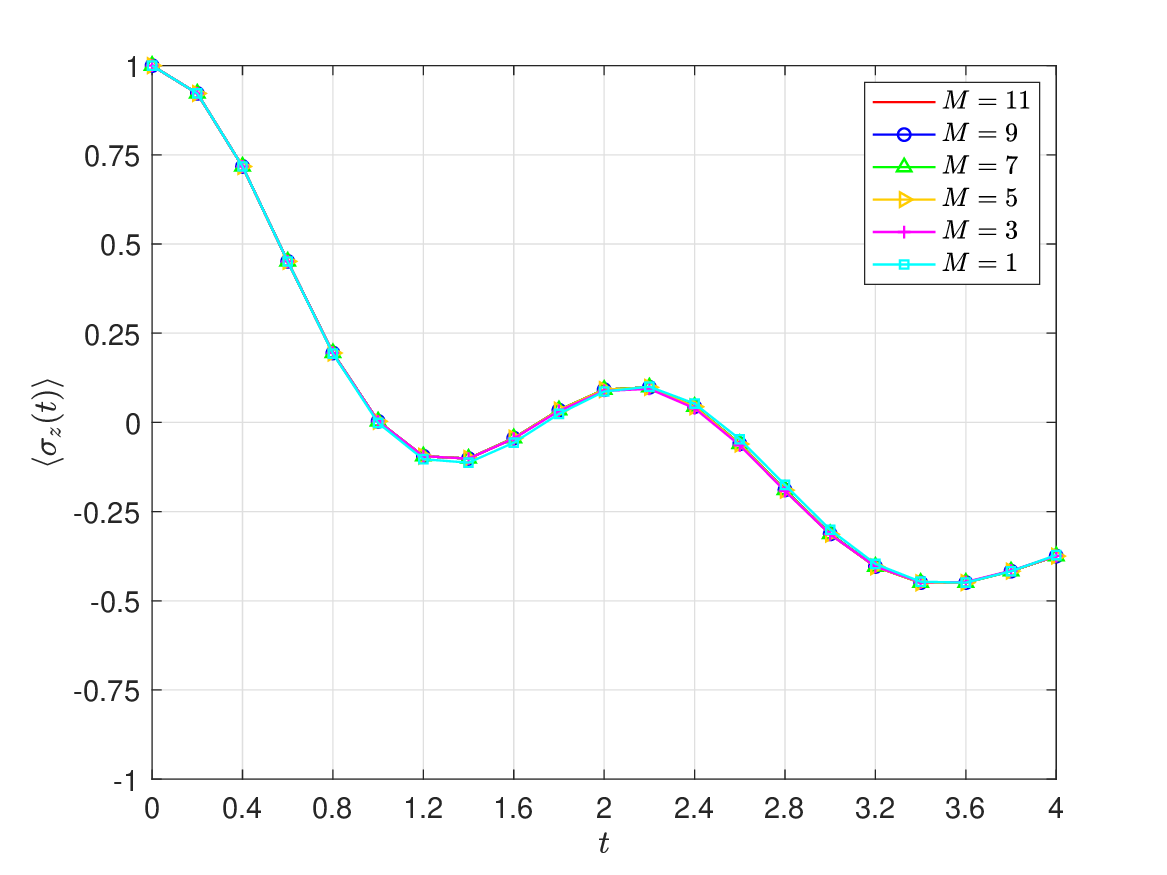}
        \caption{$\xi = 0.2$}
        \label{fig:conv_m_xi02}
    \end{subfigure}
    \hfill 
    \begin{subfigure}[b]{0.45\textwidth} 
        \centering
        \includegraphics[width=1.1\textwidth]{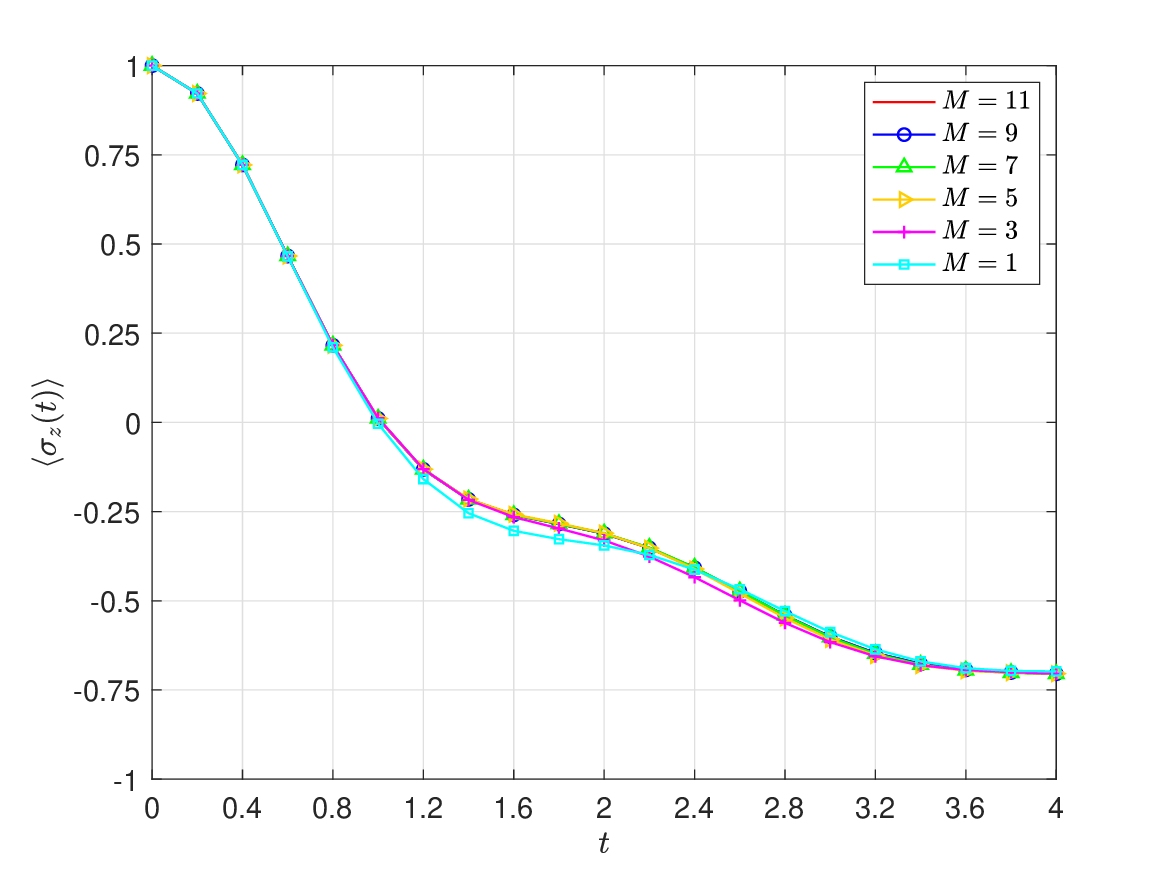}
        \caption{$\xi = 0.4$}
        \label{fig:conv_m_xi04}
    \end{subfigure}
    \vspace{0.5em}
    \begin{subfigure}[b]{0.45\textwidth} 
        \centering
        \includegraphics[width=1.1\textwidth]{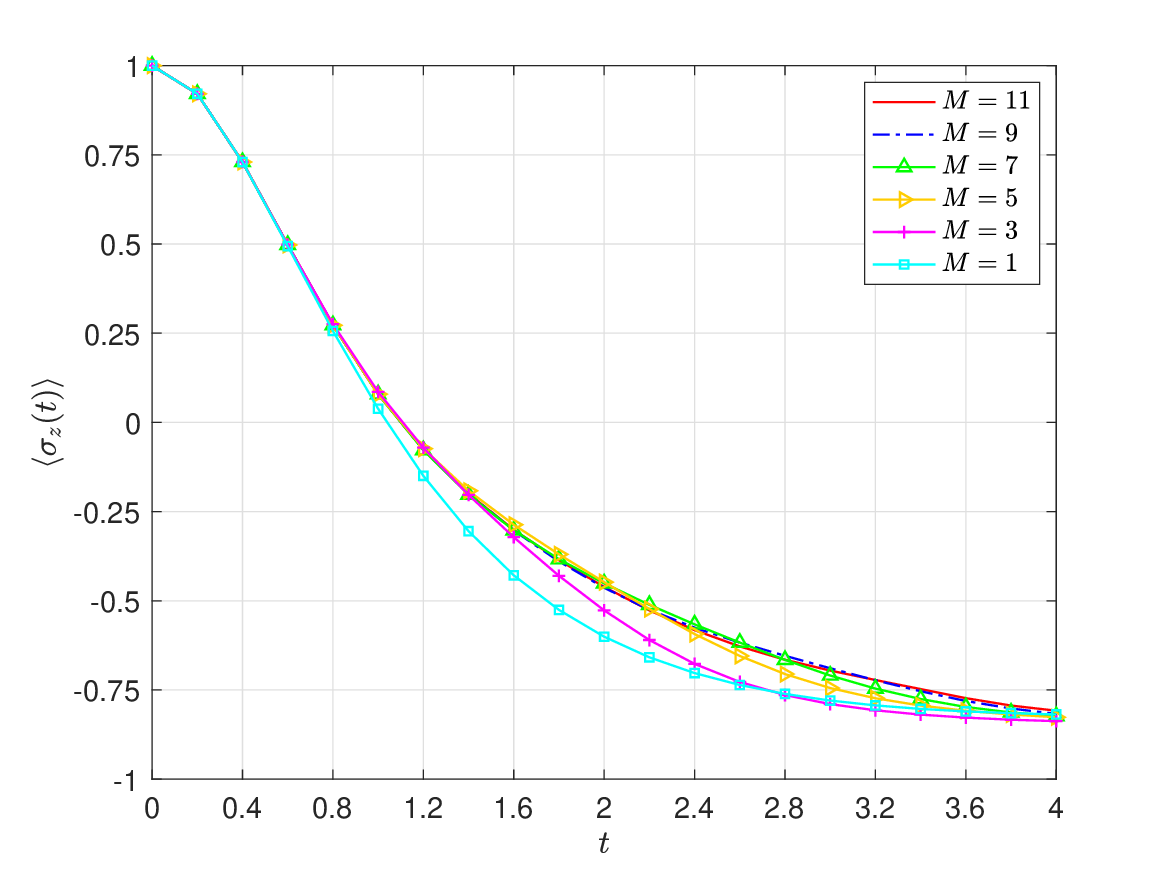}
        \caption{$\xi = 0.8$}
        \label{fig:conv_m_xi08}
    \end{subfigure}
    \hfill
    \begin{subfigure}[b]{0.45\textwidth} 
        \centering
        \includegraphics[width=1.1\textwidth]{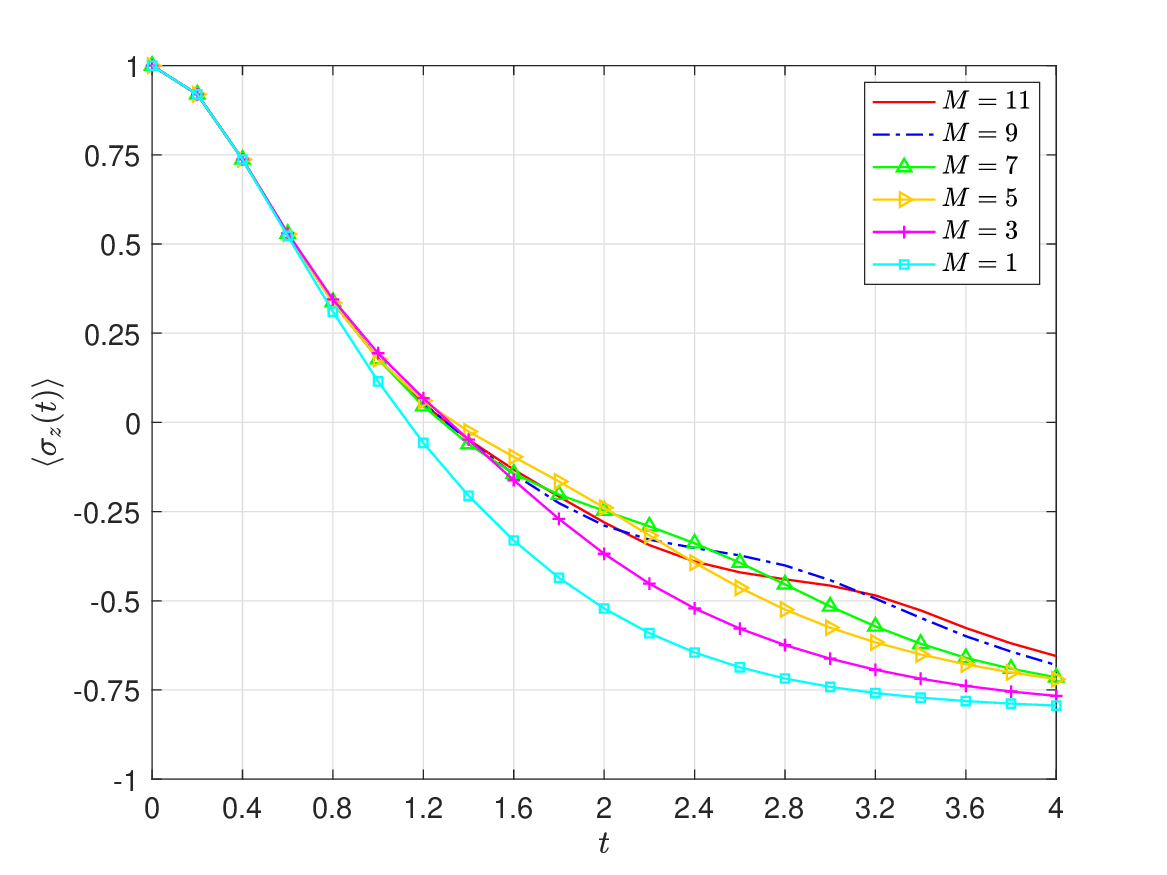}
        \caption{$\xi = 1.2$}
        \label{fig:conv_m_xi12}
    \end{subfigure}
    \caption{Convergence of inchworm method with respect to $M$ for different $\xi$.}
    \label{fig:conv_m}
\end{figure}


\subsection{Combining with the tensor transfer method}
\label{subsec_ttm}
In \Cref{subsec_bond_dim_biftt}, we have observed that the computational cost of our method relies heavily on the bond dimensions of the BIF-TTs, which in turn depend on the rank $r$ of $\mathbf{B}$.
When the number of time steps $N$ increases, the rank $r$ of $\mathbf{B}$ also increases as shown in \Cref{fig:rankB}.
Therefore, the computational cost increases rapidly for long-time simulations.
To overcome this difficulty, 
we can couple our method with the transfer tensor method (TTM) \cite{cerrillo2014nonmarkovian,rosenbach2016efficient} in the numerical simulations.

TTM approximates the dynamics of the reduced system using a sequence of dynamical maps $\mathcal{E}_k$:
\begin{equation}
\label{eq_defn_dynamical_maps}
    \rho_s(k\dt) = \mathcal{E}_k \rho_s(0),
\end{equation}
where $\mathcal{E}_k$ is a map from $\mathbb{C}^{2\times 2}$ to $\mathbb{C}^{2\times 2}$, which can be represented by a $4\times 4$ matrix.
Given any numerical solver, $\mathcal{E}_k$ can be determined by choosing $\rho_s(0)$ to be all the bases of $\mathbb{C}^{2\times 2}$ and performing simulations to obtain the corresponding $\rho_s(k\dt)$. 
The transfer tensor method assumes these dynamic maps can be recursively expressed in terms of transfer tensors $\mathcal{T}_j$:
\begin{equation}
\label{eq_transfer_tensor_to_dynamical_maps}
    \mathcal{E}_k = \sum_{j=1}^{k} \mathcal{T}_{j} \mathcal{E}_{k-j}.
\end{equation}
In practice, the sequence of transfer tensors decays to zero as $j$ increases. TTM therefore assumes a finite memory length of $K_{\max}\dt$, or $K_{\max}$ time steps in the discrete form.
With this assumption, for $k>K_{\max}$, we have
\begin{equation}
\label{eq_TTM_propagate}
    \rho_s(k\dt) = \sum_{j=1}^{K_{\max}} \mathcal{T}_j \rho_s((k-j)\dt),
\end{equation}
so that simulations up to any time can be efficiently carried out once the transfer tensors $\mathcal{T}_j$, $j = 1,\cdots, K_{\max}$ are determined.
As a summary, the application of TTM in our method contains the following three stages:
\begin{description}
    \item[Data generation:] Compute the dynamical maps within the memory length, denoted by $\mathcal{E}_j$ for $j=1,\cdots,K_{\max}$, using the method of BIF-TT.
    \item[TTM learning:] Construct transfer tensors $\mathcal{T}_{j}$ for $j=1,\cdots,K_{\max}$ via \eqref{eq_transfer_tensor_to_dynamical_maps}.
    \item[TTM propagating:] Perform long-time simulations using \eqref{eq_TTM_propagate}.
\end{description}

We first verify the TTM using an example with relatively weak coupling intensity.
By setting $\epsilon = 1$, $\xi = 0.4$ and $\beta = 5$, we can use $M = 5$ to obtain accurate results up to $t = 10$.
Using this result as a reference solution, we present the TTM results with $\dt=0.1$ and $K_{\max}=10,20$ in \Cref{fig:xi04_m5dt02_ttmdt02_inchworm}.
For $K_{\max} = 10$, the TTM results have an obvious deviation from the reference solution after $t=2$, due to the inadequacy of memory effects.
The curve corresponding to $K_{\max} = 20$ closely overlaps with the reference observable, indicating that the non-Markovian memory effects are sufficiently captured at this memory length.


For stronger couplings and longer-time simulations, computing with the whole memory length becomes unaffordable even if the TT rounding is applied. For instance, in \Cref{fig:xi08_m9dt02_ttmdt02_quapi}, we plot numerical results for the parameters $\xi=0.8$ and $\beta = 5$ with $\epsilon = 0$ and $\epsilon = 1$, requiring $M = 9$ to achieve an accurate result up to $t = 4$ according to \Cref{fig:conv_m_xi08}. For this simulation, the BIF-TTs take up 1.62GB of memory for time step $\dt = 0.1$ and $K_{\max} = 20$.
Further extension of the simulation with full memory length requires either larger memory usage or more severe compression of the tensor train, and the introduction of TTM allows us to trade off the computational cost and the error introduced by memory length truncation.
When turning to TTM, we observe that the norm of the transfer tensor at time $t = 2$ decreases to $1.18\times10^{-3}$ for $\epsilon=0$ and $1.33\times10^{-3}$ for $\epsilon=1$, proving the validity of further propagation.
We compare our solutions with the i-QuAPI method, where the memory truncation is also applied, showing good agreement of the two results.
Compared to i-QuAPI, one advantage of our method (as well as other continuous-time methods) is that it has no difficulty in dealing with multilevel systems, whereas i-QuAPI has memory cost $\mathcal{O}(n^{2K_{\max}})$ for an $n$-level system.

\begin{figure}[!ht]
    \centering
    \includegraphics[width = .5 \textwidth]{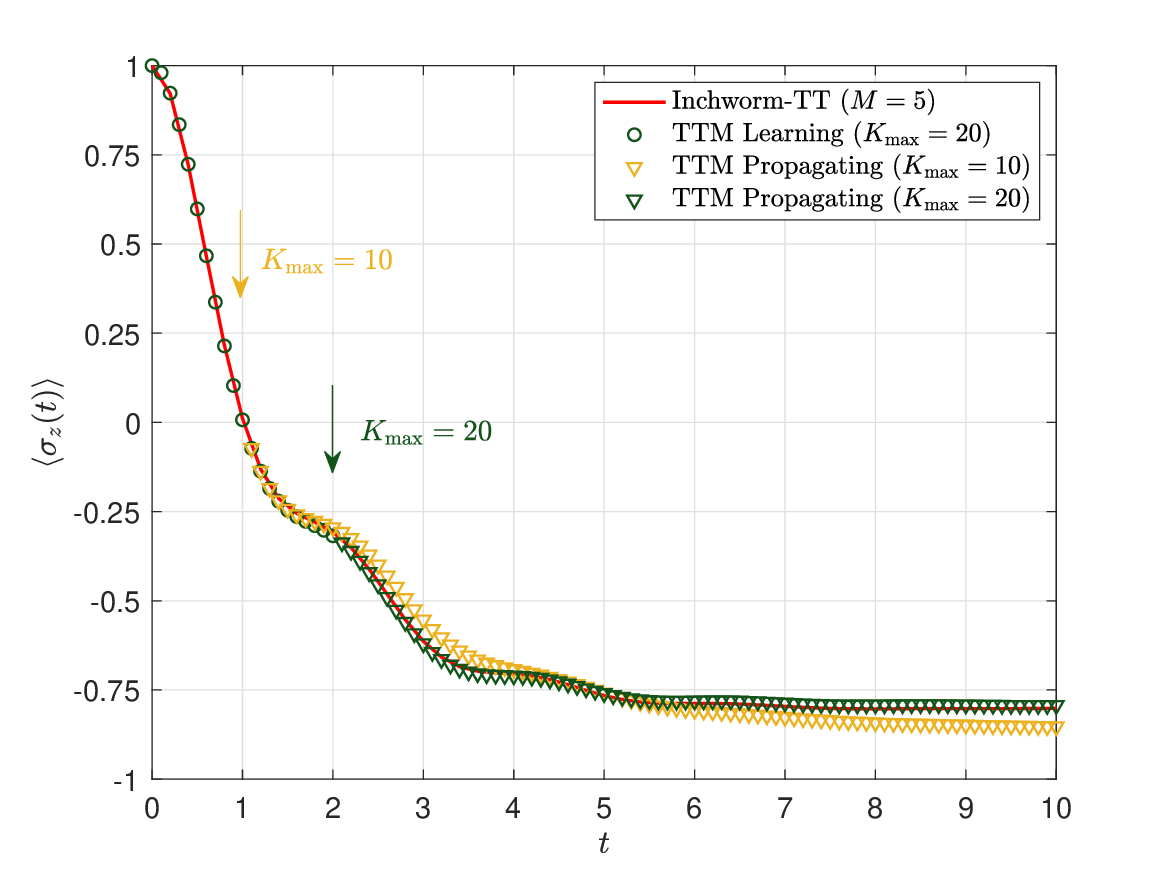}
    \caption{Numerical results using TTM with different memory lengths.}
    \label{fig:xi04_m5dt02_ttmdt02_inchworm}
\end{figure}

\begin{figure}[!ht]
    \centering
    \begin{subfigure}[b]{0.49\textwidth} 
        \centering
        \includegraphics[width=\textwidth]{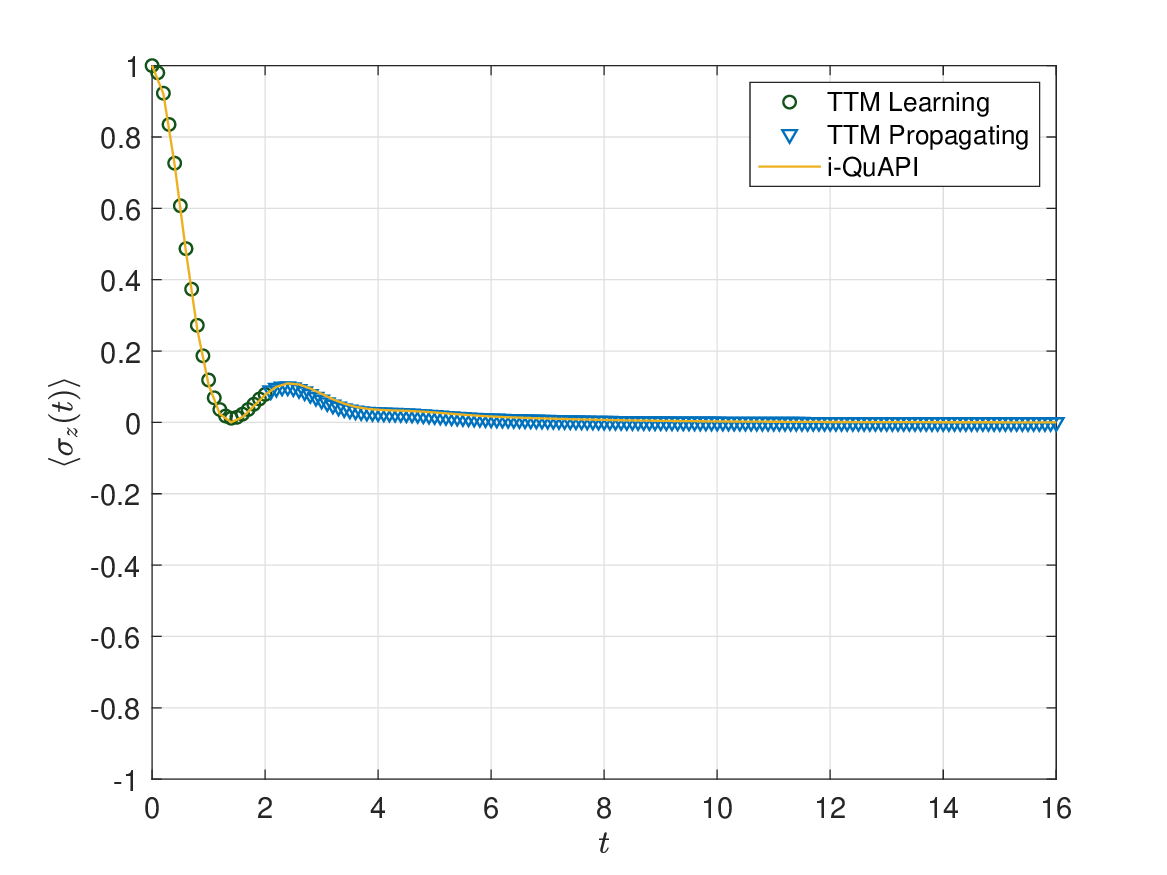}
        \caption{$\epsilon = 0$, $\xi = 0.8$}
        \label{fig:conv_m_xi08_eps0}
    \end{subfigure}
    \begin{subfigure}[b]{0.49\textwidth} 
        \centering
        \includegraphics[width=\textwidth]{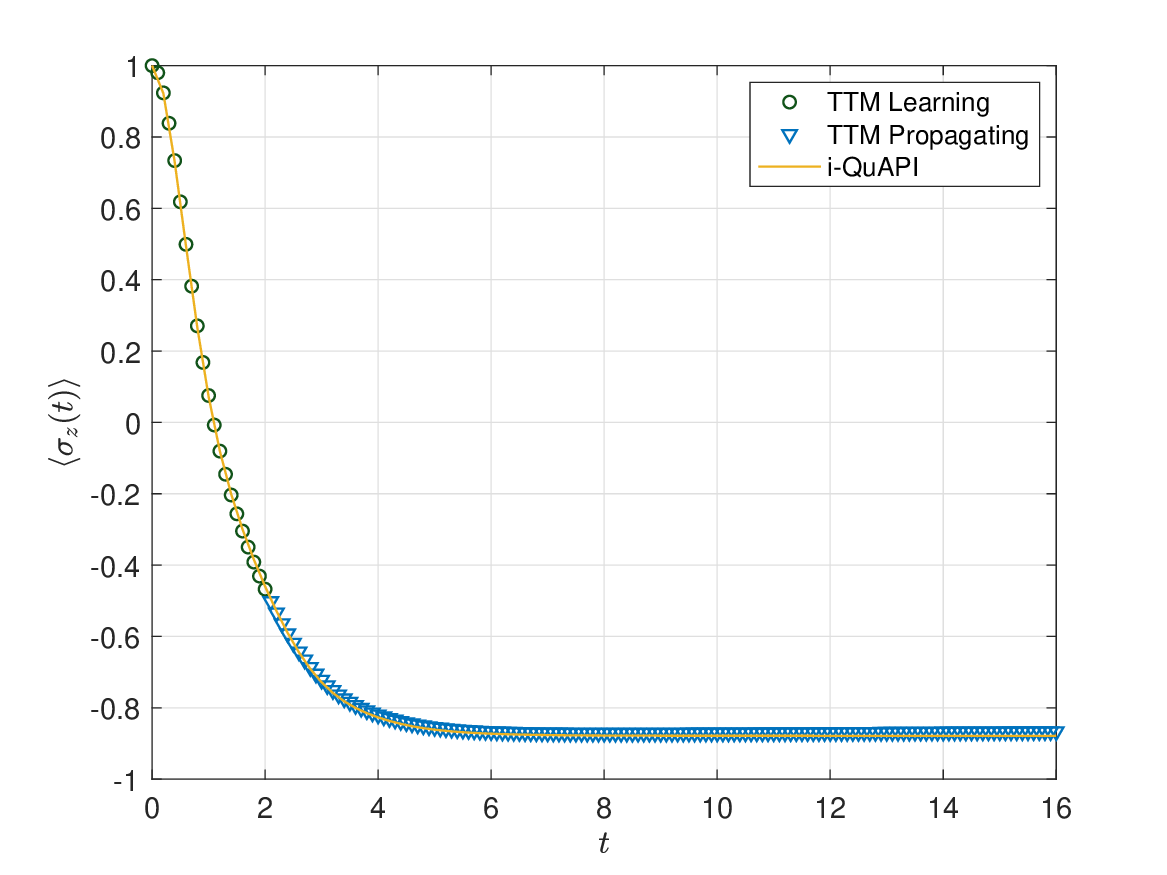}
        \caption{$\epsilon = 1$, $\xi = 0.8$}
        \label{fig:conv_m_xi08_eps1}
    \end{subfigure}
        \caption{Long-time simulation using BIF-TT coupled with TTM.}
    \label{fig:xi08_m9dt02_ttmdt02_quapi}
\end{figure}

\section{Conclusion and discussion}
\label{sec_conclusion}
In this paper, we develop a fast algorithm for efficient computation of path integrals in the simulation of open quantum system.
The key idea is to represent the bath influence functional using tensor trains, enabling iterative computation of high-dimensional integrals.
A detailed algorithm is developed for the construction of the tensor trains, and the numerical complexity is analyzed to demonstrate its efficiency.
Extensive numerical tests are carried out to show how the bond dimensions are related to various parameters for the Ohmic spectral density,
and we simulate the dynamics of the spin-boson model using the inchworm method with integrals up to 11 dimensions.

Although the introduction of BIF-TTs is the main advantage of our method since it significantly reduces the computational cost of high-dimensional integrals,
the construction of BIF-TTs remains the most expensive part in the whole process, due to the following reasons.
First, the nature of the bath influence functional causes a large number of terms in the construction of BIF-TTs even if we apply the iterative procedure discussed in \cref{sec_construct_bif_tt}.
The algorithm requires computing Hadamard products and TT summations, significantly increasing the rank of the resulting TTs.
This will dramatically increase both the memory cost and the computational cost.
Although TT rounding can effectively keep the memory cost under control, the rounding operation itself needs a large computational cost.
Recently, a fast Hadamard product algorithm for tensor trains is introduced in \cite{sun2024hatt}, which may be useful in improving the computational efficiency.
Alternatively, the TT-cross \cite{oseledets2010ttcross} algorithm may also help to develop faster algorithm for BIF-TT constructions.
These improvements will be explored in our future works.

\section*{Acknowledgements}
Zhenning Cai's work was supported by the Academic Research Fund of
the Ministry of Education of Singapore under grant A-8002392-00-00.

\appendix
\section{Connected pairings and bath influence functionals}
\label{app_pairings}
This appendix includes some details of the connected pairings and bath influence functionals.

In \cref{eq_bif_inchworm}, the set $\mathcal{Q}_n^c$ stands for all connected pairings of integers from 1 to $n$. For instance,
\begin{align*}
  \mathcal{Q}_2^c &= \Big\{\{(1,2)\}\Big\}, \qquad
  \mathcal{Q}_4^c = \Big\{\{(1,3), (2,4)\}\Big\}, \\
  \mathcal{Q}_6^c &= \Big\{\{(1,3), (2,5), (4,6)\}, \{(1,4), (2,5), (3,6)\}, \{(1,5), (2,4), (3,6)\}, \{(1,4), (2,6), (3,5)\}\Big\}.
\end{align*}
One can see why these pairings are said to be ``linked'' by drawing diagrams connecting nodes with arcs.
Examples include:
\begin{displaymath}
\begin{array}{rl}
  \{(1,3), (2,4)\}: &  \begin{tikzpicture}[baseline=0pt]
\draw plot[only marks,mark =*, mark options={color=black, scale=0.5}]coordinates {(0,0) (0.5,0) (1,0)(1.5,0)};
\draw[-] (0,0) to[bend left=75] (1,0);
\draw[-] (0.5,0) to[bend left=75] (1.5,0);
\node at (0,0.1) [label=below:\scriptsize 1]{};
\node at (0.5,0.1) [label=below:\scriptsize 2]{};
\node at (1.0,0.1) [label=below:\scriptsize 3]{};
\node at (1.5,0.1) [label=below:\scriptsize 4]{};
 \end{tikzpicture} \\
  \{(1,3), (2,5), (4,6)\}: &  \begin{tikzpicture}[baseline=0pt]
\draw plot[only marks,mark =*, mark options={color=black, scale=0.5}]coordinates {(0,0) (0.5,0) (1,0)(1.5,0) (2.0,0)(2.5,0)};
\draw[-] (0,0) to[bend left=75] (1,0);
\draw[-] (0.5,0) to[bend left=75] (2.0,0);
\draw[-] (1.5,0) to[bend left=75] (2.5,0);
\node at (0,0.1) [label=below:\scriptsize 1]{};
\node at (0.5,0.1) [label=below:\scriptsize 2]{};
\node at (1.0,0.1) [label=below:\scriptsize 3]{};
\node at (1.5,0.1) [label=below:\scriptsize 4]{};
\node at (2.0,0.1) [label=below:\scriptsize 5]{};
\node at (2.5,0.1) [label=below:\scriptsize 6]{};
 \end{tikzpicture} \\
  \{(1,4), (2,5), (3,6)\}: &  \begin{tikzpicture}[baseline=0pt]
\draw plot[only marks,mark =*, mark options={color=black, scale=0.5}]coordinates {(0,0) (0.5,0) (1,0)(1.5,0) (2.0,0)(2.5,0)};
\draw[-] (0,0) to[bend left=75] (1.5,0);
\draw[-] (0.5,0) to[bend left=75] (2.0,0);
\draw[-] (1,0) to[bend left=75] (2.5,0);
\node at (0,0.1) [label=below:\scriptsize 1]{};
\node at (0.5,0.1) [label=below:\scriptsize 2]{};
\node at (1.0,0.1) [label=below:\scriptsize 3]{};
\node at (1.5,0.1) [label=below:\scriptsize 4]{};
\node at (2.0,0.1) [label=below:\scriptsize 5]{};
\node at (2.5,0.1) [label=below:\scriptsize 6]{};
 \end{tikzpicture}
\end{array}
\end{displaymath}
The pairing $\{(1,5), (2,6), (3,4)\}$, represented diagrammatically by
\scalebox{.6}{\begin{tikzpicture}[baseline=0pt]
\draw plot[only marks,mark =*, mark options={color=black, scale=0.7}]coordinates {(0,0) (0.5,0) (1,0)(1.5,0) (2.0,0)(2.5,0)};
\draw[-] (0,0) to[bend left=75] (2.0,0);
\draw[-] (0.5,0) to[bend left=75] (2.5,0);
\draw[-] (1,0) to[bend left=75] (1.5,0);
 \end{tikzpicture}},
is not included since the pair $(3,4)$ is not linked to the other part of the diagram.
As a result, the BIF $\mathcal{L}_b^c(\bs)$ takes the form as 
\begin{equation}
    \mathcal{L}_b^c(\bs) = \begin{cases}
    B\left(s_1, s_2\right) &\text{~if~}m=2 \\
    B\left(s_1, s_3\right) B\left(s_2, s_4\right) &\text{~if~}m=4 \\
    B(s_1,s_4)B(s_2,s_5)B(s_3,s_6) + 
        B(s_1,s_4)B(s_2,s_6)B(s_3,s_5)  \\
        \quad+ 
        B(s_1,s_3)B(s_2,s_5)B(s_4,s_6) + 
        B(s_1,s_5)B(s_2,s_4)B(s_3,s_6)&\text{~if~}m=6 \\
    \cdots
\end{cases}
\end{equation}

\section{Tensor-train operations}
\label{app:tt operations}
In this section, we elaborate the tensor-train operations we use in our proposed fast numerical integration scheme.
\begin{itemize}
    \item \emph{TT sum.} Given two TTs $\tensor{X},\tensor{Y}\in \mathbb{C}^{n_1\times \cdots \times n_d}$, their sum $\tensor{Z}=\tensor{X}+\tensor{Y}$ also admits a TT form with each core given by 
    \begin{equation}
        \tensor{Z}^{(j)}(i_j) = 
        \begin{dcases}
        \begin{pmatrix}
            \tensor{X}^{(1)}(1,i_1)
            & \tensor{Y}^{(1)}(1,i_1)
        \end{pmatrix},  \text{~if~} j=1 \\
        \begin{pmatrix}
            \tensor{X}^{(j)}(:,i_j,:) & \mathbf{0} \\
            \mathbf{0} & \tensor{Y}^{(j)}{(:,i_j,:)}
        \end{pmatrix}, \text{~if~} j = 2,\cdots,d-1  \\
        \begin{pmatrix}
            \tensor{X}^{(d)}(i_d,1) \\
            \tensor{Y}^{(d)}(i_d,1)
        \end{pmatrix}, \text{~if~} j = d.
        \end{dcases}
    \end{equation}    
    By this construction, each bond dimension of $\tensor{Z}$ is the sum of the corresponding bond dimensions of $\tensor{X}$ and $\tensor{Y}$.
        \item \emph{TT Hadamard (elementwise) product.} Given two TT $\tensor{X},\tensor{Y}\in\mathbb{C}^{n_1\times \cdots \times n_d}$ as defined in \eqref{TT representation}, $\tensor{Z} = \tensor{X}\odot\tensor{Y}$ also admits a TT representation with each core formulated as 
    \begin{equation}
       \tensor{Z}^{(j)}(:,i_j,:) =  \tensor{X}^{(j)}(:,i_j,:)\otimes  \tensor{Y}^{(j)}(:,i_j,:) .
    \end{equation}
    Here $\otimes$ denotes the Kronecker product of matrices. 
    According to the formula above, this operation will increase the TT ranks from $\mathcal{O}(r)$ to $\mathcal{O}(r^2)$. The computational cost of TT Hadamard product is $\mathcal{O}(dnr^4)$.
    \item \emph{TT extensions.}
    Given a TT $\tensor{X} \in \mathbb{C}^{n_1\times \cdots \times n_d}$ with rank $(r_0,r_1,\cdots,r_d)$, this operation adds extra cores to $\tensor{X}$ to extend it to another TT $\tensor{Y} \in \mathbb{C}^{m_1\times \cdots \times m_D}$, which keeps the cores of $\tensor{X}$ in given dimension $1\leqslant k_1 < k_2 < \cdots < k_d \leqslant D$, and the rest cores are chosen such that 
    \begin{displaymath}
        \tensor{Y}(j_1,\cdots,j_D) = \tensor{X}(i_1,\cdots,i_d) \quad \text{for} \quad  i_l = j_{k_l}.
    \end{displaymath}
This is done by construct new cores in $\tensor{Y}$ with size
    \begin{equation}
        \tensor{Y}^{(j)} \in 
        \begin{cases}
            \mathbb{C}^{r_{k_{l-1}} \times n_j \times r_{k_l}} 
                &\text{~if~} j = k_{l} \text{~for some~} l\in\{1,2,\cdots,d\}, \\
            \mathbb{C}^{r_{k_{l}}\times n_l \times r_{k_l}}
                &\text{~if~} k_l < j < k_{l+1} \text{~for some~} l\in\{0,1,2,\cdots,d\}
        \end{cases}
    \end{equation}
    and the values are given by
    \begin{equation}
    \label{eq_tt_extension_values}
        \tensor{Y}^{(j)}(\alpha,i_j,\beta) =
        \begin{cases}
            \tensor{X}^{(k_l)}(\alpha,i_{k_l},\beta)
                 &\text{~if~} j = k_{l} \text{~for some~} l\in\{1,2,\cdots,d\}, \\
            \delta_{\alpha\beta} &\text{~otherwise}.
        \end{cases}
    \end{equation}
For convenience, such operation is represented by 
    \begin{equation}
        \tensor{Y} 
        = \mathbb{TT}(\tensor{X},D,[k_1,\cdots,k_d]).
    \end{equation}

\end{itemize}

\section{Proofs of \Cref{lemma_rank2,thm_low_rank,thm_rank_bound}}
\label{app_proof}
This appendix includes the proofs of lemmas and theorems in the paper.
Here is the proof of \Cref{lemma_rank2}:
\begin{proof}
    By defining $\lambda(\omega)=\coth{(\frac{\beta\omega}{2})}$, we can rewrite \eqref{eq_fixed_frequency_tpc_matrix} as follows
\begin{equation}
        \mathcal{B}_{k_1,k_2}(\omega)=\frac{1}{2}(\lambda(\omega)-1)\e^{\ii \omega \Delta k\Delta t} + \frac{1}{2}(\lambda(\omega)+1) \e^{-\ii \omega \Delta k\Delta t}
\end{equation}
Therefore, the matrix $\mathcal{B}$ can be written as the sum of two rank-1 matrices,
\begin{equation}
\label{eq_rank_2_FF_TPCM}
    \mathcal{B}(\omega)=\frac{1}{2}(\lambda(\omega)-1)\boldsymbol{x}(\omega)\boldsymbol{x}^{\dagger}(\omega) + \frac{1}{2}(\lambda(\omega)+1)\boldsymbol{y}(\omega)\boldsymbol{y}^{\dagger}(\omega)
\end{equation}
where $\boldsymbol{x}$ and $\boldsymbol{y}$ are linearly independent vectors of length $2N+1$ defined as
\begin{equation}
    \begin{split}
        &\boldsymbol{x}(\omega) = \left(\phi^{N},\ldots,\phi^{1},1,\phi^1,\ldots,\phi^N\right)^T \\
        &\boldsymbol{y}(\omega) = \left(\phi^{-N},\ldots,\phi^{-1},1,\phi^{-1},\ldots,\phi^{-N}\right)^T
    \end{split}
\end{equation}
with $\phi = \e^{\ii \omega \Delta t}$.
When $\omega \neq 0$ and $\Delta t > 0$, the vectors $\boldsymbol{x}(\omega)$ and $\boldsymbol{y}(\omega)$ are linearly independent.
Then it is clear that $\mathcal{B}$ is a rank-2 matrix.
\end{proof}

Here is the proof of \Cref{thm_rank_bound}
\begin{proof}
By representing the $k$-th element in $\boldsymbol{x}(\omega)$ as $\boldsymbol{x}_k(\omega)$ and the $k$th element in $\boldsymbol{y}(\omega)$ as $\boldsymbol{y}_k(\omega)$ where $\boldsymbol{x}, \boldsymbol{y}$ are defined in the proof of \cref{lemma_rank2}, it can be shown that, for $k_1 =1,\ldots,N$, 

    \begin{equation}
            \boldsymbol{x}_{-k_1}(\omega) =  \boldsymbol{x}_{k_1}(\omega),\quad
            \boldsymbol{y}_{-k_1}(\omega) =  \boldsymbol{y}_{k_1}(\omega),
    \end{equation}
which further yields
\begin{equation}
    \mathcal{B}_{-k_1,k_2} = \mathcal{B}_{k_1,k_2}.
\end{equation}
Therefore,
\begin{equation}
    \begin{split}
        \mathbf{B}_{-k_1,k_2}
        = \frac{1}{\pi} \int_0^\infty J(\omega) \mathcal{B}_{-k_1,k_2}(\omega) \dd \omega 
        = \frac{1}{\pi} \int_0^\infty J(\omega) \mathcal{B}_{k_1,k_2}(\omega) \dd \omega
        = \mathbf{B}_{k_1,k_2}.
    \end{split}
\end{equation}
Therefore, the rows of $\mathbf{B}$ are symmetric with respect to the middle row. 
$\mathbf{B}$ can be reduced to its row echelon form of order no more than $N+1$ using Gaussian elimination. 
Consequently, $r\leqslant N+1$.

For the case when $J$ is given by \cref{eq_spectral_density}, the matrix $\mathbf{B}$ in \eqref{eq_tpc_matrix_fftpc_matrix} can be written as the sum of $\mathcal{B}(\omega)$,
\begin{equation}
    \begin{aligned}
        \mathbf{B} & = \sum_{j=1}^L \frac{c_j^2}{2\omega_j}\mathcal{B}(\omega_j)
    \end{aligned}
\end{equation}
Given that each $\mathcal{B}(\omega_j)$ is of rank 2 in \cref{lemma_rank2}, it follows that $r\leqslant 2L$.
\end{proof}
Here is the proof of \Cref{thm_low_rank}:
\begin{proof}
By the definition of $\mathcal{B}$ \cref{eq_fixed_frequency_tpc_matrix},
 each component of $\mathcal{B}$ is a smooth function with respect to $\omega$.
Applying Taylor theorem to $\mathcal{K}(\omega) = J(\omega)\mathcal{B}(\omega)$ yields
\begin{equation}
\begin{split}
    \mathfrak{B}_1 - \mathfrak{B}_2
    &= \frac{1}{\pi} \int_{\omega_j}^{\omega_{j+1}} J(\omega) \mathcal{B}(\omega) - J(\bar{\omega}) \mathcal{B}(\bar{\omega}) \dd \omega \\
    &= \frac{1}{\pi} \int_{\omega_j}^{\omega_{j+1}}
        \mathcal{K}'(\bar{\omega}) (\omega - \bar{\omega})
        + \frac{1}{2}\mathcal{K}''(\eta(\omega)) (\omega-\bar{\omega})^2
    \dd \omega
\end{split}
\end{equation}
with $\eta(\omega)\in[\omega_{j},\omega_{j+1}]$.
Therefore,
\begin{equation}
    \Vert \mathfrak{B}_1 - \mathfrak{B}_2 \Vert_F
    \leqslant
    C \left(\omega_{j+1}-\omega_j\right)^3
\end{equation}
with $\displaystyle C = \frac{1}{24\pi} \max_{\omega\in[\omega_j,\omega_{j+1}]} \Vert \mathcal{K}''\Vert_F$.
\end{proof}

\section{An example illustrating construction of BIF-TT}
In this appendix, we show our method to construct BIF-TT for $m=8$ from the one for $m=6$.
Their relation is given by
\begin{equation}
\label{eq_L8_extend}
    \begin{split}
        \tensor{L}_8 
        =\ & \mathbb{TT}(\textbf{B},8,[1,3])
        \odot
        \mathbb{TT}(\tensor{L}_6,8,[2,4,5,6,7,8]) \\
        +\ &\mathbb{TT}(\textbf{B},8,[1,4])
        \odot
        \mathbb{TT}(\tensor{L}_6,8,[2,3,5,6,7,8]) \\
        +\ & \mathbb{TT}(\textbf{B},8,[1,5])
        \odot
        \mathbb{TT}(\tensor{L}_6,8,[2,3,4,6,7,8]) \\
        + \ & \mathbb{TT}(\textbf{B},8,[1,6])
        \odot
        \mathbb{TT}(\tensor{L}_6,8,[2,3,4,5,7,8]) \\
        + \ & \mathbb{TT}(\textbf{B},8,[1,7])
        \odot
        \mathbb{TT}(\tensor{L}_6,8,[2,3,4,5,6,8]) \\
        + \ & \mathbb{TT}(\textbf{B},8,[1,4])
        \odot
        \mathbb{TT}(\textbf{B},8,[2,7])
        \odot
        \mathbb{TT}(\textbf{B},8,[3,5])
        \odot
        \mathbb{TT}(\textbf{B},8,[6,8]) \\
        + \ & \mathbb{TT}(\textbf{B},8,[1,4])
        \odot
        \mathbb{TT}(\textbf{B},8,[2,8])
        \odot
        \mathbb{TT}(\textbf{B},8,[3,6])
        \odot
        \mathbb{TT}(\textbf{B},8,[4,7]) \\
        + \ & \mathbb{TT}(\textbf{B},8,[1,5])
        \odot
        \mathbb{TT}(\textbf{B},8,[2,7])
        \odot
        \mathbb{TT}(\textbf{B},8,[3,8])
        \odot
        \mathbb{TT}(\textbf{B},8,[4,6]) \\
        + \ & \mathbb{TT}(\textbf{B},8,[1,5])
        \odot
        \mathbb{TT}(\textbf{B},8,[2,8])
        \odot
        \mathbb{TT}(\textbf{B},8,[3,6])
        \odot
        \mathbb{TT}(\textbf{B},8,[4,7]) \\
        + \ & \mathbb{TT}(\textbf{B},8,[1,5])
        \odot
        \mathbb{TT}(\textbf{B},8,[2,8])
        \odot
        \mathbb{TT}(\textbf{B},8,[3,7])
        \odot
        \mathbb{TT}(\textbf{B},8,[4,6]) \\
        + \ & \mathbb{TT}(\textbf{B},8,[1,6])
        \odot
        \mathbb{TT}(\textbf{B},8,[2,4])
        \odot
        \mathbb{TT}(\textbf{B},8,[3,8])
        \odot
        \mathbb{TT}(\textbf{B},8,[5,7]) \\
        + \ & \mathbb{TT}(\textbf{B},8,[1,6])
        \odot
        \mathbb{TT}(\textbf{B},8,[2,8])
        \odot
        \mathbb{TT}(\textbf{B},8,[3,5])
        \odot
        \mathbb{TT}(\textbf{B},8,[4,7])
    \end{split}.
\end{equation}
Each of the first five terms on the right side actually represents the sum of four terms in \cref{eq_bif_inchworm} because $\tensor{L}_4$ for $m=4$ contains 4 terms.
By taking out common factor, the total computational cost reduces in the step of constructing the bath influence functional tensor train.
If we use the naïve method based on \ref{steps_tt3},
 there are in total 27 terms in $\tensor{L}_8$.
In \cref{eq_L8_extend}, there are 12 terms, much fewer than the original method.







\bibliographystyle{elsarticle-num}
\bibliography{myBib_abbr}

\end{document}